\documentclass[12pt]{article}
\pdfoutput=1 
\usepackage[T1]{fontenc}
\usepackage[utf8]{inputenc}
\usepackage[british]{babel}
\usepackage{hyperref}
\usepackage{siunitx}
\usepackage{multicol}
\usepackage{graphicx}
\usepackage{amsmath}
\usepackage{amssymb}
\usepackage{multirow}
\usepackage{bbm}
\usepackage{bm}
\usepackage{authblk}
\usepackage{color}
\usepackage[round]{natbib}
\usepackage{fancyhdr} 
\usepackage{subfig}
\usepackage{float}
\usepackage{textgreek}
\usepackage{mathtools}
\usepackage{paralist}
\usepackage{adjustbox}
\usepackage{bbm}
\usepackage{booktabs,ctable,threeparttable}
\usepackage{enumitem}
\usepackage{amsthm}
\newcommand{\blind}{0}
\usepackage[paper=a4paper,margin=1in]{geometry}
\newgeometry{top=0.8in,bottom=1in,right=0.8in,left=0.8in}
\newcommand{\diagdots}[3][-25]{%
	\rotatebox{#1}{\makebox[0pt]{\makebox[#2]{\xleaders\hbox{$\cdot$\hskip#3}\hfill\kern0pt}}}}
\newcommand{\isEquivTo}[1]{\overset{#1}{\sim}}
\newcommand\norm[1]{\left\lVert#1\right\rVert}

\theoremstyle{plain} \newtheorem{theorem}{Theorem}
\theoremstyle{plain} \newtheorem{proposition}{Proposition}
\theoremstyle{plain} 
\theoremstyle{plain} 
\theoremstyle{plain} \newtheorem{corollary}{Corollary}
\theoremstyle{plain}\newtheorem{remark}{Remark}

\usepackage{paralist} % Used for the compactitem environment which makes bullet points with less space between them
\DeclareMathAlphabet{\mathpzc}{OT1}{pzc}{m}{it}

 \allowdisplaybreaks

 \date{\vspace{-5ex}}
\begin{document}
	\def\spacingset#1{\renewcommand{\baselinestretch}%
		{#1}\small\normalsize} \spacingset{1}
	%%%%%%%%%%%%%%%%%%%%%%%%%%%%%%%%%%%%%%%%%%%%%%%%%%%%%%%%%%%%%%%%%%%%%%%%%%%%%%
\if0\blind
{
	\title{\bf Simultaneous inference for linear mixed model parameters with an application to small area estimation}
	\author{Katarzyna Reluga\thanks{%Katarzyna Reluga is a Research Associate at the 
		Department of Statistical Sciences, University of Toronto, Canada. E-mail: katarzyna.reluga@utoronto.ca.},\;
	Mar\'{i}a Jos\'{e} Lombard\'{i}a\thanks{%Mar\'{i}a Jos\'{e} Lombard\'{i}a is a Professor at the 
	University of A Coru\~{n}a, CITIC, Spain. E-mail: maria.jose.lombardia@udc.es.}\;
	and\;Stefan Sperlich\thanks{%Stefan Sperlich is a Professor at the 
		University of Geneva, GSEM, Switzerland. E-mail: stefan.sperlich@unige.ch.\\
	{{The authors gratefully acknowledge support from} the Swiss National Science Foundation for the project  200021-192345, MINECO grants MTM2017-82724-R, MTM2014-52876-R and PID2020-113578RB-I00, the Xunta de Galicia (Grupos de Referencia Competitiva ED431C 2020/14 and Centro de Investigaci\'on del Sistema Universitario de Galicia ED431G 2019/01). {We would like to thank} the Galician Institute of Statistics (IGE) for the transfer of the data. Finally, we thank D. Flores, W. Gonzalez Manteiga, E. L\'opez-Vizca\'{\i}no, D. Morales, T. Schmid, N. Salvati and S. Ranjbar for helpful discussions. The computations were performed at University of Geneva on the Baobab cluster.}}
	}
	\maketitle
} \fi	

\if1\blind
{
	\bigskip
	\bigskip
	\bigskip
	\begin{center}
		{\LARGE\bf  Simultaneous inference for linear mixed model parameters with an application to small area estimation\par}
	\end{center}
} \fi
\begin{abstract}
\noindent
Over the past decades, linear mixed models have attracted considerable attention in various fields of applied statistics. They are popular whenever clustered, hierarchical or longitudinal data are investigated. Nonetheless, statistical tools for valid simultaneous inference for mixed parameters are rare. This is surprising because one often faces inferential problems beyond the pointwise examination of fixed or mixed parameters. For example, there is an interest in a comparative analysis of cluster-level parameters or subject-specific estimates in studies with repeated measurements. We discuss methods for simultaneous inference assuming a linear mixed model. Specifically, we develop simultaneous prediction intervals as well as multiple testing procedures for mixed parameters. They are useful for joint considerations or comparisons of cluster-level parameters. We employ a consistent bootstrap approximation of the distribution of max-type statistic to construct our tools. The numerical performance of the developed methodology is studied in simulation experiments and illustrated in a data example on household incomes in small areas.
\end{abstract}

\noindent%
\vfill

\newpage
	
\section{Introduction}\label{sec:intro}

The family of linear mixed effects models (LMMs) developed by \cite{henderson1950} has been extensively applied in the statistical analysis of clustered and longitudinal data
%\citep{tuerlinckx2006statistical, Jiang2007}
% Tuerlinckx et al 2006 is on generalised linear mixed models, not on linear mixed models
\citep{Jiang2007,verbekeMolenberghs2000} as well as for the treatment-level analysis in medicine \citep{FrancqEtal2019}. This modelling framework arises naturally in many fields such as environmental sciences, economics, medicine, etc. Under LMM one supposes that the extra between-cluster variation (or between-subject variation in longitudinal studies) is captured by cluster-specific random effects. %For longitudinal data, referring to subjects with repeated measurement, the subjects are those 'clusters'. 
Cluster-level parameters might be the most relevant part of the statistical analysis. In particular, they can be %cluster-specific 
modelled by random effects themselves, or more frequently, by mixed effects which are often linear combinations of fixed and random effects. Mixed parameters are particularly appealing in, among others, animal husbandry, ecology and small area estimation (SAE) \citep[see, for example, the monograph of][and the review of \citeauthor{TzavidisEtal2018}, \citeyear{TzavidisEtal2018}]{rao2015small}. %for a recent review.
The latter critically observed that although the resulting mixed parameter estimates
"are a set of numbers of identical definition and simultaneous interest" and that one should thus consider a simultaneous rather than a point estimation problem, the topic of "ensemble properties of small
area estimates (...) has been
largely overlooked". %\cite{TzavidisEtal2018} % As steps in this direction they only 
They mention benchmarking and rank estimation as examples in this direction. This is in line with our observation regarding related literature on Bayesian hierarchical models in which the authors examine constrained Bayes and triple-goal estimation \citep{gosh92,ShenLouis98}. %Note, however, %that most of 
However, most of this literature hardly considers LMM.    
To the best of our knowledge, beyond such constrained and rank estimation, simultaneous inference for mixed parameters is still missing. %, in particular for linear mixed model parameters. 
This is surprising given the utility of such inference in applied domains, for example, within public health centres carrying out studies on demographic groups, or when statistical offices report to policy makers for resource distribution.  
The existing pointwise inference or joint estimation of mixed parameters is not less 
relevant or useful; nevertheless simultaneous inference would provide a framework for formulating statistically valid statements about a set of mixed parameters.

%For example, 
Numerous national and regional governments as well as international organisations conduct studies on socio-economic conditions in order to implement targeted policy interventions. The European Union, World Bank and statistical institutes regularly draft reports on economic development and poverty across countries, regions and provinces. 
When looking at such regional estimates, practitioners often aim to simultaneously assess and compare them. Nevertheless, existing methods are often not suitable to carry out such assessments or comparisons on a sound statistical basis. 
As soon as one begins to formulate a comparative statement about the situation in several areas simultaneously, the area-wise (or cluster-wise) analysis is rendered statistically invalid by an additional variability arising from the joint consideration. Consider cluster-wise prediction intervals (CPI) for mixed parameters; the coverage probabilities of $100(1-\alpha)$ intervals refer to the mean across all clusters. This implies that, by construction, about $100\alpha$ percent of the provided intervals (sometimes more) do not contain the true parameter. In other words, each time a statistical institute publishes its estimates for all areas with prediction intervals, the latter fail in at least $100\alpha \%$ to contain the true value. The same holds true for multiple comparisons via testing. The aim of this paper is to develop statistical tools that fill this gap. The investigation of such methods is not only theoretically appealing, but also relevant for practitioners. 

We develop simultaneous prediction intervals (SPIs) and multiple testing (MT) procedures to disprove or support simultaneous hypotheses about certain  characteristics. More specifically, our main proposal is to use a max-type statistic for a set of mixed parameters. We then employ a bootstrap procedure to consistently approximate the distribution of this statistic. The latter permits us to recover a critical value to construct an operational SPI or conduct MT procedures. Despite the unquestionable
utility of such tools in the context of LMM, to the best of our knowledge, we are the first who investigate their
theoretical and empirical properties. Furthermore, we compare the performance of our method with
alternative simultaneous inference techniques that we adapt from regression and nonparametric
curve estimation. Last but not least, we analytically derive simultaneous intervals based on the volume-of-tube formula of \cite{weyl1939volume} to approximate the tail probabilities. Our mathematical investigation demonstrates that his technique is not operational in our context.

Our methods are different from those considered by \cite{sun1999confidence} or \cite{maringwa2008application} within the framework of longitudinal studies. They propose to apply, respectively, the volume-of-tube formula and Monte Carlo (MC) sampling to construct simultaneous bands for linear combinations of fixed effects only. In contrast, we investigate a more complex problem of examining mixed effects. Our proposal also differs from the derivation of \cite{krivobokova2010simultaneous} who employ a mixed model representation for penalized splines to construct uniform bands for one-dimensional regression curves. Contrary to us, the authors can use a simplified version of the volume-of-tube formula.  %suitable for one-dimensional problems. 
Our results are distinct from those of \cite{ganesh2009simultaneous} who constructs simultaneous Bayesian credible intervals for a linear combination of area-level parameters under the model of \cite{fay_herriot}. We consider a more general inferential problem in a broder class of LMM within the frequentist framework. Furthermore, our contribution to the area of MT is a practical methodology used under LMM for the first time. Employment of the max-type statistic might be considered as a complement to the study of \cite{kramlinger2018marginal} who examined chi-square statistics for constructing MT and confidence sets for mixed parameters. In the classical linear regression literature, %(without random effects), 
max-type and chi-square statistics have been considered as complements, and are both well established in the practitioners' toolbox. We believe that this is equally valid for mixed parameters. The former are more popular for SPI, %confidence sets as they can provide SPI% in Tukey's sense, 
whereas chi-square statistics are widely recognized for MT. %It is worth mentioning that 
Finally,
\cite{reluga2021} consider simultaneous inference for empirical best predictors under generalised linear mixed models, whereas our paper seeks to address simultaneous inference under LMM. %for the empirical best unbiased predictor (EBLUP). 
Our study examines the statistical properties of SPI in contrast to the literature that  investigates CPI.  Starting from the work of \cite{cox1975} and \cite{morris1983parametric}, researchers proposed numerous methods based on analytical derivations \citep[e.g.][]{basu2003empirical,kubokawa2010corrected,yoshimori2014second} and resampling  \cite[e.g.][]{hall2006parametric,chatterjee2008parametric}. However, CPI and SPI serve different purposes and are not alternatives to each other. 

%The remainder of the paper is organized as follows. 
In Section~\ref{sec:LMM} we introduce the modelling framework and the parameter of interest. The construction of SPI and the MT procedure making use of the max-type statistic is outlined in Section~\ref{sec:SPI}. In Section \ref{sec:bootstrap_SPI} we introduce bootstrap-based SPI and MT and prove their consistency. Section \ref{sec:other_methods} contains potential alternatives which we adapted to our setting. We investigate the finite sample performance of our method in Section \ref{sec:simulations}, and apply it to study
the household income in Galicia in Section \ref{sec:data_example}. Section \ref{sec:conclusions} contains final remarks and conclusions. Technical details are deferred to Appendix \ref{sec:appendix} and the supplementary material. The latter also includes the discussion of extensions of our method. 

%%%%%%%%%%%%%%%%%%%%%%%%%%
\section{Linear mixed model inference} \label{sec:LMM}

Consider a classical LMM formulation $\bm{y}=\bm{X\beta}+\bm{Z u}+\bm{e}$, where $\bm{X}$, $\bm{Z}$ are known, full column rank matrices for a fixed and a random part, $\bm{\beta}$ is a vector of fixed effects, $\bm{u}$ is a vector of random effects, and $\bm{e}$ denotes stochastic errors. It is common to assume $\bm{u}$ and $\bm{e}$ to be mutually independent with $\bm{u}\isEquivTo{ind} N_{q}{(\bm{0},\bm{G})}$ and 	$\bm{e}\isEquivTo{ind}  N_{n}{(\bm{0},\bm{R})}$. More specifically, consider a LMM with a block diagonal covariance matrix (LMMb):
\begin{equation}\label{eq:LMM_b}
	\bm{y}_d=\bm{X}_d\bm{\beta}+\bm{Z}_d\bm{u}_d+\bm{e}_d,\quad d=1,\dots,D,
\end{equation}
where $n_d$ is the number of units in the $d^{th}$ cluster (or area), $\bm{y}_d\in \mathbb{R}^{n_d}$, $\bm{X}_d \in \mathbb{R}^{n_d\times(p+1)}$ and $\bm{Z}_d\in\mathbb{R}^{n_d\times q_d}$. Here, $D$ is the number of clusters, $\bm{\beta}\in\mathbb{R}^{p+1}$ an unknown %common 
vector of regression coefficients, $\bm{u}_d\isEquivTo{ind}  N_{q_d}{(\bm{0},\bm{G}_d)}$ and 	$\bm{e}_d\isEquivTo{ind} \prescript{}{n_d}{(\bm{0},\bm{R}_d)}$, $n=\sum_{d=1}^D n_d$. We assume that 	$\bm{G}_d=\bm{G}_d(\bm{\theta})\in\mathbb{R}^{q_d\times q_d}$ and $\bm{R}_d=\bm{R}_d(\bm{\theta})\in\mathbb{R}^{n_d\times n_d}$ depend on variance parameters $\bm{\theta}=(\theta_1,...,\theta_h)^t$. LMM can be easily retrieved applying the notation of %introduced by
\cite{prased_rao}. %, p.\ 168. 
Under this setup, suppose that the variance-covariance $\bm{V}$ is nonsingular $\forall \theta_i$, $i=1,\dots,h$ with 
%\begin{equation*}\label{eq:moments_LMM}
$\mathbb{E}(\bm{y})=\bm{X\beta}$ and $\mathbb{V}\mathrm{ar}(\bm{y})=\bm{R} + \bm{Z G}\bm{Z}^t=\bm{V}(\bm{\theta})\eqqcolon\bm{V}$.
%\end{equation*} 
Two important examples of LMM that are extensively used, especially in SAE, are the \textit{nested error regression model} (NERM) of \cite{battese1988error}, and the \textit{Fay-Herriot model} (FHM) of \cite{fay_herriot}. The former is defined as
\begin{equation}\label{eq:NERM}
	y_{dj}=\bm{x}^t_{dj}\bm{\beta}+u_d+e_{dj},\quad d=1,\dots,D,\quad j=1,\dots,n_d,
\end{equation}
where $y_{dj}$ is the quantity of interest for the $j^{th}$ unit in the $d^{th}$ cluster, $\bm{x}_{dj}=(1,x_{dj1},\dots,x_{djp})^t$, $u_d\isEquivTo{iid}  N(0,\sigma^2_u)$ and $e_{dj}\isEquivTo{iid}  N(0,\sigma^2_e)$ for $d=1,\dots,D$, $j=1,\dots,n_d$. Here $\bm{y}_d=(y_{d1},\dots,y_{dn_d})$, $\bm{X}_d=\mathrm{col}_{1\leqslant j\leqslant n_d}\bm{x}^t_{dj}$, $q_d=1$, $Z_d=\bm{1}_{n_d}$ with $\bm{1}_{n_d}$ a $n_d$ vector of ones, $\bm{e}_d=(e_{d1},\dots,e_{dn_d})^t$, $\bm{\theta}=(\sigma_{e}^2,\sigma_u^2)^t$, $\bm{R}_d(\bm{\theta})=\sigma_{e}^2\bm{I}_{n_d}$ with $\bm{I}_{n_d}$ the $n_d\times n_d$ identity matrix and $\bm{G}_d(\bm{\theta})=\sigma_{u}^2$. In contrast, the FHM is often referred to as an {\sl area-level model} and consists of two levels. The model at level 1, called the sampling model, assumes that the direct estimators $y_d$ of a cluster mean $\mu_d^F$ are design unbiased, and satisfy  $y_{d}=\mu_d^F +e_{d}$, $e_{d}\isEquivTo{iid} N(0,\sigma^2_{e_{d}})$, $d=1,\dots,D$. Under FHM, the sampling variance $\sigma^2_{e_{d}}=\mathbb{V}ar(y_d|\mu_d^F)$ is supposed to be \textit{known} for each cluster $d$. On the other hand, the linking model at level 2 is $\mu_d^F=\bm{x}^t_{d}\bm{\beta}+u_d$, $ u_d\isEquivTo{iid}  N(0,\sigma^2_u)$, $d=1,\dots,D$,
where $\bm{x}_{d}=(1,x_{d1},\dots,x_{dp})^t$ is a $(p+1)$-vector of cluster-level auxiliary variables. Observe that the FHM can be rewritten as a LMM with $n_d=q_d=1$, $Z_d=1$, $\bm{\theta}=\sigma^2_u$,  $\bm{R}_d(\sigma^2_u)=\sigma^2_{e_{d}}$, that is
\begin{equation}\label{eq:FH}
	y_{d}=\bm{x}^t_{d}\bm{\beta}+u_d +e_{d},\quad d=1,\dots,D. \end{equation} 
Due to the data availability, the FHM is more frequently used in practice. While cluster-level information can be easily obtained (for example, using open access internet repositories), this is clearly not the case for units.

Assuming LMM, one is often interested in a simultaneous or comparative inference for general mixed parameters  
\begin{equation}\label{eq:mu}
	\mu_d=\bm{k}^t_{d}\bm{\beta}+\bm{m}^t_{d}\bm{u}_d,\quad d=1,\dots,D,
\end{equation}
with $\bm{k}_{d}\in \mathbb{R}^{p+1}$ and $\bm{m}_{d}\in \mathbb{R}^{q_d}$ known. %These parameters are essential in many domains, such as SAE. 
In our article %example below, 
$\mu_d$ is a cluster conditional mean, but other parameters %of interest 
can be explored as well. \cite{henderson} developed the best linear unbiased predictor (BLUP) of a linear combination of random %effects $\bm{u}$  %$\bm{\beta}$ 
and fixed effects
when %the covariance matrix 
$\bm{V}$ is completely known. Applying his idea one obtains BLUP estimator for \eqref{eq:mu} %, that is
%\begin{equation}\label{eq:mu_hat_tilde} 
%	\tilde{\mu}_d\coloneqq\tilde{\mu}_d(\bm{\theta})=\bm{k}^t_{d}\tilde{\bm{\beta}}+\bm{m}^t_{d}\tilde{\bm{u}}_d, 
%	\quad d=1,\dots,D,
%\end{equation}
that is $\tilde{\mu}_d\coloneqq\tilde{\mu}_d(\bm{\theta})=\bm{k}^t_{d}\tilde{\bm{\beta}}+\bm{m}^t_{d}\tilde{\bm{u}}_d$, %$d=1,\dots,D$, 
where $\bm{\theta}=(\theta_1,\dots,\theta_h)^t$, $\tilde{\bm{\beta}}=\tilde{\bm{\beta}}(\bm{\theta})=(\bm{X}^t\bm{V}^{-1}\bm{X})^{-1}\bm{X}^t\bm{V}^{-1}\bm{y}$, $\tilde{\bm{u}}_d=\tilde{\bm{u}}_d(\bm{\theta})=\bm{G}_d \bm{Z}^t_d \bm{V}_d^{-1}(\bm{y}_d-\bm{X}_d \tilde{\bm{\beta}})$. 
In practice $\bm{\theta}$ is usually unknown, hence one uses $\hat{\bm{\theta}}\coloneqq\hat{\bm{\theta}}(\bm{y})$ which yields the EBLUP
\begin{equation}\label{eq:mu_hat_e}
	\hat{\mu}_d := \hat{\mu}_d(\hat{\bm{\theta}})=
	\bm{k}^t_{d}\hat{\bm{\beta}}+\bm{m}^t_{d}\hat{\bm{u}}_d,\quad d=1,\dots,D,
\end{equation}
with $\hat{\bm{\beta}}=\hat{\bm{\beta}}(\hat{\bm{\theta}})$, $\hat{\bm{u}}=\hat{\bm{u}}(\hat{\bm{\theta}})$ and $\hat{\bm{\theta}}=(\hat{\theta}_1,\dots,\hat{\theta}_h)^t$. Having assumed certain conditions on the distributions of random effects and errors, as well as the variance components $\bm{\theta}$ (see Appendix \ref{sec:RC}), \cite{kackar_unbias} proved that the two-stage procedure provides an %(marginally) 
unbiased estimator for $\mu_d$.

To construct a studentized max-type statistic, it is important to assess the variability of prediction. The most common measure of uncertainty is the mean squared error $\mathrm{MSE}(\hat{\mu}_d)=\mathbb{E}(\hat{\mu}_d-\mu_d)$. Here, $\mathbb{E}$ denotes the expectation with respect to model \eqref{eq:LMM_b}. We can decompose the MSE into  
\begin{equation}\label{eq:MSE}
	\mathrm{MSE}(\hat{\mu}_d )=\mathrm{MSE}(\tilde{\mu}_d )+\mathbb{E}\left(\hat{\mu}_d 
	-\tilde{\mu}_d \right)^2+2\mathbb{E}\left\{(\tilde{\mu}_d -\mu_d)(\hat{\mu}_d -\tilde{\mu}_d )\right\}  , 
\end{equation}
where $\mathrm{MSE}(\tilde{\mu}_d )$ accounts for the variability when the variance components $\bm{\theta}$ are known. Assuming LMMb and $\bm{b}_d^t=\bm{k}^t_d-\bm{o}_d^t\bm{X}_d$ with $\bm{o}^t_d=\bm{m}^t_d\bm{G}\bm{Z}^t_d\bm{V}^{-1}_d$, the $\mathrm{MSE}(\tilde{\mu}_d)$ reduces to
\begin{equation}\label{eq:MSE_b_first_term}
	\bm{m}^t_d(\bm{G}_d-\bm{G}_d\bm{Z}^t_d \bm{V}_d^{-1}\bm{Z}_d\bm{G}_d)\bm{m}_d+\bm{b}_d^t\left(\sum_{d=1}^{D}\bm{X}^t_{d}\bm{V}^{-1}_d\bm{X}_d\right)^{-1}\bm{b}_d   =: g_{1d}(\bm{\theta}) + g_{2d}(\bm{\theta}),
\end{equation}
where $g_{1d}$ accounts for the variability of $\tilde{\mu}_d$ once $\bm{\beta}$ is known, and $g_{2d}$ for the estimation of $\tilde{\bm{\beta}}$. The second term in \eqref{eq:MSE} is intractable, but there exists a vast literature which deals with the estimation of it \citep[see][for a review]{rao2015small}. The third term disappears under normality of errors and random effects; it is rarely considered. Following \cite{chatterjee2008parametric}, we suggest the construction of a SPI using $\bm{g}_1(\hat{\bm{\theta}})=( g_{11}(\hat{\bm{\theta}}),\dots,g_{1D}(\hat{\bm{\theta}}))^t$, where $g_{1d}(\hat{\bm{\theta}})$ is defined in \eqref{eq:MSE_b_first_term} with $\bm{\theta}$ replaced by a consistent estimator. In fact, simulations studies in \cite{reluga2020thesis} indicate %along various simulation results it seems 
that alternative measurements of variability do not improve the performance of SPI based on the max-type statistic. %our methods .   

%%%%Contribution%%%%%
\section{SPI and MT for mixed parameters using max-type statistics}\label{sec:SPI}
%SPI
We concentrate on the construction of SPI and MT procedures for the mixed parameter in \eqref{eq:mu} considering a confidence region $\mathcal{I}_{1-\alpha}=\bigtimes_{d=1}^{D}\mathcal{I}_{d,1-\alpha}$ such that $P(\mu_d \in \mathcal{I}_{1-\alpha}\; \forall d\in[D])=1-\alpha$, $[D]=\{1,\dots,D\}$. This is equivalent to finding a critical value $c_{S_0}(1-\alpha)$ which satisfies
\begin{equation*}\label{eq:S_proba}
	\alpha={P}\left(\left\lvert  \frac{\hat{\mu}_d -
		\mu_d}{\hat{\sigma}(\hat{\mu}_d )} \right\rvert \geqslant c_{S_0}(1-\alpha) \; \text{for some } d\in[D]\right)   
	={P}\left(\max_{d=1,\dots, D}\left\lvert  \frac{\hat{\mu}_d -
		\mu_d}{\hat{\sigma}(\hat{\mu}_d )} \right\rvert \geqslant c_{S_0}(1-\alpha) \right),
\end{equation*}
where we denote by $\hat{\sigma}(\hat{\mu}_d )$ the estimated variability of $\hat{\mu}_d$ (for example, the estimated square root of $\mathrm{MSE}(\hat{\mu}_d)$). The critical value $c_{S_0}(1-\alpha)$ is in fact the $(1-\alpha)^{th}$-quantile of the studentized statistic	
\begin{equation}\label{eq:s}
	S_0\coloneqq	\max_{d=1,\dots,D}\left\lvert S_{0d}  \right\rvert,
	\mbox{  where}
	\quad S_{0d}=\frac{\hat{\mu}_d -
		\mu_d}{\hat{\sigma}(\hat{\mu}_d )}, \quad c_{S_0}(1-\alpha)\coloneqq \inf\{t\in \mathbb{R}: P(S_0\leqslant t)\geqslant 1 -\alpha\}.
\end{equation}
It follows that with probability $1-\alpha$, a region defined as
\begin{equation*}\label{eq:unif_band_S}
	\mathcal{I}^{S}_{1-\alpha}=
	\bigtimes_{d=1}^D \mathcal{I}^{S}_{d,1-\alpha}, \quad\text{where}\quad\mathcal{I}^{S}_{d,1-\alpha}=
	\left\{\hat{\mu}_d  \pm c_{S_0}(1-\alpha) \hat{\sigma}(\hat{\mu}_d ) \right\}, 
\end{equation*}
covers all mixed parameters. Since the probability density function (pdf) of $S_0$ is right skewed, we suggest to consider its upper quantile and construct symmetric $\mathcal{I}^{S}_{d,1-\alpha}$, $d\in [D]$. This approach can be regarded as a variation of the studentized maximum modulus method of %\cite{tukey1952various,
\cite{tukey1953u}. At this point, we formally define CPI to circumvent all possible doubts concerning its relation to SPI. Let 
$c_{d}(1-\alpha)\coloneqq \inf\{t\in \mathbb{R}: P(S_{0d}\leqslant t)\geqslant 1 -\alpha\}$. CPI is defined as 
\begin{equation*} \label{eq:ch2_iCI}
	\mathcal{I}^{CPI}_{d,1-\alpha} =  \left\{\hat{\mu}_d \pm c_{d}(1-\alpha) \times \hat{\sigma}(\hat{\mu}_d)\right\} \quad \forall d\in [D],
\end{equation*}
which covers $\mu_d$ with probability $1-\alpha$. %In fact, 
Due to the central limit theorem, the most common choice is $c_{d}(1-\alpha)=\Phi^{-1}(1-\alpha)$, that is, a quantile from a normal distribution. %MT
Furthermore, due to the correspondence between interval estimation and hypothesis testing, our methodology is %readily 
applicable for the latter. Consider a following pair of hypotheses:
\begin{equation}\label{eq:test_proc}
	H_0:\bm{A\mu}=\bm{h}\quad vs.\quad H_1:\bm{A\mu}\neq\bm{h},
\end{equation}
where $\bm{A}\in \mathbb{R}^{D'\times D}$ with $D'\leqslant D$ and $\bm{h}\in \mathbb{R}^{D'}$ is a vector of constants. A test based on a max-type statistic $t_{H}$ rejects $H_0$ at the $\alpha$-level if $t_{H}\geqslant c_{H_0}(1-\alpha)$ with $c_{H_0}(1-\alpha)\coloneqq \inf\{t\in \mathbb{R}:P(S_{H_0}\leqslant t)\geqslant 1-\alpha \}$,  
\begin{equation}\label{eq:mult_test_quant}
	t_{H}\coloneqq \max_{d=1,\dots, D'}\left \lvert t_{H_d} \right\rvert, \
	S_{H_0}\coloneqq	\max_{d=1,\dots,D}\left\lvert S_{H_0d}  \right\rvert, \
	t_{H_d} =	\frac{\hat{\mu}^H_d-h_d}{\hat{\sigma}(\hat{\mu}^H_d)}	\;\text{and}\;
	S_{H_0d}=\frac{\hat\mu^H_d - \mu_d^H}{\hat{\sigma}(\hat\mu^H_d )},
\end{equation}
where $\bm{\mu}^H=(\mu^H_1,\dots,\mu^H_{D'})^t=\bm{A}\bm{\mu}\in \mathbb{R}^{D'}$ and $\hat{\bm{\mu}}^H$ its estimated counterpart.
In other words, $\bm{h}\notin \mathcal{I}_{1-\alpha}^{H_0}$ 
%\begin{equation*}
%\mbox{ with  }\quad 
%\mathcal{I}_{1-\alpha}^{H_0}=
%\bigtimes_{d=1}^D %\mathcal{I}_{d,1-\alpha}^{H_0}  ,
%\quad\text{ where }\quad 
%\mathcal{I}_{d,1-\alpha}^{H_0}= %\left\{\hat{\mu}^H_d  \pm %c_{H_0}(1-\alpha) %\hat{\sigma}(\hat{\mu}^H_d )\right\}.
%\end{equation*}
with $\mathcal{I}_{1-\alpha}^{H_0}=
\bigtimes_{d=1}^D \mathcal{I}_{d,1-\alpha}^{H_0}$, where $ 
\mathcal{I}_{d,1-\alpha}^{H_0}= \left\{\hat{\mu}^H_d  \pm c_{H_0}(1-\alpha) \hat{\sigma}(\hat{\mu}^H_d )\right\}$.
In practice, a standard problem is to test for statistical differences between various clusters with respect to some characteristic. Our test is based on a single step procedure and exhibits a weak control of a family-wise error (FWER). If one aims at testing multiple hypotheses with a strong control of FWER, the step-down technique of \cite{romano2005exact} could be implemented. As this is beyond the scope of this paper, details and related simulation results are deferred to the supplementary material.

%Bootstrap procedure
\section{Bootstrap-based SPI and MT procedure} \label{sec:bootstrap_SPI}

It is challenging to estimate the distribution of $S_{0}$ in \eqref{eq:s} %$S_{H_0}$ in \eqref{eq:mult_test_quant} 
and to recover critical values because, among others, mixed effects $\mu_d$, $d=1,...,D$, are unknown.Nevertheless, an almost straightforward way to approximate critical value $c_{S_0}(1-\alpha)$ is to use a parametric bootstrap procedure which circumvents a direct application of the normal asymptotic distribution
\citep{gonzalez2008bootstrap}. It can also provide faster convergence \citep{hall2006parametric, chatterjee2008parametric}. 
Let $B$ be the number of bootstrap samples $(\bm{y}^{*(b)},\bm{X}, \bm{Z})$. The bootstrap analogue of expression in \eqref{eq:s} is
\begin{equation}\label{eq:s_crit_bbot}
	S^{*(b)}_B\coloneqq\max_{d=1,\dots,D}\left\lvert S^{*(b)}_{Bd} \right\rvert, \quad
	S^{*(b)}_{Bd} =\frac{\hat{\mu}^{*(b)}_d-\mu^{*(b)}_d} {\hat{\sigma}^*(\hat{\mu}^{*(b)}_d)}, \quad b=1,\dots, B.
\end{equation}
The critical value can be consistently approximated by the $(1-\alpha)^{th}$-quantile of \eqref{eq:s_crit_bbot}, i.e., $c_{BS}(1-\alpha)\coloneqq \inf\{t^*\in \mathbb{R}: P(S^*_B\leqslant t^*)\geqslant 1 -\alpha\}$.Consequently, the bootstrap SPI is defined as	
\begin{equation}\label{eq:unif_band_boot}
	\mathcal{I}^{BS}_{1-\alpha}=
	\bigtimes_{d=1}^D \mathcal{I}^{BS}_{d,1-\alpha},
	\quad \text{where}\quad  \mathcal{I}^{BS}_{d,1-\alpha}=\left\{\hat{\mu}_d  \pm c_{BS}(1-\alpha) \hat{\sigma}(\hat{\mu}_d )\right\}.
\end{equation}
Our choice of $\hat{\sigma}(\hat{\mu}_d)=\sqrt{g_{1d}(\hat{\bm{\theta}})}$ is motivated by the asymptotic analysis of \cite{chatterjee2008parametric}. Validity of the above bootstrap method is shown by adapting Theorem 3.1 of these authors (henceforth Theorem CLL, provided in the supplementary material) and combining it with
some results from the extreme value theory.  
\begin{proposition}\label{prop:con_max}
	Suppose that the assumptions of Theorem CLL and regularity conditions R.1-R.7 from Section \ref{sec:RC} hold. Then \begin{equation*}
		\sup\limits_{q\in\mathbb{R}}\left \lvert P^* (
		S^*_{B}\leqslant q)   -
		P (S_{0}\leqslant q)
		\right\rvert=o_P(1).
	\end{equation*}
\end{proposition}
An important implication of Proposition \ref{prop:con_max} is the coverage probability of $\mathcal{I}_{1-\alpha}^{BS}$.
\begin{corollary}\label{cor:convergence_interval}
	Under Proposition \ref{prop:con_max} it holds that 
	\begin{equation*}
		P\left( \mu_d \in \mathcal{I}^{BS}_{1-\alpha}\;\forall d\in[D]\right)\xrightarrow[]{D\rightarrow \infty} 1-\alpha .
	\end{equation*}
\end{corollary}
According to the theoretical developments for max-type statistics, the Kolmogorov distance defined in Proposition \ref{prop:con_max}  converges to $0$ at best at polynomial rate $(\log(\cdot))^{c_2}/n^{c_3}$, where $(\cdot)$ is the number of parameters for which we wish to obtain the maximum (in our case $D$), and $c_2, c_3$ some constants, see e.g. \cite{chernozhukov2013gaussian}. We are not aware of results for max-type statistics that one could employ to obtain second order correctness without such $\log(\cdot)$ term. Observe that in our setting $n\rightarrow\infty$ is equivalent to $D\rightarrow \infty$, because we assumed that $n_d$ is bounded (see Appendix \ref{sec:RC}). Certainly, our result would still hold if both, $n_d$ and $D$ grow. In case of a fixed $D$, we would replace Proposition \ref{prop:con_max} using explicitly a variation of studentized maximum modulus distribution \citep[cf.][]{stoline1979tables}.

%%%%%%%%%% Test
Due to the relation between interval estimation and tests, the methodology developed for SPI can be used to find a critical value for MT procedure in  \eqref{eq:test_proc}. In particular, consider slightly modified bootstrap statistics
\begin{eqnarray*}  &&
	S^{*(b)}_{BH_0}\coloneqq\max_{d=1,\dots,D}\left\lvert S^{*(b)}_{BH_0d} \right\rvert, \quad
	S^{*(b)}_{BH_0d} =\frac{\hat{\mu}^{*H(b)}_d-\mu^{*H(b)}_d} {\hat{\sigma}^*(\hat{\mu}^{*H(b)}_d)},
\end{eqnarray*}
with 
%\begin{equation*}
$\bm{\mu}^{*H(b)}=(\mu^{*H(b)}_1,\dots,\mu^{*H(b)}_{D'})^t\coloneqq\bm{A}\bm{\mu}^{*(b)} \in \mathbb{R}^{D'}$, and its estimated versions 
\begin{equation*}
	\hat{\bm{\mu}}^{*H(b)}=(\bm{a}^t_1(\bm{k}^t_{1}\hat{\bm{\beta}}^{*(b)}+\bm{m}^t_{1}\hat{\bm{u}}_1^{*(b)}),\dots,
	\bm{a}^t_D (\bm{k}^t_{D}\hat{\bm{\beta}}^{*(b)}+\bm{m}^t_{D}\hat{\bm{u}}_D^{*(b)}) )^t\coloneqq\bm{A}\hat{\bm{\mu}}^{*(b)},
\end{equation*}
where $\bm{a}_d\in\mathbb{R}^D$ are the rows of $\bm{A}$. These are applied to find a bootstrap approximation for the critical value $c_{H_0}(1-\alpha)$ of our test, namely $c_{BH_0}(1-\alpha)\coloneqq \inf\{t\in \mathbb{R}:P(S^*_{BH_0}\leqslant t)\geqslant 1-\alpha \}$. It is worth mentioning that we do not need to generate bootstrap samples under $H_0$ to obtain the critical values of our test.

%%%%%%%%%%%%%%%%%%%%%%%%%%%%%%%%%%%%
In Section \ref{sec:LMM} we defined NERM and FHM as popular examples of LMM. We describe a parametric bootstrap procedure that yields promising results when constructing SPI under these models.
Under NERM and FHM, we use simplified versions of $g_{1d}$ in \eqref{eq:MSE_b_first_term} derived by \cite{prased_rao} as the estimators of $\hat{\sigma}^2 (\hat{\mu}_d)$: $g^N_{1d}(\hat{\bm{\theta}})=\hat{\sigma}^2_{u}/(\hat{\sigma}^2_{u}+\hat{\sigma}^2_{e}/n_d)(\hat{\sigma}^2_{e}/n_d)$ for NERM and $g^F_{1d}(\hat{\bm{\theta}})=\hat{\sigma}^2_{u}\sigma^2_{e_d}/(\hat{\sigma}^2_{u}+\sigma^2_{e_d})$ for FHM. Under NERM the bootstrap algorithm is 
\begin{enumerate}
	\setlength\itemsep{-0.5em}
	\item From the original sample, obtain consistent estimators $\hat{\bm{\beta}}$ and $\hat{\bm{\theta}}=(\hat{\sigma}^2_e, \hat{\sigma}^2_u)$.
	\item Generate $D$ independent copies of $W_1\sim N(0,1)$. Construct $\bm{u}^*=(u^*_1,u^*_2,\dots,u^*_D)$ with 
	$u^*_d=\hat{\sigma}_uW_1$, $d=[D]$. 
	\item Generate $n$ independent copies of $W_2\sim N(0,1)$. Construct $\bm{e}^*=(e^*_1,e^*_2,\dots,e^*_n)$ with $e^*_j=\hat{\sigma}_{e}W_2$, $j=[n]$.  
	\item Create a bootstrap sample $\bm{y}^*=\bm{X}\hat{\bm{\beta}}+\bm{u}^*+\bm{e}^*$.
	\item Fit the model to the bootstrap sample and obtain bootstrap estimates $\hat{\bm{\beta}}^*$, $\hat{\bm{\theta}}^*=(\hat\sigma^{2*}_e,\hat\sigma^{2*}_u)$,
	$\mu_{dj}^*$ and $\hat{\mu}_{dj}^*$.
	\item Repeat Steps 2-5 $B$ times. Calculate $S^{*(b)}_B$, $b=1,\dots,B$, using $g^{N*(b)}_{1d}(\hat{\bm{\theta}}^{*(b)})$ to  
	obtain $c_{BS}(1-\alpha)$ and $\mathcal{I}^{BS}_{1-\alpha}$. 
	% as defined in Section \ref{sec:bootstrap_SPI}. 
\end{enumerate}
Here, $g^{N*(b)}_{1d}(\hat{\bm{\theta}}^{*(b)}) = \hat{\sigma}^{2*(b)}_{u}/(\hat{\sigma}^{2*(b)}_{u} + \hat{\sigma}^{2*(b)}_{e}/n_d)(\hat{\sigma}^{2*(b)}_{e}/n_d)$ is the bootstrap equivalent of $g^N_{1d}(\hat{\bm{\theta}})$. To implement the analogous bootstrap under FHM, we need to slightly modify step 1 defining $\hat{\bm{\theta}}=\hat{\sigma}^2_u$ and  $g^{F{*(b)}}_{1d}(\hat{\bm{\theta}}^{*(b)})=\hat{\sigma}^{2*(b)}_{u}\sigma^2_{e_d}/(\hat{\sigma}^{2*(b)}_{u}+\sigma^2_{e_d})$. Additionally, we need to replace step 3 by
\begin{enumerate}%[topsep=0pt]
	\item[3.'] Generate $D$ independent copies of a variable  $W_2\sim N(0,1)$. Construct vector $\bm{e}^*=(e^*_1,e^*_2,\dots,e^*_D)$ with elements 
	$e^*_d={\sigma}_{e_d}W_2$, $d=[D]$. 
\end{enumerate}

The parametric bootstrap algorithm can be modified to accommodate more complex models, for example with spatial or temporal correlation, %and structures 
by adapting accordingly the process of generating errors and random effects. An indisputable advantage of the bootstrap approach is its generality. As soon as we can mimic a data generating process for the assumed model, it can be implemented and applied to construct SPI and carry out MT for any kind of estimator. In addition, bootstrap SPI are relatively robust to model misspecifications (see results in Table \ref{tab:simulations_chi_dist}), in particular when the number of units in each cluster grows. This is in alignment with related remarks of \cite{jiang1998asymptotic}. On the other hand, it is usually more computer intensive than an analytical derivation.

We conclude this section with a practical extension of our results. In the testing problem (\ref{eq:test_proc}) we have already allowed for a scenario where only $D' < D$ hypotheses were considered, even though all data were used to estimate fixed parameters and predict random effects. Similarly, one might be interested in the construction of SPI with a joint coverage probability for a subset of $D'<D$ clusters. Without loss of generality, we assume that our goal is to construct SPI for the first $D'$ cluster-level mixed parameters. Then, in the definition of $S_{0}$ in \eqref{eq:s} and $S^*_B$ \eqref{eq:s_crit_bbot} one replaces $\max_{d=1,\dots, D}$ by $\max_{d=1,\dots, D'}$ and proceeds along the same lines as for $D$ areas. If $D'=O(D)$, we can evoke the same results from the extreme value theory as in case of Proposition \ref{prop:con_max} to prove a result similar to Corollary \ref{cor:convergence_interval}, that is:
%in all expressions. Our procedure would else stay the same, producing somewhat narrower SPI %for which we can state:

\begin{corollary}\label{cor:subset_interval}
	Let $D'<D$, $d\in[D']$, $D'=O(D)$. Consider $S^{*(b)}_{B'}\coloneqq\max_{d=1,\dots,D'}\left\lvert S^{*(b)}_{Bd} \right\rvert$, $c_{B'S}(1-\alpha)\coloneqq \inf\{t^*\in \mathbb{R}: P(S^*_{B'}\leqslant t^*)\geqslant 1 -\alpha\}$ where $S^{*(b)}_{Bd}$ as defined in \eqref{eq:s_crit_bbot}, and 
	%\begin{equation}\label{eq:unif_band_boot}
	$\mathcal{I}^{B'S}_{1-\alpha}=
	\bigtimes_{d=1}^{D'} \mathcal{I}^{B'S}_{d,1-\alpha}$ with $\mathcal{I}^{B'S}_{d,1-\alpha}=\left\{\hat{\mu}_d  \pm c_{B'S}(1-\alpha) \hat{\sigma}(\hat{\mu}_d )\right\}$. Then, under  Proposition \ref{prop:con_max} it holds that 
	\begin{equation*}
		P\left( \mu_d \in \mathcal{I}^{B'S}_{1-\alpha}\;\forall d\in[D']\right)\xrightarrow[]{D'\rightarrow \infty} 1-\alpha .
	\end{equation*}
\end{corollary}

%%%%%%%%%%%%%%%%%%%%%%%

\section{Alternative methods for SPI and MT}\label{sec:other_methods}

Although, to the best of our knowledge, we are the first %As said, to the best of our knowledge 
who introduce SPI and MT procedures for mixed parameters, various approaches have been put forward to tackle the problem of simultaneous confidence bands for linear regression surfaces. For example,  Bonferroni t-statistics are a straightforward tool to compare a set of fixed parameters. Furthermore, other authors, such as \cite{working1929applications}, \cite{scheffe1953method}, \cite{sun1994simultaneous} or \cite{beran1988balanced}, to mention a few, developed equally important methodologies for the simultaneous inference of fixed parameters. In the rest of this section, we adapt some of the methods from the linear regression and nonparametric curve estimation to our setting. One could thus treat them as alternative approaches to our proposal. 

\subsection{The volume-of-tube procedure}\label{sec:vot_sec}

Consider LMMb defined in \eqref{eq:LMM_b}. One way to obtain BLUP estimates for $\bm{\beta}$ and $\bm{u}$ is to solve the mixed model equations of \cite{henderson1950}:
\begin{equation}
	\label{eq:mom_eq}
	\begin{bmatrix}
		\bm{X}^t\bm{R}^{-1}\bm{X}    & \bm{X}^t\bm{R}^{-1}\bm{Z}  \\
		\bm{Z}^t\bm{R}^{-1}\bm{X}    & \bm{Z}^t\bm{R}^{-1}\bm{Z} + \bm{G}^{-1}  \\
	\end{bmatrix}
	\begin{bmatrix}
		\tilde{\bm{\beta}}\\
		\tilde{\bm{u}}
	\end{bmatrix}
	=
	\begin{bmatrix}
		\bm{X}^t\bm{R}^{-1}\bm{y}\\
		\bm{Z}^t\bm{R}^{-1}\bm{y}
	\end{bmatrix}
\end{equation}
which can be re-expressed in the following simplified form:
% . \end{equation*} This can be expressed as 
%\begin{equation}
%\qquad \Leftrightarrow \qquad 
\begin{equation}
	\label{eq:mom_eq2}
	\bm{K}\tilde{\bm{\phi}}=\bm{C}^t\bm{R}^{-1}\bm{y}, \quad \text{where} \quad
	\bm{K}=\bm{C}^t\bm{R}^{-1}\bm{C}+\bm{G}^{+}, 
	\quad 
	\bm{G}^{+}=\begin{bmatrix}\bm{0}_{(p+1)\times (p+1)} & \bm{0}_{(p+1)\times D}\\\bm{0}_{D\times (p+1)} & \bm{G}^{-1}_{D\times D}\end{bmatrix},
\end{equation}
with $\tilde{\bm{\phi}}=\left(\tilde{\bm{\beta}}^t,\tilde{\bm{u}}^t\right)^t$, $\bm{C}=\left[\bm{X}\:\bm{Z}\right]$. %$\bm{K}=\bm{C}^t\bm{R}^{-1}\bm{C}+\bm{G}^{+}$ with $\bm{G}^{+}=\begin{bmatrix}\bm{0}_{(p+1)\times (p+1)} & \bm{0}_{(p+1)\times D}\\\bm{0}_{D\times (p+1)} & \bm{G}^{-1}_{D\times D}\end{bmatrix}$. 
From \eqref{eq:mom_eq} and \eqref{eq:mom_eq2} we obtain a straightforward formula for the estimates $\tilde{\bm{\phi}}=\bm{K}^{-1}\bm{C}^t\bm{R}^{-1}\bm{y}$ and $\hat{\bm{\phi}}=\tilde{\bm{\phi}}(\hat{\bm{\theta}})$. For some $\bm{x}=\left(1,x_1,\dots,x_{p}\right)^t$ with $x_1,\dots,x_p\in \mathcal{X}\subset \mathbb{R}^{p}$, $\bm{z}=\left(z_{1},\dots,z_q\right)^t\in\mathcal{Z}\subset \mathbb{R}^q$ and $\bm{c}=(\bm{x}^t,\bm{z}^t)^t\in\mathcal{X}\times\mathcal{Z}\eqqcolon\mathcal{C}$ one has $\bm{x}^t\tilde{\bm{\beta}}+\bm{z}^t\tilde{\bm{u}}= \bm{c}^t\tilde{\bm{\phi}}\equiv\bm{l}(\bm{x},\bm{\theta})^t\bm{y}=\sum_{i=1}^{n}l_i(\bm{x},\bm{\theta})y_i$ where $\bm{l}(\bm{x},\bm{\theta})^t=(l_1(\bm{x},\bm{\theta}),l_2(\bm{x},\bm{\theta}),\dots,l_n(\bm{x},\bm{\theta}))
=\bm{c}^t(\bm{C}^t\bm{R}^{-1}\bm{C}+\bm{G}^{+})^{-1}\bm{C}^t\bm{R}^{-1} $ i.e., $\bm{l}(\bm{x},\bm{\theta})$ is an $n$-vector. In addition, BLUP fitted values are $\tilde{\bm{y}}=\bm{L}\bm{y}$ where $\bm{L}=\bm{C}(\bm{C}^t\bm{R}^{-1}\bm{C}+\bm{G}^{+})^{-1}\bm{C}^t\bm{R}^{-1}$ which can be considered as a ridge regression formulation of the BLUP. Having reformulated the LMMb, we extend the approach of  \cite{sun1999confidence} and \cite{krivobokova2010simultaneous}. To simplify the notation, let $\bm{R}=\sigma^2_e\bm{I}_n$ which leads to  $\bm{l}(\bm{x},\bm{\theta})^t=\bm{c}^t\left(\bm{C}^t\bm{C}+\sigma_{e}^{2}\bm{G}^{+}\right)^{-1}\bm{C}^t$. Assuming normality for errors and random effects one obtains 
\begin{equation}\label{eq:Z}
	Z=\frac{\bm{c}^t\left(\tilde{\bm{\phi}}-\bm{\phi}\right)}{\sqrt{\mathbb{}\mathbb{V}\mathrm{ar} \left\{\bm{c}^t\left(\tilde{\bm{\phi}}-\bm{\phi}\right)\right\} }}
	=\frac{\bm{c}^t\left(\tilde{\bm{\phi}}-\bm{\phi}\right)}{\sqrt{\sigma_e^2 \bm{c}^t \left(\bm{C}^t\bm{C}+\sigma_{e}^{2}\bm{G}^{+}\right)^{-1} \bm{c} }}\sim\mathrm{N}(0,1).
\end{equation}
We conclude that $Z$ is a nonsingular Gaussian random variable with mean 0 and variance 1. Consequently, the following expressions can be retrieved from equation \eqref{eq:Z}:
\begin{align*}\label{eq:l_M_e_M}
	\bm{l}_M(\bm{x},\bm{\theta})\coloneqq\left(\bm{C}^t\bm{C}+\sigma^{2}_e\bm{G}^{+}\right)^{-1/2} \bm{c}, 
	& & 
	\norm{\bm{l}_M(\bm{x},\bm{\theta})}^2=\bm{c}^t \left(\bm{C}^t\bm{C}+\sigma^{2}_e\bm{G}^{+}\right)^{-1} \bm{c},  
	& & \bm{e}_M(\bm{x},\bm{\theta})\coloneqq\left(\bm{C}^t\bm{C}+\sigma^{2}_e\bm{G}^{+}\right)^{1/2}\left(\tilde{\bm{\phi}}-\bm{\phi}\right).
\end{align*}
The problem of finding a $(1-\alpha )$ SPI for $\hat{\mu}_d $ \ $\forall d\in [D]$ boils down to the choice of a critical value $c_{VT}(1-\alpha)$ such that
\begin{equation}\label{eq:alpha_first}
	\alpha=P\left(\left\lvert\hat{\mu}_d  - \mu_d\right\lvert \geqslant c_{VT}(1-\alpha)\hat{\sigma}_{e}\norm{\bm{l}_M(\bm{x},\hat{\bm{\theta}})},\text{ for some }\bm{c}\in \mathcal{C}\right).
\end{equation}
Let $\bm{l}_M=\bm{l}_M(\bm{x},\bm{\theta})$, $\hat{\bm{l}}_M=\bm{l}_M(\bm{x},\bm{\hat{\theta}})$, $\bm{e}_M=\bm{e}_M(\bm{x},\bm{\theta})$, $\hat{\bm{e}}_M=\bm{e}_M(\bm{x},\hat{\bm{\theta}})$ and $\lambda_M=\hat{\bm{e}}_M-\bm{e}_M$. We state Proposition \ref{prop:P_tube} for $p=2$, whereas other cases are considered in the supplementary material. 
\begin{proposition}\label{prop:P_tube}
	\setcounter{proposition}{2}
	Suppose that $\sigma^2_e$ is estimated by some consistent estimator. Define $\mathcal{Q}= \bm{l}_M/\norm{\bm{l}_M}$, $\xi=\inf\limits_{\bm{c}\in\mathcal{C}}||\hat{\bm{l}}_M||/||\bm{l}_M||$ and 
	$\eta=\sup\limits_{\bm{c}\in\mathcal{C}} |(\hat{\bm{l}}_M-\bm{l}_M)^t\bm{e}_M + \hat{\bm{l}}^t_M(\hat{\bm{e}}_M-\bm{e}_M) |/ \sigma_e||\bm{l}_M||$. 
	%\begin{align*}\label{eq:constants}
	%\mathcal{Q}= \frac{\bm{l}_M}{\norm{\bm{l}_M}},& &
	%& & \xi=\inf\limits_{\bm{c}\in\mathcal{C}}\frac{||\hat{\bm{l}}_M||}{||\bm{l}_M||} & & \text{and}  
	%& & \eta=\sup\limits_{\bm{c}\in\mathcal{C}}\frac{\left|\left(\hat{\bm{l}}_M-\bm{l}_M\right)^t\bm{e}_M + \hat{\bm{l}}^t_M(\hat{\bm{e}}_M-\bm{e}_M) \right|}{ \sigma_e||\bm{l}_M||}.
	%\end{align*}
	We assume that $\exists$ $\xi_{0}>0$, $\eta_{0}>0$ such that $P(\xi\leqslant\xi_{0})=o(\alpha)$ and $P(\eta\leqslant\eta_{0})=o(\alpha)$ as $n\rightarrow\infty$ and $\alpha\rightarrow 0$. Thus one can approximate \eqref{eq:alpha_first} as follows: 
	\begin{equation}
		\label{eq:vot_alpha}
		\alpha \leqslant Q\{c_{VT}(1-\alpha), \eta_0, \kappa_0, \xi_{0}\}  + 2\mathcal{E} P\left\{|t_v|> c_{VT}(1-\alpha) \xi_0\right\}, 
	\end{equation}
	where $t_{\nu}$ is a t-distributed random variable with $\nu$ degrees of freedom, $\kappa_0=\int_{\bm{c}\in\mathcal{C}}^{}\norm{\mathcal{Q}'(\bm{x})}d{\bm{x}}$ the volume of the manifold $\mathcal{M}=\left\{\mathcal{Q}(\bm{c}),\bm{c}\in \mathcal{C}\right\}$, and $\zeta_0$ the boundary area of $\mathcal{M}$. Finally, $\mathcal{E}$ is the Euler-Poincar\'{e} characteristic of $\mathcal{M}$. 
\end{proposition}
Due to its limited practical relevance, the proof of Proposition \ref{prop:P_tube} 
and a spelled-out expression for function $Q$ in \eqref{eq:vot_alpha} are deferred to our supplementary material. By assuming $\bm{R}=\sigma^2_e\bm{I}_n$ our proposal does not suffer from the loss of generality if we can write $\bm{R}=\sigma^2_e\bm{R}_{e}$, where $\bm{R}_{e}$ is some positive definite matrix. Then it follows that $\bm{l}(\bm{x},\bm{\theta})=\bm{c}^t(\bm{C}^t\bm{R}^{-1}\bm{C}+\bm{G}^{+})^{-1}\bm{C}^t\bm{R}^{-1}=\bm{c}^t(\bm{C}^t\bm{R}_e^{-1}\bm{C}+\sigma^2_e\bm{G}^{+})^{-1}\bm{C}^t\bm{R}_e^{-1}$, and we can still use the distribution of the ratio of $\hat{\sigma}_e/\sigma_e$ which is essential in the derivation of the confidence bands. In case of FHM, $\bm{R}=\mathrm{diag}(\sigma^2_{e_d})$ is assumed to be known, but in practice we could estimate it from other sources. Thus the derivation could be applied as well.  

Having retrieved the critical value, one could construct the volume-of-tube SPI using $\mathcal{I}^{VT}_{1-\alpha}=
\bigtimes_{d=1}^D \mathcal{I}^{VT}_{d,1-\alpha}$, where  
%\quad \text{where}\quad 
$\mathcal{I}^{VT}_{d,1-\alpha}= \left\{\hat{\mu}_d  \pm  c_{VT}(1-\alpha)\hat{\sigma}_{e}||\hat{\bm{l}}_M||\right\}$.
%\begin{equation*}\label{eq:unif_band_VoT}
%\mathcal{I}^{VT}_{1-\alpha}=
%\bigtimes_{d=1}^D %\mathcal{I}^{VT}_{d,1-\alpha}, 
%\quad \text{where}\quad 
%\mathcal{I}^{VT}_{d,1-\alpha}= \left\{\hat{\mu}_d  \pm  c_{VT}(1-\alpha)\hat{\sigma}_{e}||\hat{\bm{l}}_M||\right\}.
%\end{equation*}
The approximation in Proposition \ref{prop:P_tube} is conservative, i.e., the coverage probability is higher than the nominal $1-\alpha$; %(although still lower than for Sheff\'{e}'s bands); 
it approaches $1-\alpha$ as $\alpha\rightarrow 0$ and $n\rightarrow\infty$. Similarly as in in Section \ref{sec:bootstrap_SPI}, the latter assumption is equivalent to $D\rightarrow\infty$ in our setting. The results are valid under different asymptotic regimes too, if we are able to consistently estimate the  variance parameters, in particular $\sigma^2_e$. We immediately see that %the approximation formulas 
function $Q$ in \eqref{eq:vot_alpha} contains several constants. Numerical approximation of $\kappa_0$ which describes the geometry of the manifold $\mathcal{M}$ may not pose a major problem, but it is not clear how to estimate $\xi_0$ and $\eta_0$ under LMMb. Some ideas were derived for simpler one-dimensional models. \cite{sun1999confidence} proposes a derivative and a perturbation method to estimate constant $\xi_{0}$, while \cite{sun1994simultaneous} suggest estimating $\eta_0$ nonparametrically. It is unclear, though, how to extend their implementations to the LMM setting. Bootstrap approximation can be regarded as an alternative. However, in this case it would be easier to use bootstrap directly as described in Section \ref{sec:bootstrap_SPI}. Finally, the application of the volume-of-tube formula results in two sources of errors; from the approximation itself and from the estimation of the constants, making the approximation less reliable.

\subsection{Monte Carlo procedure}\label{sec:MC_SPI}
To %get around 
deal with the problem of approximating unknown constants in \eqref{eq:vot_alpha},  
consider mixed model equations in \eqref{eq:mom_eq} and \eqref{eq:mom_eq2} in Section \ref{sec:vot_sec}. When using LMM for spline regression, \cite{ruppert2003semiparametric} proposed a simple numerical approach to construct confidence bands of one-dimensional nonparametric curves by the empirical approximation of \eqref{eq:Z}, that is:
\begin{equation}\label{eq:rand_norm_app}
	\begin{bmatrix}
		\hat{\bm{\beta}}-\bm{\beta} \\
		\hat{\bm{u}}-\bm{u} 
	\end{bmatrix}  %\isEquivTo{approx} 
	\approx
	N\left\{ \bm{0},\left(\bm{C}^t\hat{\bm{R}}^{-1}\bm{C}+\hat{\bm{G}}^{+}\right)^{-1}\right\}.
\end{equation}
Likewise, we apply expression \eqref{eq:rand_norm_app} to simulate the distribution of $S_0$ in \eqref{eq:s}, and set
\begin{equation*}\label{eq:hat_s}
	S_0=\max_{d=1,\dots,D}\left\lvert S_{0d} \right\rvert
	\approx \max_{d=1,\dots,D}  \frac {\left|\bar{\bm{c}}_d^t
		\begin{bmatrix}
			\hat{\bm{\beta}}-\bm{\beta} \\
			\hat{\bm{u}}_d-\bm{u}_d 
		\end{bmatrix} \right|}{\hat{\sigma}(\hat{\mu}_d )} \eqqcolon 	
	\max_{d=1,\dots,D}\left\lvert S_{MCd} \right\rvert=S_{MC}, 
\end{equation*}
where $\bar{\bm{c}}_d=(\bm{k}_d^t,\bm{m}_d^t)^t$. Afterwards, we draw $K$ realisations from normal distribution in  \eqref{eq:rand_norm_app}, estimate the critical value $c_{S_0}(1-\alpha)$ by the $([(1-\alpha)K]+1)^{th}$ order statistic of $S_{MC}$ and construct MC SPI as follows
\begin{equation}\label{eq:unif_band_MC}
	\mathcal{I}^{MC}_{1-\alpha}=
	\bigtimes_{d=1}^{D}
	\mathcal{I}^{MC}_{d,1-\alpha}
	\quad \text{where}\quad  \mathcal{I}^{MC}_{d,1-\alpha}=
	\left\{\hat{\mu}_d \pm c_{MC}(1-\alpha) \hat{\sigma}(\hat{\mu}_d )\right\}.
\end{equation}
We can similarly obtain a critical value for MT. The consistency of $\mathcal{I}^{MC}_{1-\alpha}$ follows from equation \eqref{eq:Z} which is a standard result for mixed models. % and $Z=N(0,1)+O(n^{-1/2})$. GAR of \cite{chernozhukov2013gaussian}
The same results from the extreme value theory  as in the proof of Proposition \ref{prop:con_max} might be invoked to prove the consistency for the maxima. Monte Carlo SPI are easy to implement and less computer intensive than bootstrap. Yet, %as we will see, 
they are less robust to departures from the normality
of errors and random effects (cf. Section \ref{sec:simulations}).

\subsection{Bonferroni procedure}\label{sec:bonfer}

Classical simultaneous inference has been considered via Bonferroni procedure. If all statistics $(\tilde{\mu}_d-\mu_d)/\sigma(\tilde{\mu}_d)$ were independent Gaussian pivots, the critical value to construct SPI or MT could be selected as $c_{BO}(1-\alpha)=\Phi^{-1}(1-\alpha/2D)$. One may use quantiles from the normal instead of the t-distribution, because the number of mixed parameters is allowed to grow to infinity such that the latter distribution converges to the former, cf.\ the high-dimensional regression setting in \cite{chernozhukov2013gaussian}.
Having retrieved the value of interest, a Bonferroni SPI is defined as 
\begin{equation}\label{eq:unif_band_BO}
	\mathcal{I}^{BO}_{1-\alpha}=
	\bigtimes_{d=1}^D
	\mathcal{I}^{BO}_{d,1-\alpha},\quad\text{where}\quad
	\mathcal{I}^{BO}_{d,1-\alpha}=
	\left\{\hat{\mu}_d  \pm c_{BO}(1-\alpha) \hat{\sigma}(\hat{\mu}_d )\right\}.
\end{equation}
While the same critical value might be used in MT procedure \eqref{eq:test_proc}, it provides a weak control of FWER. Using Bonferroni's methodology, we do not approximate the true distribution of statistic $S_0$ in equation \eqref{eq:s}. Hence, we work with $\hat{\sigma}(\hat{\mu}_d) = \sqrt{\mathrm{mse}(\hat{\mu}_d)}$ which is an estimated version of MSE defined in \eqref{eq:MSE}. An application of this procedure %remains still relatively 
is simple and does not require almost any computational effort. It will be our benchmark under  
asymptotic independence of parameters. However, the results of \cite{romano2005exact} confirm that the method of Bonferroni performs poorly for correlated random variables, a problem that is even aggravated when allowing for spatio- and/or temporal dependencies, see our discussion in Section \ref{sec:conclusions}. %But
Nevertheless, similarly as bootstrap SPI, Bonferroni bands are fairly robust to the distributional departures from normality of errors and random effect if the number of units in each cluster grows to infinity.

\subsection{Beran procedure}

\cite{beran1988balanced} developed a procedure to obtain balanced simultaneous intervals with an overall coverage probability $1-\alpha$ within the context of models without random effects. His technique is based on so called \textit{roots} and bootstrapping to approximate their cumulative distribution functions (cdfs). We can follow Beran's methodology and evaluate its performance under LMM. Suppose that $S_d$ is a root and consider $\max_{d=1,\dots, D}S_d$ as defined in equation \eqref{eq:s}. Let $F_{S_d}$ and $F_{S}$ be their respective cdfs. Furthermore, we denote with $F^{-1}_{S_d}(a) $ and $F^{-1}_{S}(a)$ the largest $a^{th}$ quantiles of $F_{S_d}$ and $F_{S}$.  \cite{beran1988balanced} suggested bootstrap approximations $F^*_{S_d}$ and $F^*_{S}$ to obtain $D$ critical values defined as $c_{BEd}(1-\alpha)=F^{*-1}_{S_d}\{F^{*-1}_{S}(1-\alpha)\}$. Further details can be found in his paper. The SPI is then
\begin{equation}\label{eq:unif_band_Be}
	\mathcal{I}^{BE}_{1-\alpha}=
	\bigtimes_{d=1}^D 
	\mathcal{I}^{BE}_{d,1-\alpha},
	\quad \text{where} \quad
	\mathcal{I}^{BE}_{d,1-\alpha}
	=\left\{\hat{\mu}_d  \pm c_{BEd}(1-\alpha) \hat{\sigma}(\hat{\mu}_d )\right\}.
\end{equation} 
The critical values $c_{BEd}(1-\alpha)$ are not directly applicable for MT. Moreover, Beran's method is as computer intensive as bootstrap SPI, but in comparison to the former it might provide a poorer coverage rate as its convergence in sup-norm is not guaranteed, cf.\ results in Section \ref{sec:simulations}. Last but not least, it is not necessarily robust to the distributional departures from normality of errors and random effects.

%%%%%%%%%%%%%%%%%
\section{Simulation experiments}\label{sec:simulations}

We carry out simulations to examine finite sample properties of bootstrap (BS), Monte Carlo (MC), Beran (BE) and
Bonferroni (BO) SPIs as well as to evaluate the empirical power of MT procedures under various scenarios. In particular, we analysed them under NERM and FHM. %being prominent examples of LMM in SAE and longitudinal studies. 
As far as the former is concerned, we set $x_{dj1}=1$, $x_{dj2}\sim U(0,1)$ $\forall$ $d\in[D]$ and $j\in[n_d]$, whereas under the FHM we set $x_{d1}=1$, $x_{d2}\sim U(0,1)$ $\forall d\in [D]$ with $\bm{\beta}=(1,1)^t$ in both models. The number of simulation runs is $I=2500$, each with $B=1000$ bootstrap samples. The covariates are fixed in all simulation runs. We considered small to medium numbers of clusters with $D\in \{15,30,60,90\}$.

When NERM is considered, we first set $n_d=5$ $\forall d\in [D]$, $e_{dj}\sim N(0,\sigma^2_e)$, $u_{d}\sim N(0,\sigma^2_u)$ such that the intraclass correlation coefficient $\mathrm{ICC}=\sigma^2_u/(\sigma^2_u+\sigma^2_e)$ equals 1/3, 1/2 or 2/3 (see the first column of Table \ref{tab:simulations}). Then we relax the modelling assumptions by allowing $e_{dj}$ and $u_d$ to deviate from normality to become heavy-tailed or asymmetric. Namely, we draw them from centred chi-square distribution with 5 degrees of freedom, student-t distribution with 6 degrees of freedom and skewed student-t distribution with 5 degrees of freedom and the skewness parameter equal to 1.25. We then rescale them to variances $\sigma^2_{e}$ and $\sigma^2_{u}$ indicated in parentheses in Table \ref{tab:simulations_chi_dist}. Furthermore, we allow the number of units to grow with the number of clusters, cf. \cite{jiang1998asymptotic}. %Otherwise, the SPI would diverge which is compatible with the remark of \cite{jiang1998asymptotic}. Fortunately, we see from the right plot that their distribution is very well described by the skewed t-distribution $st_{5,1.25}(2)$. 
The unusual choice of the skewed t-distribution is motivated by the data example in Section \ref{sec:data_example}. In particular, it aims to mimic the pdf of estimated errors in our application, see the middle panel of Figure \ref{fig:kd_qq_R}. %; for further deviations see the supplement. 
We consider a scenario with $D=52$ and $n_d=100$, i.e., the number of areas in the data example with $n_d=100$ being close to the median of the number of units across counties. We also evaluate the performance of our method for a smaller sample size with $D=26$ and $n_d=50$. Since the results hardly differ when estimating $\bm{\theta}$ using restricted maximum likelihood (REML) or the method of moments, we skip the latter. %Additional numerical results are available from the authors.  

We apply a similar setting as in \cite{datta2005measuring} in the simulation study with FHM. We suppose that random effects and errors are independent, centred and normally distributed with unknown variance $\sigma_u^2=1$ and known $\sigma^2_{e_d}$ as follows. Each fifth part of the total number of clusters is assigned to a different value for $\sigma^2_{e_d}$; in Scenario 1: $0.7$, $0.6$, $0.5$, $0.4$, $0.3$ and in Scenario 2: $2.0$, $0.6$, $0.5$, $0.4$, $0.2$. That is, we consider the case of known heteroscedasticity for errors. Variance $\sigma_u^2 $ is estimated using REML, Henderson's method \citep{prased_rao} and the method of \cite{fay_herriot}. We present results only for the former as other methods perform similarly for
SPI and MT. %Note that 
All simulated scenarios are almost optimal settings for the Bonferroni procedure as the mixed parameter estimates are asymptotically independent. Therefore we can take it as a benchmark (cf.\ comments in Section \ref{sec:bonfer}). 

We use three criteria to evaluate the performance of different methods to construct SPI: the empirical coverage probability (ECP), the average width (WS), and the average variance of widths (VS) 
\begin{equation*}
	\begin{split}
		\mathrm{ECP} &=\frac{1}{I}\sum_{k=1}^{I}\bm{1} \{ \mu^{(k)}_d\in \mathcal{ I}^{P}_{1-\alpha} \,\, \forall d \in [D] \}, \quad 
		\text{where}\quad \text{P = BS, MC, BE or BO}, \\   
		\mathrm{WS}&=\frac{1}{DI} \sum_{d=1}^{D}\sum_{k=1}^{I}\rho^{(k)}_d,
		\rho^{(k)}_d=2c^{(k)}_{P}(1-\alpha)\hat{\sigma}^{(k)}(\hat{\mu}_d ), \quad \text{where}\quad \text{P = BS, MC, BE or BO}, \\
		\mathrm{VS}&=\frac{1}{D(I-1)} \sum_{d=1}^{D} \sum_{k=1}^{I} 
		\left(\rho^{(k)}_d-\bar{\rho}_d \right)^2,\;
		\bar{\rho}_d=\sum_{k=1}^{I} \rho^{(k)}_d/I.
	\end{split}
\end{equation*}	
ECP is the percentage of times all cluster-level parameters are inside their SPI. On the other hand, WS is calculated for each cluster over the widths of the intervals from $I$ simulations, and averaged over all clusters to obtain an aggregated indicator. Lower values of WS are preferable. Finally, for assessing their variability, we compute the variance of widths over the simulations, and average them over all clusters  (VS). We prefer lower values of VS, as they would indicate that the length of intervals is stable. 

Last but not least, in practice, $c_{BS}(1-\alpha)$ and $c_{BE}(1-\alpha)$ are approximated by $[\{(1-\alpha)B\}+1]^{th}$ order statistics of the empirical bootstrap distribution. In addition, to construct $\mathcal{I}^{MC}_{1-\alpha}$ in \eqref{eq:unif_band_MC} we can use $\bm{g}_{1}$ or the variance expression from the denominator in \eqref{eq:Z}. Since the numerical differences were negligible, we present results only for the latter.

Table \ref{tab:simulations} shows the numerical results of our criteria to compare the performance of different methods when errors and random effects are normally distributed. Under these scenarios, BS attains the nominal level of 95\% even for a small number of clusters ($D=15$). Yet, due to the overestimation of variability of the cluster parameters, this method suffers from an overcoverage when ICC$=1/3$ for $D=15$ and $D=30$. Furthermore, although our simulations constitute a nearly optimal design for the Bonferroni method, BO exhibits almost always undercoverage. MC has worse performance, the convergence to the nominal level is slower with ECP oscillating around 94\% only for $D=60$ and $D=90$ when ICC$=2/3$ or ICC$=1/2$. It does not attain the nominal coverage under the third scenario. Moreover, BE diverges which might by explained by the lack of convergence in sup-norm of the Beran's procedure. The second part of Table \ref{tab:simulations} summarizes results for WS and VS. As expected, the width increases with growing $D$. Nonetheless, the speed of this increase is moderate; with a growing number of areas, the SPI has to cover more parameters, but the estimate of variability decreases \citep[for more details, see][]{reluga2021}.
%Notice that for BS the width hardly increases when $D$ grows from $15$ to $90$. 
When we consider VS, we conclude that BS is more variable than other methods for $D=15$, but this difference decreases for increasing $D$. In SAE, undercoverage is often considered a more severe type of error than overcoverage, partly due to the difficulties to detect and alleviate it \citep{yoshimori2015numerical}. On the other hand, overcoverage is often a result of an excessive variability in small samples which is illustrated in Table \ref{tab:simulations}. Having this in mind, we conclude that BS is the most satisfactory method. 
%It also works well when we have only few units in each cluster. 

Table \ref{tab:simulations_subset} shows the performance of SPI constructed for a subset of $D'=D/5$ clusters. It illustrates finite sample performance of Corollary \ref{cor:subset_interval}. Since the simulations under other scenarios led to the same conclusions, we only consider $\sigma^2_u = \sigma^2_e = 1$. %The subset of $D'$ areas was first selected randomly and then fixed for all simulation runs, that is, in each simulation run we took . 
In each simulation run we constructed SPI for $D'$ areas, but used all data to compute variances, fixed and random effects. The ECP is similar as in Table \ref{tab:simulations}. In contrast, the widths of SPI are narrower than those in Table \ref{tab:simulations}, because they are constructed to cover simultaneously only $D'<D$ mixed parameters. This empirical study confirms a practical relevance of our proposal. In fact, it shows that one can construct reliable SPI or conduct MT for an arbitrary subset of cluster-level parameters.

While the vast majority of the SAE literature relies heavily on the normality of random effects and errors, especially regarding MSE estimation and CPI construction,  this assumption may be violated in practice. Thus, we conduct a robustness study regarding departures from normality of errors and/or random effects. The first part of the empirical results of our criteria under this setting is  presented in Table \ref{tab:simulations_chi_dist}. Further simulation results are deferred to the supplementary material. First, but not surprisingly, the overall performance of all methods is worse than in Table \ref{tab:simulations}, especially for asymmetric $\chi^2$ distributed errors. % in Table \ref{tab:simulations_chi_dist}. 
Second, BS and BO are still superior to all other methods. In addition, in case of chi-square distributed departures, the coverage is higher for ICC$=1/3$ with $D=15$ due to overestimated variability, then it drops for $D=30$ and $D=60$, and increases for $D= 90$ in accordance with asymptotic theory. Importantly, under the scenario which mimics the data application, i.e., with skewed t-distributed errors and normal random effects, the ECP is close to the nominal level. However, we must conclude that the considered SPIs do not attain the nominal coverage probability if errors exhibit more severe deviations from normality than we observe in our application, irrespective of the presence of deviations from normality of random effects. 
The issue of undercoverage might be alleviated by the use of more sophisticated bootstrapping scheme requiring different theoretical derivations which is beyond the scope of this paper \citep[cf.][]{reluga2020thesis}. Simulations therein and in our supplementary material confirm that the deviation from normality of random effects hardly affects the coverage of SPI.
A similar conclusion was drawn by \cite{mcculloch2011misspecifying} in the study of the bias of estimated fixed effects and EBLUPs. When it comes to the right hand side of Table \ref{tab:simulations_chi_dist}, WS decreases with a growing sample size due to the increase of $n_d$. Even though the critical values increase with the growing number of clusters, $\hat{\sigma}^2(\hat{\mu}_d)$ decreases at a faster rate. For this reason the average width of intervals is decreasing too.

\begin{figure}[htb]
	\centering	\includegraphics[width=0.75\textwidth]{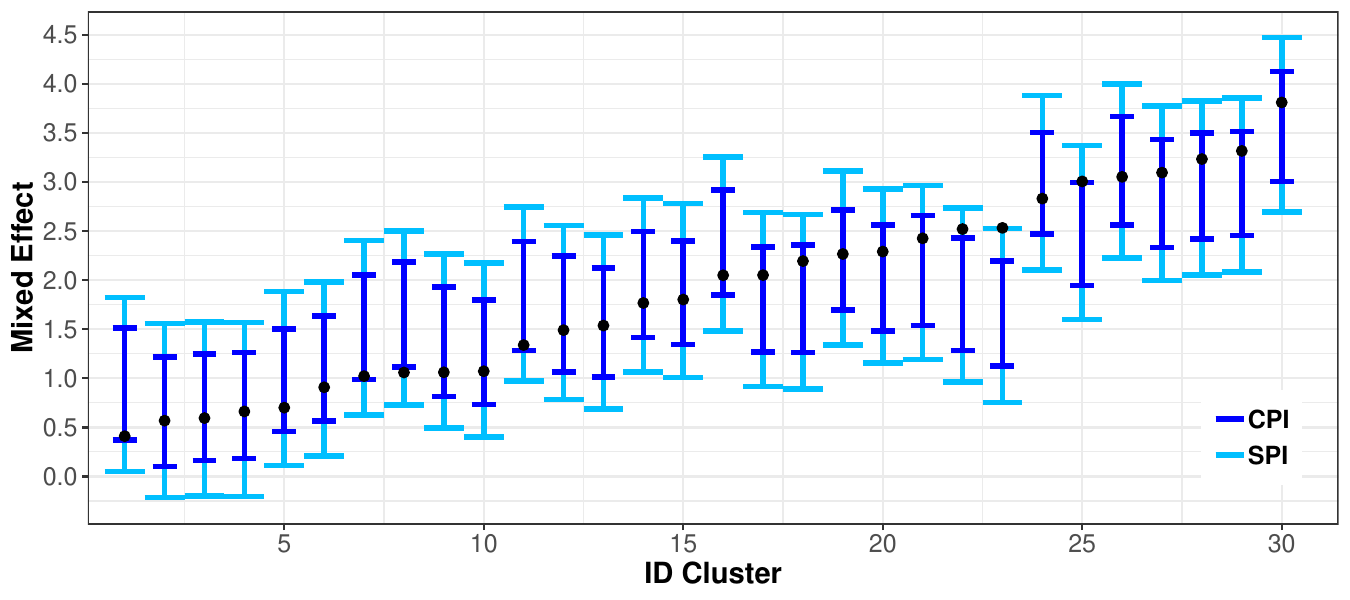}  \vspace{-3mm}
	\captionof{figure}{95\% CPI and bootstrap SPI for mixed effect means, %when variance components are estimated using REML,
		$e_{dj}\sim N(0.5)$, $u_d\sim N(1)$ and $D=30$. Black dots are true mixed parameters.} 	\label{fig:me}
\end{figure}

Let us revisit the differences between CPI and SPI. Figure \ref{fig:me} displays 95\% bootstrap SPI in light blue and CPIs of  \cite{chatterjee2008parametric} in dark blue. The critical values for CPIs have been calculated using parametric bootstrap, cf.\ \cite{chatterjee2008parametric} for details. In comparison to CPI, SPI covers all clusters with a certain probability. Black dots represent the true mixed parameters $\mu_{d}$. Out of thirty, three  cluster-level parameters (eighth, twenty-second and twenty-third) are clearly outside of their CPI, and another four (first, seventh, eleventh and twenty-fifth) are on their boundary. It does not happen by chance or by the simulation design, but by the construction of CPIs: for $100(1-\alpha)$\% CPI, about $100\alpha$\% of the true mixed parameters are not covered by their intervals. Figure \ref{fig:me} illustrates even more severe case with 10\% of the true parameters not covered by CPIs. In contrast, SPI contains all of the true mixed parameters. Moreover, SPI is not excessively wide compared to CPI. In fact, SPI seems just as wide as necessary; twenty-third cluster-level mean is right at the boundary. Undoubtedly, CPI and SPI are methodologically different and constructed to cover
distinct sets with a certain probability. One can thus argue that their direct comparison is flawed
and should not be investigated. We do not claim otherwise; rather, Figure \ref{fig:me} serves as an illustration of a practical relevance of SPI as a valid tool for comparing mixed parameters across clusters. Moreover, Figure \ref{fig:me}  demonstrates that the cluster-wise inference can lead to erroneous conclusions once applied to perform joint statements or comparisons. %of cluster-level parameters.

\begin{figure}[htb]
	\vspace{-4mm}
	\centering
	\subfloat{\includegraphics[width=0.3\textwidth]{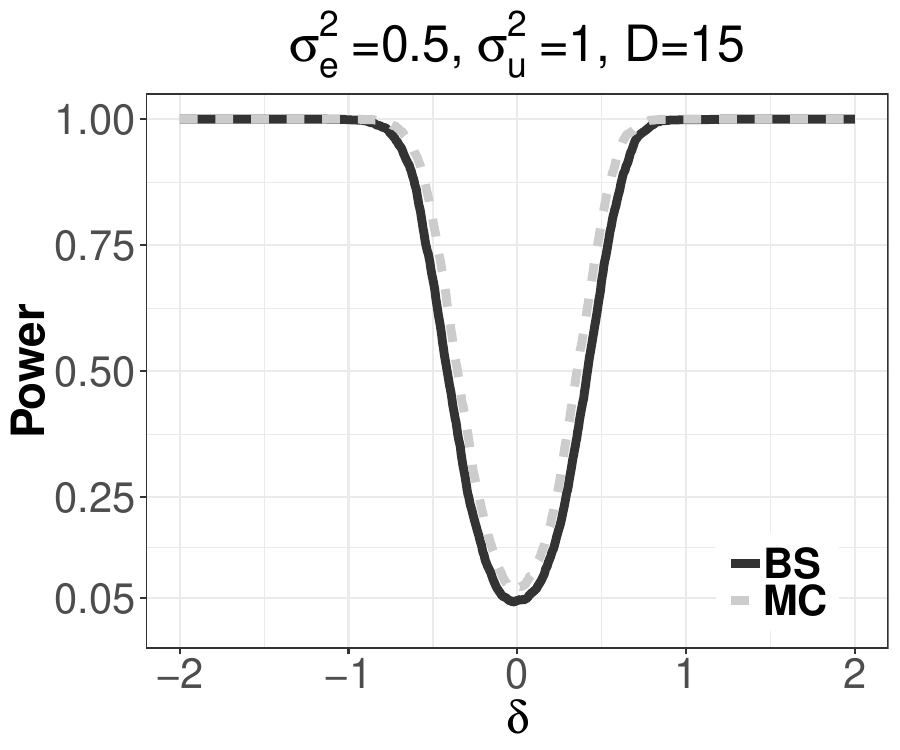}}
	\subfloat{\includegraphics[width=0.3\textwidth]{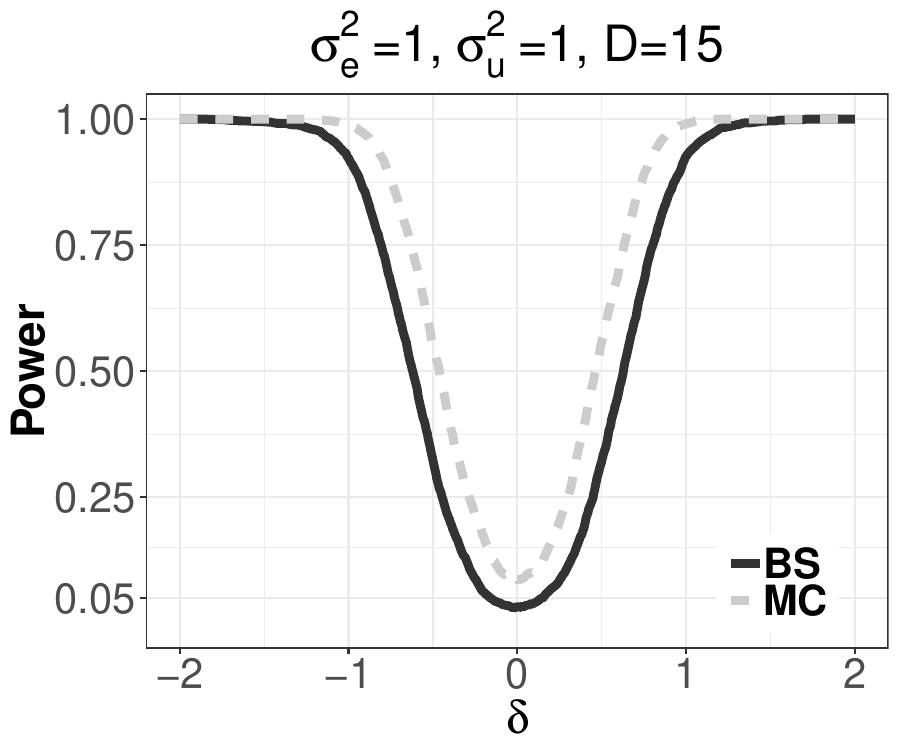}}
	\subfloat{\includegraphics[width=0.3\textwidth]{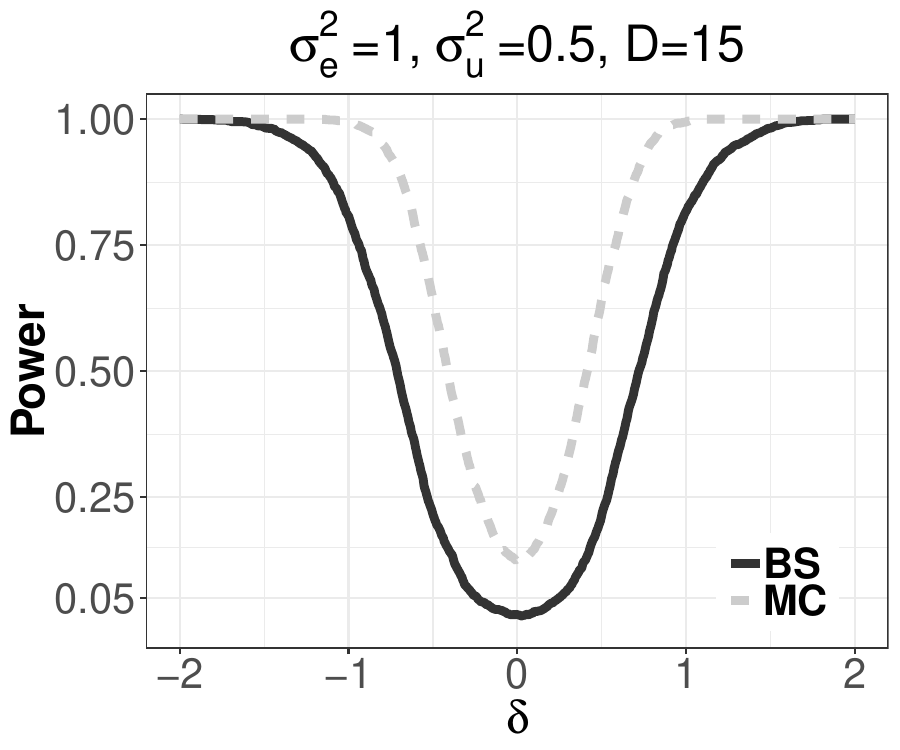}}\\
	\vspace{-4mm}
	\subfloat{\includegraphics[width=0.3\textwidth]{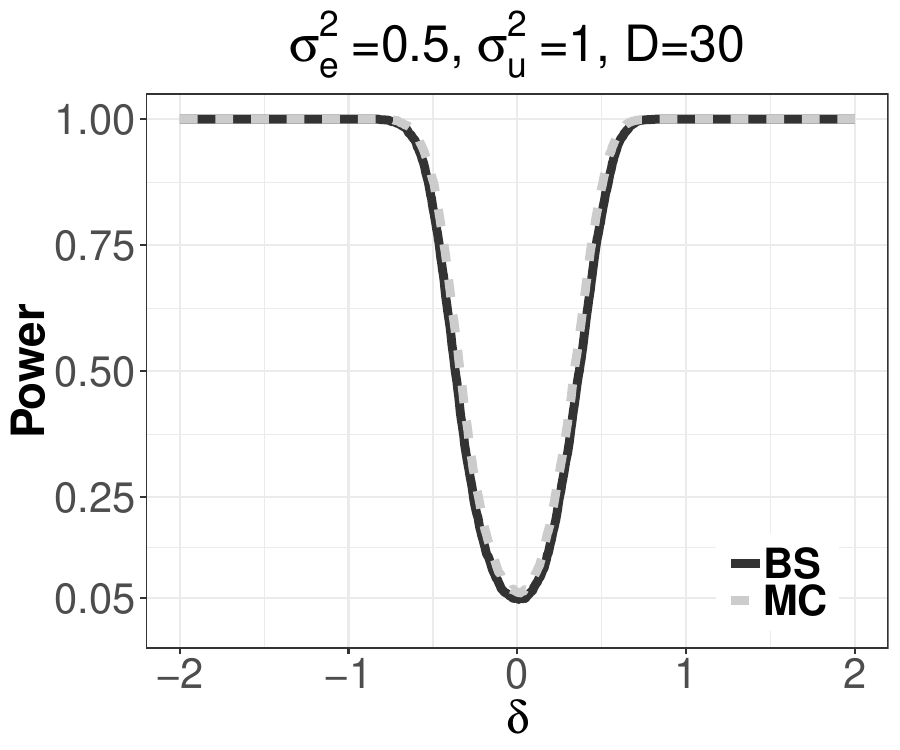}}
	\subfloat{\includegraphics[width=0.3\textwidth]{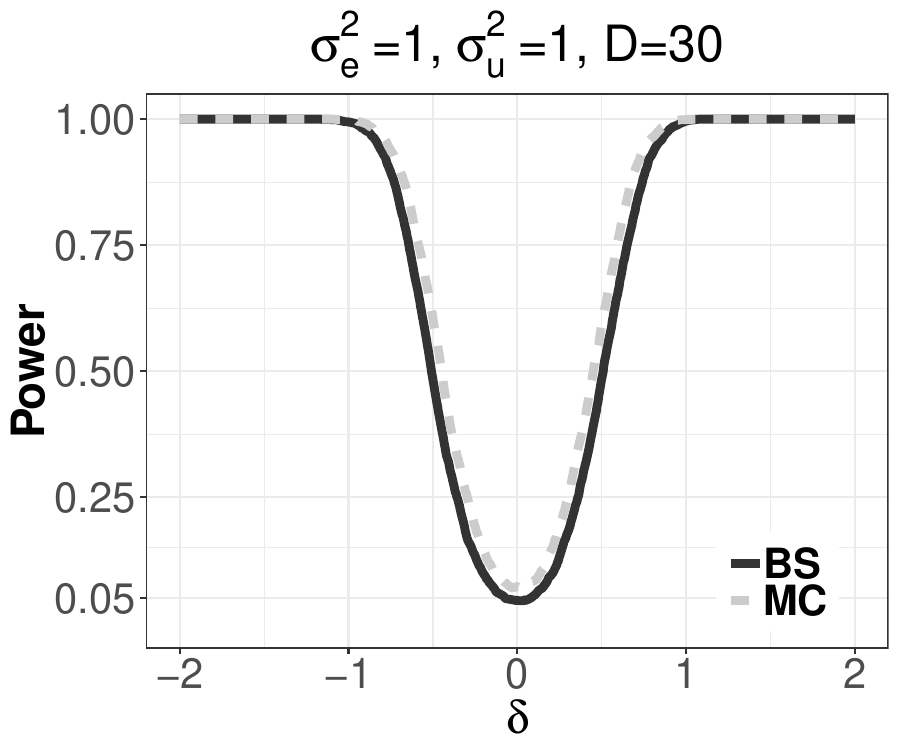}}
	\subfloat{\includegraphics[width=0.3\textwidth]{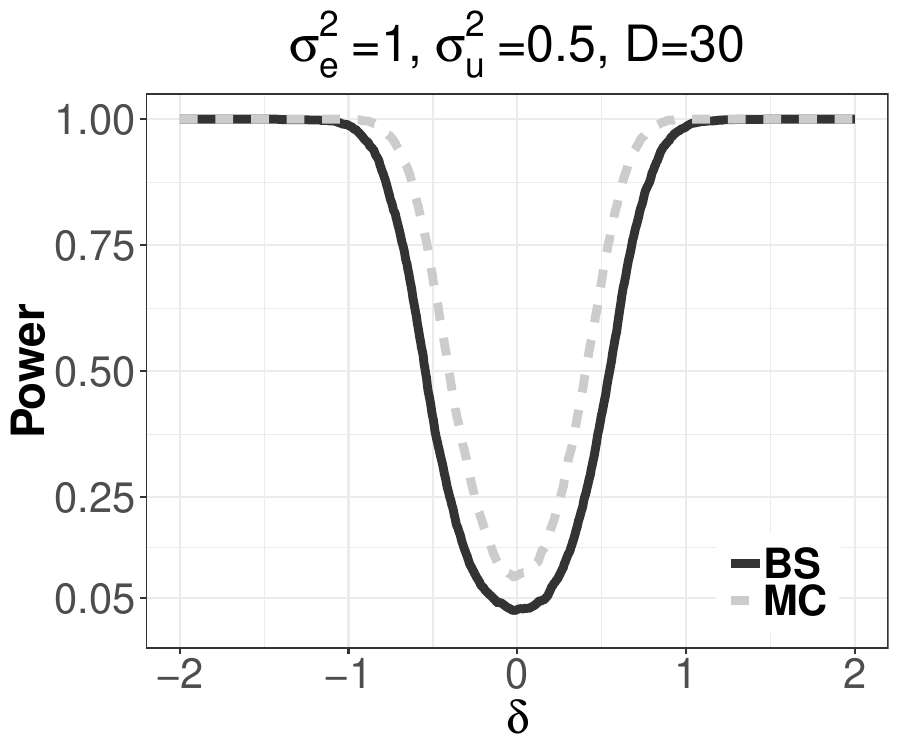}}\\
	\vspace{-4mm}
	\subfloat{\includegraphics[width=0.3\textwidth]{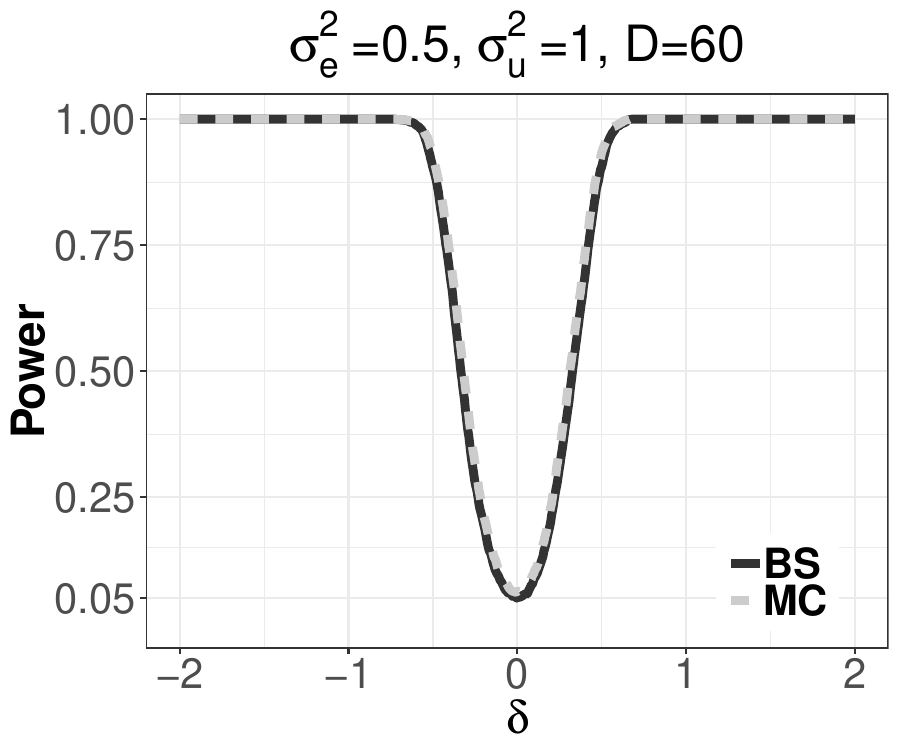}}
	\subfloat{\includegraphics[width=0.3\textwidth]{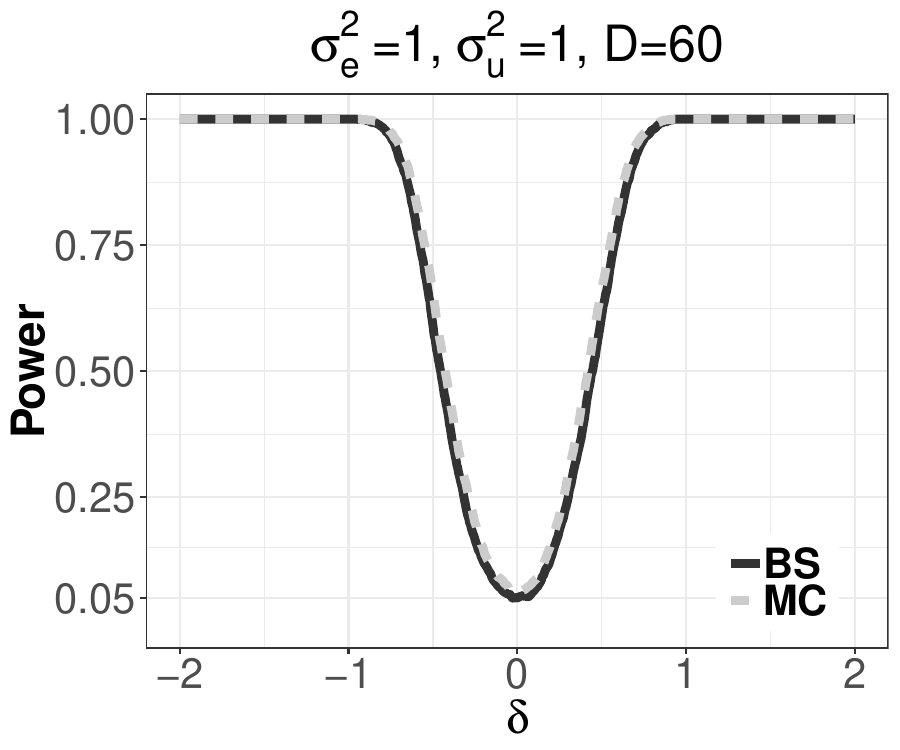}}
	\subfloat{\includegraphics[width=0.3\textwidth]{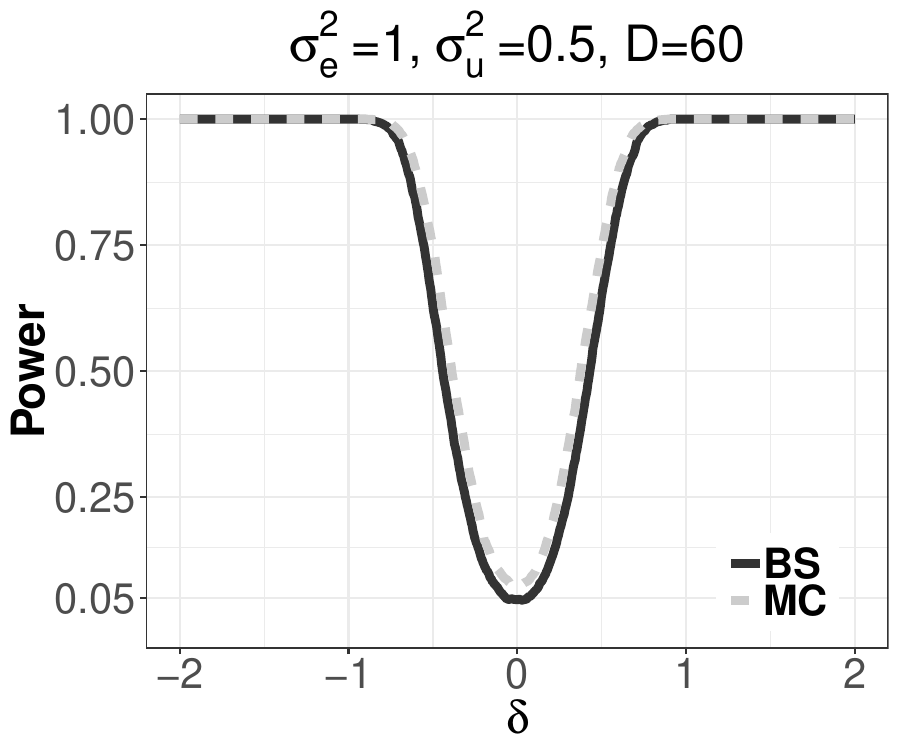}}\\
	\vspace{-4mm}
	\subfloat{\includegraphics[width=0.3\textwidth]{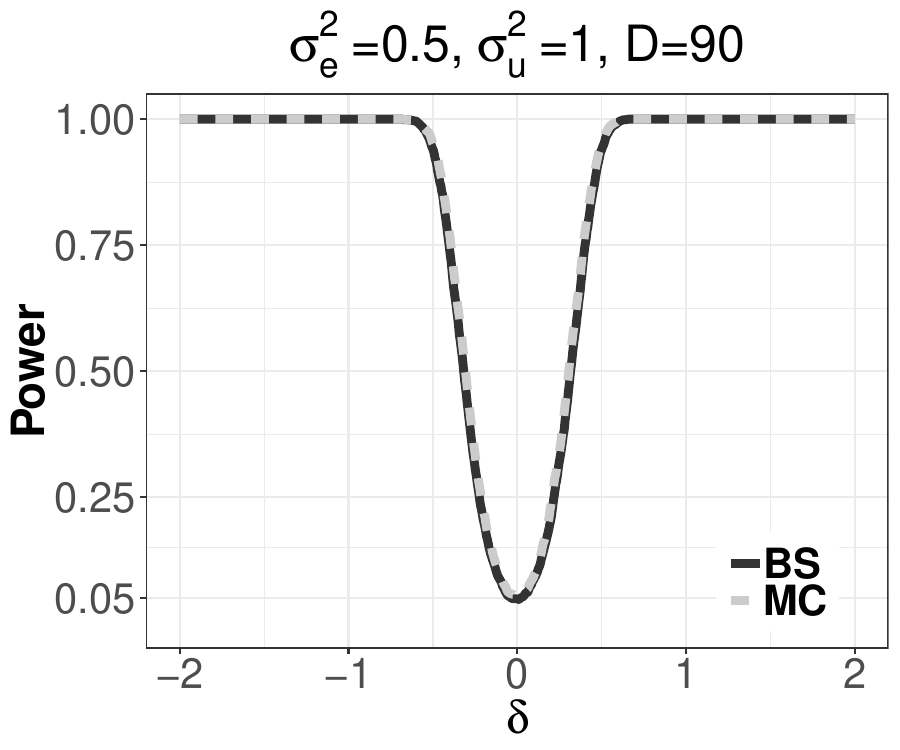}}
	\subfloat{\includegraphics[width=0.3\textwidth]{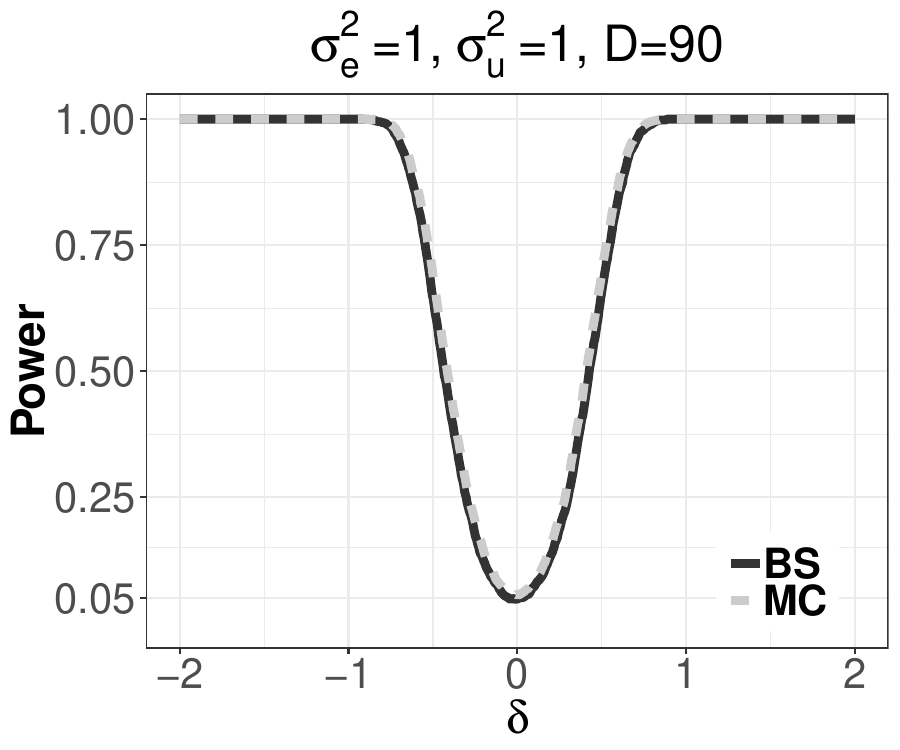}}
	\subfloat{\includegraphics[width=0.3\textwidth]{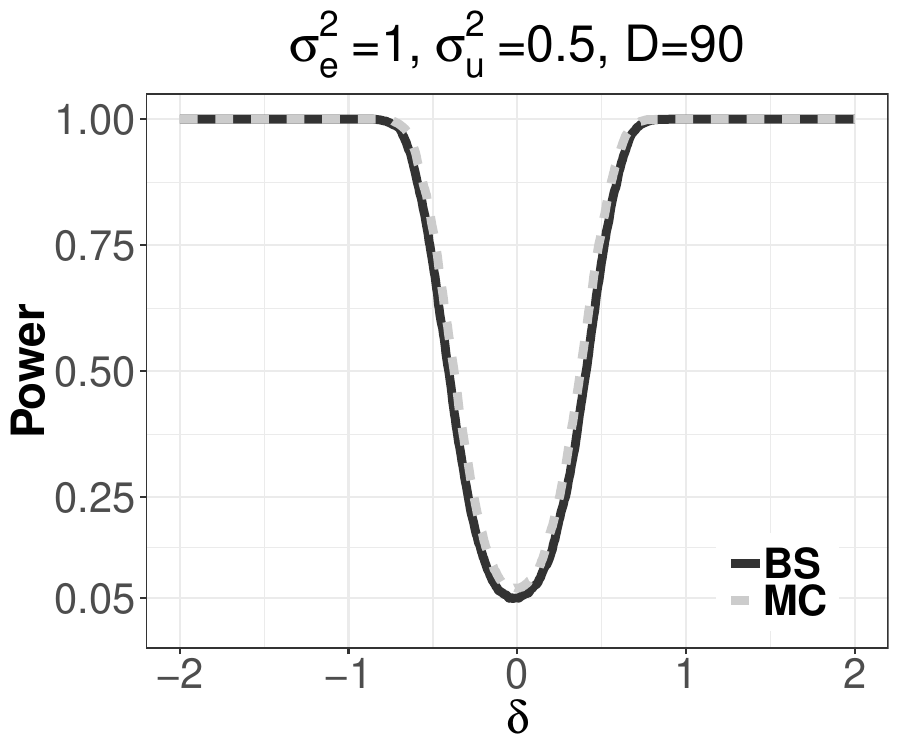}}
	\captionof{figure}{Power of MT  $H_0:\bm{\mu}=\bm{h}$ vs $H_1: \bm{\mu}=\bm{h}+\bm{1}_{D}\delta$ for BS- and MC-based multiple tests (MT).} 
	\label{fig:test}
\end{figure}

Regarding MT, Figure \ref{fig:test} displays the empirical power of bootstrap and MC based max-type tests for $H_0:\bm{\mu}=\bm{h}$ vs $H_1: \bm{\mu}=\bm{h}+\bm{1}_{D}\delta$. In the simulations, we simply set $\bm{h}:=\bm{\mu}$ under $H_0$, whereas under under $H_1$ we added a constant $\delta\in[-2,2]$ to each element of $\bm{h}$. As expected, ICC influences the Type II error -- the curves are the steepest and almost not distinguishable for ICC$=2/3$. The bootstrap test performs significantly better when ICC$=1/2$ and ICC$=1/3$ for small and medium $D$. In contrast, MC based tests do not attain the nominal level under $H_0$ for small sample sizes. For larger $D$, the curves almost coincide under all three scenarios.

Let us finally turn to the analysis under FHM. Since the performance of MT under FHM leads to similar conclusions as for NERM, we restrict ourselves to present ECP, WS and VS for different SPIs. Table \ref{tab:simulationsFH} displays the results. Bootstrap SPI suffers from overcoverage for small $D$, similarly to Bonferroni's SPI. The overcoverage is probably caused by the same reasons as for NERM. Surprisingly, Bonferroni's intervals fail to achieve the nominal level for larger numbers of clusters. Beran's SPI diverge, whereas MC SPI exhibits undercoverage for $D=60$ and $D=90$.   

%%%%%%%%%%%%%%%%%%%%%%%%%%%%%%%%%%%%%%%
\section{Application to the household income data of Galicia}\label{sec:data_example}
We consider the household income data of the Structural Survey of Homes of Galicia (SSHG) which contains many potentially correlated covariates. It is of great interest for the Galician Institute of Statistics (IGS), and the regional government alike, to study the household income across counties (\textit{comarcas}), e.g., to adjust regional policies and resource allocations. 
The IGS provides direct design-based estimates and/or EBLUP of the average household income accompanied by their variability measures or area-specific confidence intervals. However, the joint consideration of county-level parameters or comparisons between them is often important too. We start from the classical design- and model-based area-wise analysis. Afterwards we complete it  with the simultaneous inference for counties in Galicia. 

The SSHG contains data on 23628 individuals within 9203 households which were collected in 2014 and published in 2015. It comprises information about the total income as well as different characteristics on individual and household level. The variable of interest is the monthly household income. This variable was obtained by taking the twelfth of the total yearly income which consists of paid work, own professional activity and miscellaneous benefits. Following \cite{jose2018selection}, we consider covariates age, education level, type of household, and variables indicating financial difficulties of the household at the end of a month. 
Galicia is divided into four provinces (A Coruña, Lugo, Ourense and Pontevedra) which are further divided into 53 counties, the small areas that constitute our clusters. There are eighteen counties in A Coruña, thirteen in Lugo, twelve in Ourense and ten in Pontevedra. 
As the SSGH does not contain data from the county Quiroga in Lugo, we limit the study to the remaining 52 counties. Table \ref{tab:Desc_Stats} displays descriptive statistics of the number of units across the counties of each province. The model based approach is motivated by the scarcity of data (in some counties less than 20 observations were collected).

Even though the SSHG does not produce official estimates of totals $Y^{dir}_d$, $X^{dir}_{di}$ and means $\bar{Y}^{dir}_d$, $\bar{X}^{dir}_{di}$ at the county level, we calculated them using: 
\begin{equation}\label{eq:covariates}
	\hat{Y}^{dir}_d=\sum_{j\in \mathcal{R}_d} w_{j}y_{j},\quad\hat{\bar{Y}}^{dir}_d=\hat{Y}^{dir}_d/\hat{N}^{dir}_d,\quad
	\hat{X}^{dir}_{di}=\sum_{j\in \mathcal{R}_d} w_{j}x_{ji},\quad 
	\hat{\bar{X}}^{dir}_{di}=\hat{X}^{dir}_{di}/\hat{N}^{dir}_d
	\quad \text{and} \quad
	\hat{N}^{dir}_d=\sum_{j\in \mathcal{R}_d} w_{j},
\end{equation}
where $\hat{N}^{dir}_d$ stands for the estimate of the county size $N^{dir}_d$,  $\mathcal{R}_d$ is the sample in county $d$ and $w_j$ is an official calibrated sample weight. In addition, we have $w_j=1/\pi_j$ where $\pi_j\neq0$ is the first-order inclusion probability.
We used the same design-based direct variance estimator as \cite{jose2018selection}, that is:
\begin{equation}\label{eq:direct_var}
	\widehat{var}(\hat{\bar{Y}}^{dir}_d)=\frac{1}{(\hat{N}^{dir}_d)^2} \sum_{j\in \mathcal{R}_d} w_{j} (1- w_{j})\left(y_{j}-\hat{\bar{Y}}^{dir}_d\right)^2.
\end{equation}
Furthermore, we calculated the coefficient of variation (CV) of direct estimates at the county level. In twelve counties CV $> 10\%$, and in three of them CV $>15\%$. %Note that the CV are used by statistical offices to decide about their reliability.  
A direct estimate is considered official, and thus publishable, if its CV is lower than a certain threshold set by a statistical office. For example, the Office for National Statistics in the UK sets this threshold to 20\% for the labour force statistics \citep{jose2018selection}. Although the CV of our estimates does not exceed this threshold, it is still high enough to consider a model-based framework.

\begin{figure}[htb]
	\centering
	\includegraphics[width=0.85\textwidth]{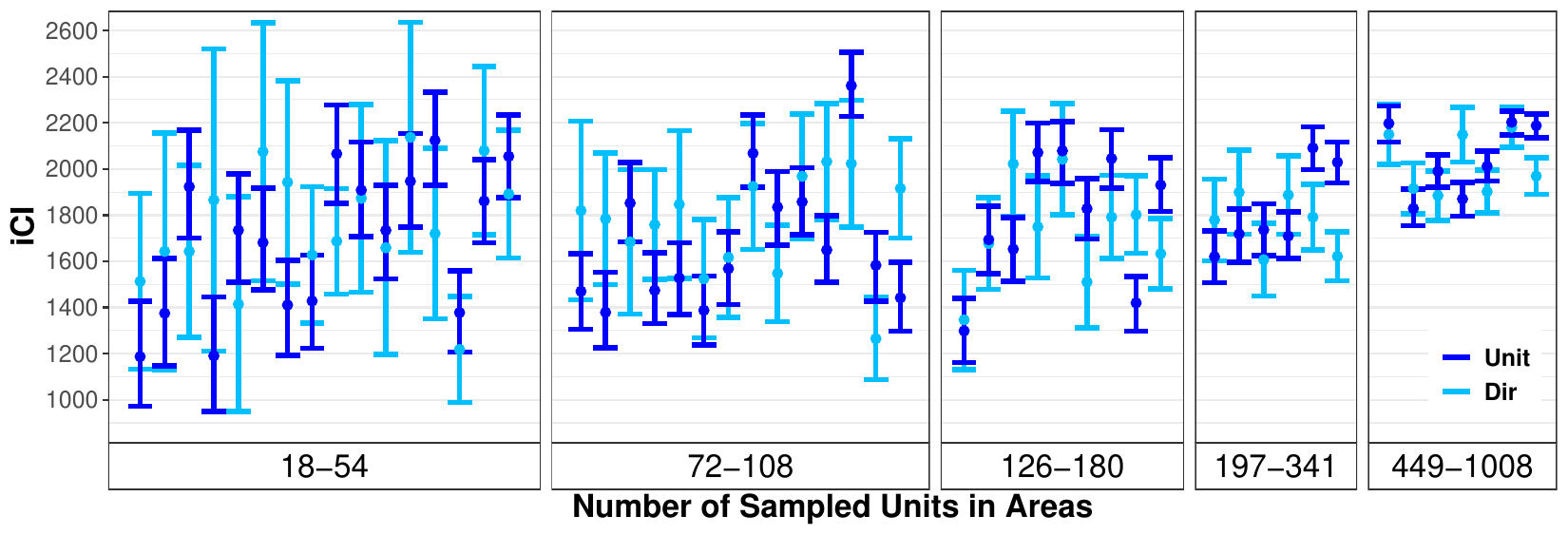}
	\captionof{figure}{Design and model-based $95\%$ CPI.} \label{fig:interval_est_ind_dir}
\end{figure}

We constructed design and model-based point and CPI estimates of monthly incomes. We employed \eqref{eq:covariates} and \eqref{eq:direct_var} for the design-based estimation. Within the model-based framework, we considered  \eqref{eq:mu} as a target parameter with $\bm{k}_d=\bar{\bm{X}}^{dir}_d$, $\bm{m}_d=1$ and calculated  EBLUP in \eqref{eq:mu_hat_e} by $\hat{\mu}_d=\hat{\bar{\bm{X}}}^{dir}_d\hat{\bm{\beta}}+\hat{\bm{u}}_d$ $\forall d\in [D]$. Since SSHG contains information on the household level, we can fit NERM to these data.
Figure \ref{fig:interval_est_ind_dir} shows design-and model-based point estimates of monthly household incomes together with 95\% CPIs. Model-based CPIs were constructed using the parametric bootstrap  \citep[cf.][]{chatterjee2008parametric}. We can use CPIs to compare different methods for the same cluster-level parameter, but not to make comparisons across different counties. For a better presentation, we divided the plot into five panels based on the number of units in each county. First, we can see that the widths of both direct- and model-based intervals decrease with increasing sample size. Second, the widths of direct CPIs are much larger that their model-based counterparts. In fact, direct estimates for certain areas (for example, the fourth and sixth area in the first panel) are too wide to make any informative conclusion. This confirms the necessity of a model-based framework. 

Table \ref{tab:desc_stat} displays the covariates with their standard deviations as well as the estimated coefficients with standard errors and p-values. We performed a variable selection in two stages. First, we selected a subset of covariates that exhibited the highest Spearman's rank correlation %coefficient 
with the household income. Afterwards, we applied a generalised AIC which uses a quasi-likelihood with generalised degrees of freedom, see  \cite{lombardia2017mixed}. The algorithm selected covariates describing  characteristics of the household and characteristics of the head of household. The estimates of variance parameters are $(\hat\sigma_e^2,\hat\sigma_u^2)=(758558.60, 19746.24)$.

\begin{figure}[htb]
	\centering
	\vspace{-5mm}		\subfloat{\includegraphics[width=0.3\textwidth]{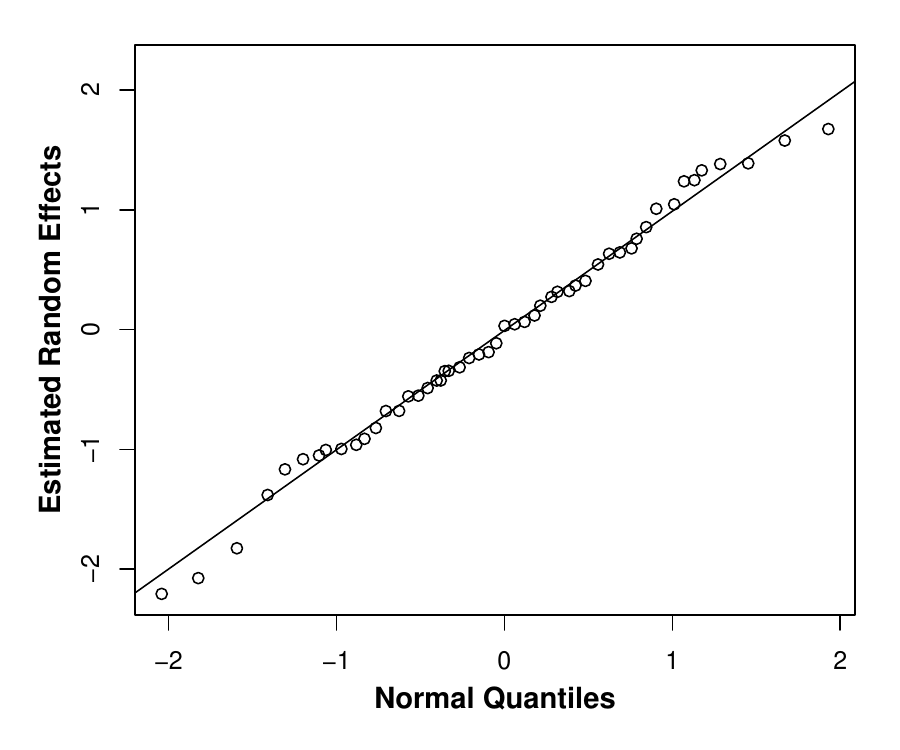}}
	\subfloat{\includegraphics[width=0.3\textwidth]{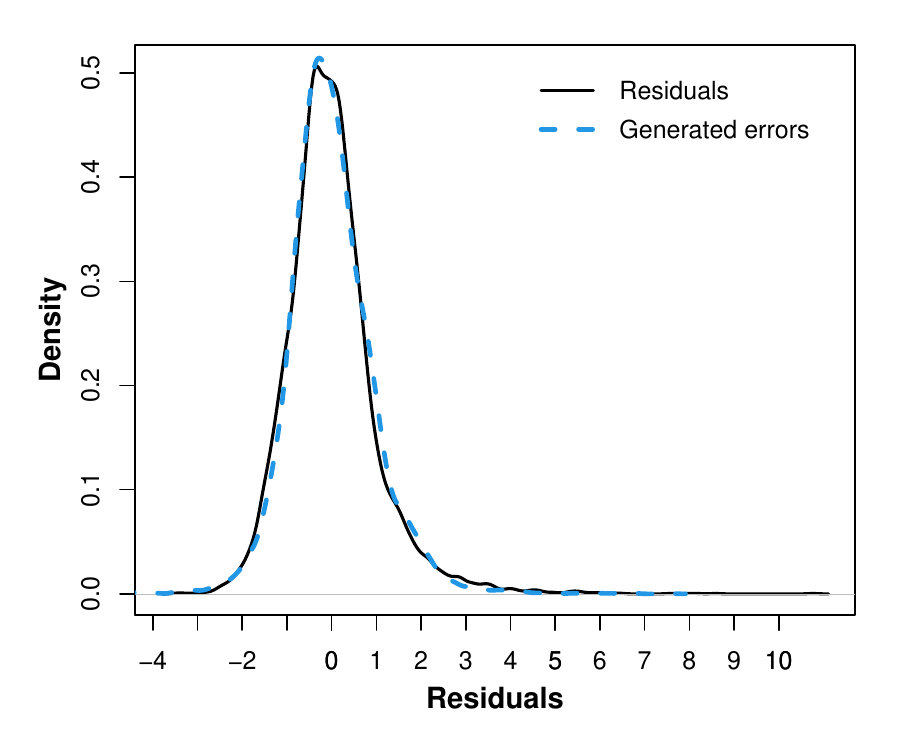}}
	\subfloat{\includegraphics[width=0.3\textwidth]{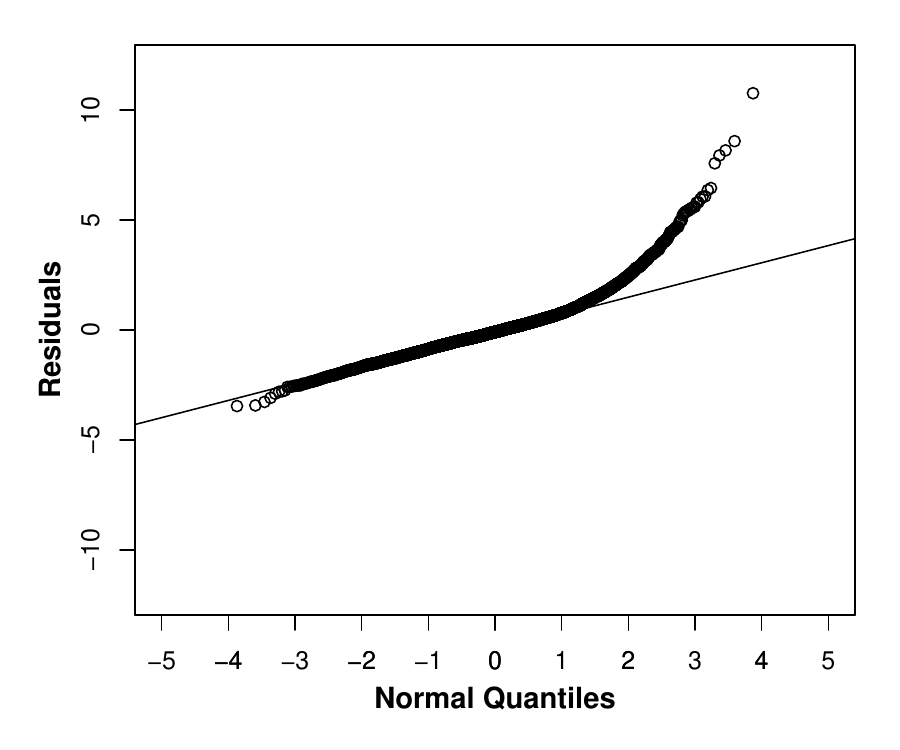}}
	\vspace{-3mm}
	\captionof{figure}{
		REML empirical Bayes estimates of random effects: (left) QQ plot; Cholesky REML residuals: (middle) kernel density estimation and (right) QQ plot.} \label{fig:kd_qq_R}
\end{figure}
%\textbf{Benchmarks: \\
%Typ: typ.6 and typ.7 : combined level composed of one centre and the other (typ.6) and several centres (typ.7).\\
%Difficulties: no difficulty dif.h1.\\
%Ten: ten.h3 and ten.h4: combined level composed of rental (ten.h3) and ceded or another (ten.h4).\\
%Education: higher education edu.ps.h3\\
%Age: combined level of below 45 and above 65}
It is well known that income data are right skewed. Unsurprisingly, our dependent variable exhibits this feature too. It is therefore popular to consider log-income or more sophisticated transformations. Since the naive back-transformation of the dependent variable could cause a serious bias due to the Jensen inequality, different estimation and inference methods were recently suggested by \cite{RojasEtal2020} and references therein. In fact, log transformation of household income in our data example did not help to overcome the problem of skewness. We thus decided to proceed with the household income on the original scale, and assess the sensitivity of our method to the departures from normality of errors and/or random effects after fitting LMM. If the undertaken inference is not compromised be these departures, we shall continue with further analysis. We carried out statistical tests and analysed diagnostic plots. The left panel of Figure~\ref{fig:kd_qq_R} displays a diagnostic plot of \cite{lange1989assessing} using standardised empirical Bayes estimates of the random effects in a weighted normal QQ plot; it supports the adequacy of the normality assumption. Moreover, the p-values of Kolmogorov-Smirnov and Shapiro-Wilk tests, which are $0.997$ and $0.944$, confirm this conclusion. Regarding the normality of errors, the middle and the right panels of Figure~\ref{fig:kd_qq_R} present Cholesky residuals  \citep{jacqmin2007robustness}. The uncorrelated Cholesky residuals are constructed by multiplying $\bm{y}-\bm{X}\hat{\bm{\beta}}$ by the Cholesky square root of the variance matrix. A right tail is visible in both panels. The p-values of Kolmogorov-Smirnov and Shapiro-Wilk tests are $<0.000$. Nevertheless, in the middle panel of Figure \ref{fig:kd_qq_R} we can see that the kernel density estimate of the pdf of the skewed errors has a long, but not a thick tail. Our simulation results in Table \ref{tab:simulations_chi_dist} indicate that such departure is not problematic for our bootstrap-based SPI. In fact, SPI is quite robust to these departures, and a good coverage probability is still provided in comparison to other methods. Last but not least, we did not assess the robustness of bootstrap CPIs to the departure from normality of errors, because they are not the topic of this article; we present them for illustrative reasons. %In addition, we have chosen %and have chosen the presumably most popular method for constructing them.
%to assess the validity of, or deviation from, the normality of errors and random effects.
%is about the distributions of the errors and random effects, and how much these affect the undertaken inference, recall our related study in Section \ref{sec:simulations}.
%{\color{blue}  However, the question that actual matters, is about the distributions of the errors and random effects, and how much these affect the undertaken inference, recall our related study in Section \ref{sec:simulations}. 
%Moreover, household income distributions are certainly less skewed than those of individuals; in our case to an extreme that the log-income distribution was actually more problematic.}
%Figure \ref{fig:kd_qq_R} presents the respective QQ plots and densities. 
\begin{figure}[htb]
	\centering	\includegraphics[width=0.85\textwidth]{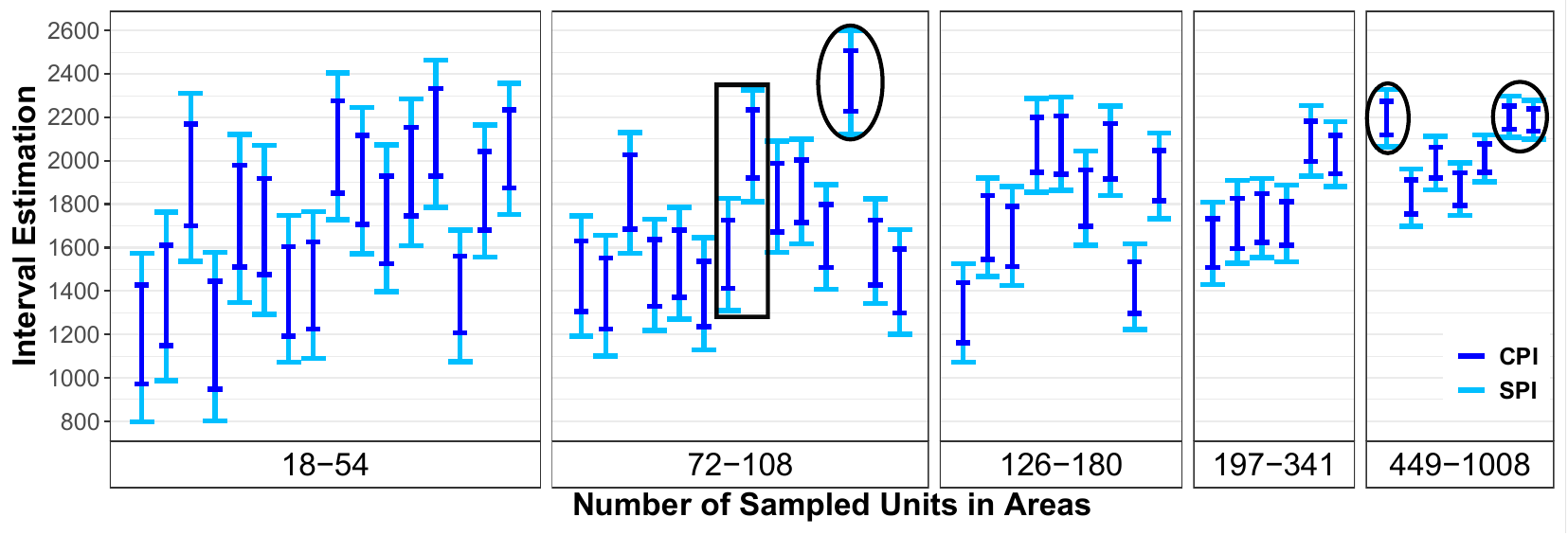}
	\captionof{figure}{$95\%$ bootstrap CPI and SPI for the county-level averages of the log of household income in Galicia.} 
	\label{fig:interval_est}
\end{figure}

Figure \ref{fig:interval_est} displays bootstrap CPI as developed by \cite{chatterjee2008parametric}, together with BS SPI for the county-level averages of monthly household income in Galicia. We can see a lot of variability over the estimates. 
%We know already from Figure \ref{fig:me} in the simulations that the cluster-wise prediction intervals are inadequate to address a joint consideration or comparisons between the counties.
%First, while we are pretty confident that the SPI contain all true values, we know almost for sure that several are outside their CPI.
Evaluating the results of CPI (blue) versus SPI (light blue), it is apparent that the cluster-wise prediction intervals are not adequate to address either a joint consideration or a comparison of the counties. If we consider, for example, the counties of A Fisterra and Noia (7th and 8th regions of the second panel in the black rectangle), the CPIs indicate significantly different incomes,
whereas the SPIs do not support this claim. Moreover, there are other counties (practically in
each panel) for which CPIs would insinuate significant differences whereas statistically valid SPIs do not confirm this conclusion. Nevertheless, SPIs are not unnecessarily wide for practical use. We detect significant and valid differences between several interval estimates. 
\begin{figure}[htb]
	\centering
	\subfloat{\includegraphics[width=0.6\textwidth]{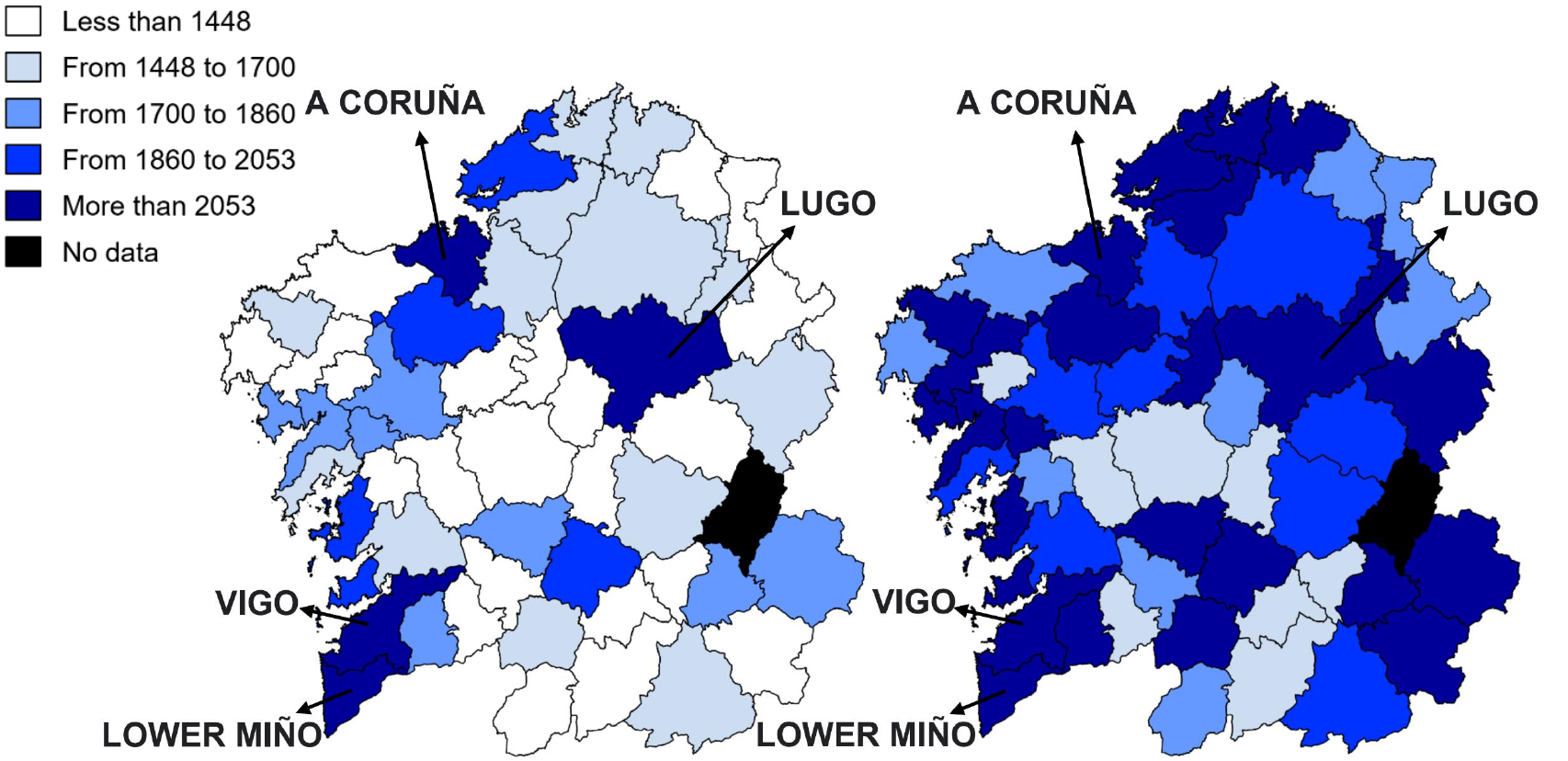}}
	\subfloat{\includegraphics[width=0.28\textwidth]{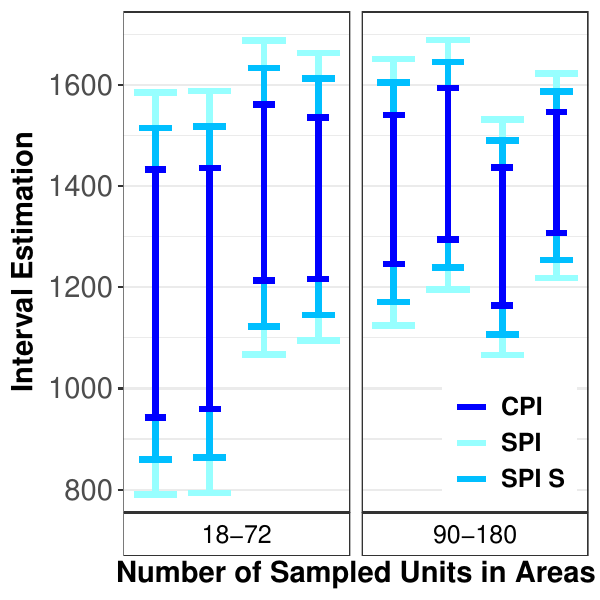}}  \\
	\captionof{figure}{$95\%$ bootstrap SCI for the county-level averages of the household income 	in Galicia: (left) lower boundary, (middle) upper boundary; SPI, SPI S and CPI for a subset of the poorest areas (right).} 
	\label{fig:maps}
\end{figure}

Figure \ref{fig:maps} presents maps with lower and upper limits of bootstrap SPI. The boundaries are classified into one of five categories which were built using $0.2$, $0.4$, $0.6$ and $0.8$ quantiles of the point estimates. We observe a substantial variation of average household income over the counties. Lower and upper boundaries of the interval estimates for the counties of A Coru\~na, Lugo, Vigo (with a large number of units) and Lower Mi\~no are classified into the richest category; they are indicated with ellipsoid in the second and the last panel of Figure \ref{fig:interval_est}. In contrast, there is a group of eight counties (three in the centre, one in the west and four in the south of Galicia) which are classified to the poorest category in the left panel and the second poorest category in the middle panel. Right panel of Figure \ref{fig:maps} presents CPIs and two different types of SPIs for eight poorest counties. In particular, SPI refers to the interval estimate constructed for 52 counties (estimates for all 52 counties are plotted in Figure \ref{fig:interval_est}). In contrast, SPI S refers to the interval estimate constructed for a subset of eight poorest counties, but  using all data to estimate fixed parameters and predict random effects (cf. Corollary \ref{cor:subset_interval} and simulations in Table \ref{tab:simulations_subset}). As expected, SPI S is still wider than CPI, but much narrower than SPI. Moreover, SPI S permits for a valid comparative inference for a subset of these poorest areas. In fact, we can conclude that the differences in the average household income are not statistically significant. Finally, CPI should not be used to make maps in Figure \ref{fig:maps}, as this would suggest that we were allowed to compare them.

Finally, it would be interesting to investigate whether the monthly income of the households which reported difficulties in coming to the end of the month is significantly different from the monthly income of the households which did not struggle with this issue. For example, in touristic areas households spend more such that they might face some difficulties without being poorer. The outcome of such test might then be used to develop a more targeted policy. % to improve living conditions. 
Our MT procedure %max-type statistic 
might be readily applied %in the hypothesis testing 
to support or disprove the %above mentioned 
hypothesis of no difference in monthly income between two mentioned types of households. 
To test this hypothesis we take clusters created from a cross-section of counties and difficulty status. Therefore we apply our developed methodology to %52 counties and 
$2\times52=104$ counties by difficulty status. More specifically, we consider $\bm{\mu}\in \mathbb{R}^{104}$ and test $H_0:\bm{A\mu}=\bm{0}_{104}$ versus $H_1:\bm{A\mu}\neq\bm{0}_{104}$, where $\bm{A}\in \mathbb{R}^{52\times104}$ with rows that are composed of $104$-dimensional vectors $\bm{a}$ with a $1$ on the $2d-1$ place, $-1$ on the $2d$ place but $0$ otherwise, where $d$ stands for a particular county. The test statistic is $t_{H} =\max_{d=1,\dots, D}|t_{H_d}|=7.569$ whereas the critical value is $c_{BH_0}(1-\alpha)=4.220$. %3.473859% 
That is, we clearly reject $H_0$ of no difference. The rejection of the null hypothesis is consistent with the %outcome of 
variable selection procedure which has suggested to find this test outcome. 
%{\color{red} 
%Note that this was just an illustrative example. You could argue that our variable selection would already suggest to find this test outcome. } 

%%%%%%%%%%%%%%%%%%%%%%%%%%%%%%%%%%%%%%%%%%%%%%%%%%
\section{Conclusions}\label{sec:conclusions}

We introduce a practical method to construct SPI and MT procedures for mixed parameters under LMM. We illustrate its use and relevance in simulation studies and a data application within the framework of SAE. We theoretically derive two techniques based on bootstrap approximation of the distribution of the max-type statistic and the volume-of-tube formula. However, we proved that the latter is not directly operational. We further discussed various alternatives %that could be adapted to the considered problem 
and assessed their empirical performance in the simulation study. %applied in practice. %For all operational methods we have carried out an empirical comparison. 
Though slightly conservative for very small samples, the bootstrap-based SPI yield the most satisfactory results in our simulations. 
Moreover, it is quite robust to certain deviations from the normality assumptions. %, however, not to severe deviations.
In addition, our bootstrap
based max-type statistic is readily applicable for testing multiple statistical hypotheses which was
illustrated by our simulation studies.

%It is clear that 
Accounting for the joint coverage probability (or the Type I error) for several or all cluster-level parameters makes SPI wider than CPI. However, only SPI are statistically valid for joint statements or comparisons between cluster-level parameters. Moreover, if one conducted studies with several surveys, SPI would contain all true
parameters in $100(1-\alpha)\%$ of all studies, whereas CPI would not cover about $D\alpha$ of them
in each survey. Our tools are equally applicable to any subset of the clusters while using all data for estimation and prediction.

Our method can be extended to account for more complex data structures such as LMM with spatial and/or temporal dependencies \citep{PratesiSalvati2008, MoralesSantamaria2019} or highly skewed response variables \citep{MouraNevesBritz2017}. Furthermore, our general idea could be also applied for a comparative analysis of nonlinear indicators of inequality or poverty obtained after the transformation of the dependent variable \citep{RojasEtal2020} or benchmarked estimators under restrictions \citep{UgarteMilitinoGoicoai2009}. 
Yet, for these cases the estimation of the variance components or MSE would have to be adjusted. Furthermore, within more complex modelling frameworks one would need to extend the asymptotic theory. In addition, different bootstrap schemes might be necessary to mimic the data-generation process in a suitable way \citep{FieldPangWelsh2008}.  
The above-mentioned extensions are beyond the scope of this paper, but are an open field for future research.

%%%%%%%%%%%%%%%%
\appendix\

\section{Appendix}\label{sec:appendix}
\subsection{Regularity conditions}\label{sec:RC}

\begin{enumerate}%[topsep=0pt]
	\setlength\itemsep{-0.5em}
	\item[R.1] $\bm{X}_d$ and $\bm{Z}_d$ are uniformly bounded such that $\sum_{d=1}^{D}\bm{X}^t_d \bm{V}_d^{-1}\bm{X}_d=\left\{O(D)\right\}_{(p+1)\times (p+1)}$.
	\item[R.2] Covariance matrices $\bm{G}_d$ and $\bm{R}_d$ have a linear structure in $\bm{\theta}$.
	\item[R.3] Convergence: $D\rightarrow \infty$, $\sup_{d\geqslant1}n_d<<\infty$ and $\sup_{d\geqslant1}q_d<<\infty$.
	\item[R.4] To ensure the nonsingularity of $\bm{\Sigma}_{\bm{\theta}}$, $ 0<\inf_{d\leqslant1}\sigma^2_{e_d} \leqslant \sup_{d\leqslant1}\sigma^2_{e_d}<\infty$ %$\mathbb{V}\mathrm{ar}(e_{dj})>0$ 
	and $\sigma^2_u\in(0,\infty)$.
	\item[R.5] $\bm{b}_d^t=\bm{k}^t_d-\bm{o}_d^t\bm{X}_d$ with $b_{di}=O(1)$ for $i=1,\dots, p+1$.
	\item[R.6] $\{\frac{\partial}{\partial \theta_j}\bm{o}_d^t\bm{X}_d\}_i=O(1)$ for $j=1,\dots,h$ and $i=1,\dots, p+1$.
	\item[R.7] $\hat{\bm{\theta}}$ satisfies: $(i)$ $\hat{\bm{\theta}}-\bm{\theta}=O_p(D^{-1/2})$, 
	$(ii)$ $\hat{\bm{\theta}}(\bm{y})=\hat{\bm{\theta}}(-\bm{y})$ and $(iii)$ $\hat{\bm{\theta}}(\bm{y}+\bm{Xr})=\hat{\bm{\theta}}(\bm{y})$ for any $\bm{r}\in \mathbb{R}^{p+1}$.
\end{enumerate}
Furthermore, we will evoke Assumptions 1-4 from \cite{chatterjee2008parametric}, which are quite technical but largely irrelevant in practice. For the sake of completeness, they are provided in our supplementary material.

\subsection{Proof of Proposition 1}\label{sec:prop_1}

We concentrate on the consistency of the bootstrap SPI. The consistency of MT procedure follows straightforwardly with some changes of notation due to the correspondence between tests and interval estimates \citep[for more details, see Corollary 2 in][]{reluga2021}. To demonstrate Proposition \ref{prop:con_max}, we make use of the result in Theorem CLL of \cite{chatterjee2008parametric}. In this section, we use some notation from their paper if it is not in conflict with ours. Consider $S_{0d}$ in \eqref{eq:s} and $S^*_{Bd}$ in \eqref{eq:s_crit_bbot} and let $\mathcal{L}_d(q) = P (S_{0d} \leqslant q)$ and $\mathcal{L}^*_d(q) = P^* (S^*_{Bd} \leqslant q)$, where $P^*(\cdot)$ stands for a probability measure induced by a parametric bootstrap. Under suitable regularity conditions \cite{chatterjee2008parametric} proved that $\mathcal{L}_d(q) = \Phi(q) + \gamma(q,\bm{\beta},\bm{\theta}, n) + O(n^{-1})$ where $\gamma(\cdot)$ is some smooth function. Furthermore, $\mathcal{L}^*_d(q)$ admits almost identical, equally short expansion with $\bm{\beta}$, $\bm{\theta}$ replaced by $\hat{\bm{\beta}}$, $\hat{\bm{\theta}}$ . Since $\mathcal{L}_d(q)\approx \Phi(q)$, we can follow the same steps as \cite{reluga2021} to prove the consistency of SPI. In particular, observe that 
\begin{equation}\label{eq:gumbel}
	P(S_0\leqslant q) = P\{ \max_{d=1,\dots, 2D} 
	(-S_{01}, \dots,-S_{0D},S_{01}, \dots,S_{0D})\leqslant q \}= \prod_{d=1}^{2D}\mathcal{L}_d(q)\approx\prod_{d=1}^{2D}\Phi(q).
\end{equation} 
The same arguments follow for $S^*_B$ with $P$ replaced by $P^*$. Moreover, the cdf of the standardised maxima of the normal distribution is in the domain of attraction of the Gumbel law. %Thus, 
If we notice that the last term in \eqref{eq:gumbel} can be approximated by the Gumbel distribution if suitably standardised, the consistency of SPI follows by applying Poyla's theorem which relates the convergence in law with sup-norm convergence \citep[see][for more details]{reluga2021}.

\begin{remark}\label{remark:GLS_vs_OLS}
	\cite{chatterjee2008parametric} estimated fixed parameters using an ordinary least squares method, whereas in our paper we used a generalized least squares. As pointed out by the authors, an asymptotic expansion still holds as soon as the weighting matrices are smooth functions of $\bm{\theta}$, which we assume in R.2 above.
\end{remark}

\begin{remark}\label{remark:smoothing_effect}
	In their original proof, 
	\cite{chatterjee2008parametric} considered a modified version of $S_{0d}$, that is $S_{0d} = \hat{\sigma}_T^{-1}(T-\hat{\mu}_T)$ where $T=\bm{f}^t(\bm{X\beta}+\bm{Zu})$ for a given fixed vector $\bm{f}$, and $\hat{\mu}_T = \bm{f}^t(\bm{X\hat{\beta}}+\bm{Z\hat{u}})$. This modification leads to some smoothing effects and results in a faster convergence rate. Nevertheless, as pointed out in Remark 6 in their paper, the analysis of the area-specific mixed effects leads to a sightly slower convergence, but is equally valid, and can be carried out along the same lines.
\end{remark}

\begin{table}[htb]
	\centering 											\setlength{\tabcolsep}{5pt}
	\begin{adjustbox}{max width=\textwidth}  
	\begin{tabular}{|cc cccc cccc|}
		\hline
		&    & \multicolumn{4}{c}{ECP (in \%)}   & \multicolumn{4}{c|}{WS (VS)}                                \\
		& $D:n_d$ & BS  & MC   & BE   & BO   & BS           & MC            & BE            & BO            \\\hline
		& 15:5 & 95.4 & 92.9 & 93.6 & 93.8 & 1.876 (0.031) & 1.754 (0.022) & 1.803 (0.026) & 1.794 (0.024) \\
		$\sigma^2_e=0.5$ & 30:5 & 95.2 & 93.9 & 92.5 & 94.4 & 1.947 (0.015) & 1.890 (0.013) & 1.871 (0.013) & 1.910 (0.013) \\
		$\sigma^2_u=1$   & 60:5 & 94.9 & 93.7 & 88.7 & 94.2 & 2.041 (0.008) & 2.011 (0.007) & 1.936 (0.007) & 2.023 (0.007) \\
		& 90:5 & 95.2 & 94.4 & 84.4 & 94.9 & 2.101 (0.006) & 2.079 (0.005) & 1.926 (0.005) & 2.088 (0.005) \\ \hline\hline
		& 15:5 & 96.7 & 91.2 & 93.8 & 94.4 & 2.695 (0.113) & 2.358 (0.046) & 2.488 (0.052) & 2.488 (0.049) \\
		$\sigma^2_e=1$   & 30:5 & 95.5 & 92.8 & 92.5 & 94.4 & 2.671 (0.027) & 2.552 (0.024) & 2.567 (0.024) & 2.608 (0.024) \\
		$\sigma^2_u=1$ & 60:5 & 95.0   & 93.7 & 89.1 & 94.5 & 2.774 (0.014) & 2.719 (0.012) & 2.631 (0.012) & 2.750 (0.012) \\
		& 90:5 & 95.2 & 94.2 & 83.2 & 94.8 & 2.850 (0.010) & 2.811 (0.009) & 2.614 (0.008) & 2.833 (0.009) \\\hline\hline
		& 15:5 & 98.3 & 87.3 & 92.3 & 96.5 & 2.816 (0.205) & 2.156 (0.065) & 2.362 (0.046) & 2.488 (0.087) \\
		$\sigma^2_e=1$ & 30:5 & 97.3 & 90.6 & 94.2 & 94.8 & 2.641 (0.050) & 2.346 (0.032) & 2.469 (0.023) & 2.485 (0.022) \\
		$\sigma^2_u=0.5$ & 60:5 & 95.3 & 92.7 & 89.7 & 94.5 & 2.616 (0.012) & 2.513 (0.015) & 2.478 (0.012) & 2.577 (0.012) \\
		& 90:5 & 95.0 & 93.0 & 83.9 & 94.6 & 2.663 (0.010) & 2.597 (0.010) & 2.441 (0.008) & 2.643 (0.009)\\\hline
	\end{tabular}	
	\end{adjustbox}
	\caption{ECP (in \%), WS and VS under NERM with normal errors and random effects. The nominal coverage probability is $95\%$.}
	\label{tab:simulations}
\end{table}

\begin{table}[htb]
	\centering 										\setlength{\tabcolsep}{4.5pt}
	\begin{adjustbox}{max width=\textwidth}
	\begin{tabular}{|ccccccccccc|}\hline
		&    &  &\multicolumn{4}{c}{ECP (in \%)}   & \multicolumn{4}{c|}{WS (VS)}                                   \\
		& $D:n_d$ & $D'$ & BS  & MC   & BE   & BO   & BS           & MC            & BE            & BO            \\\hline
		& 15:5 & 3 &95.3 & 94.9 & 91.4 & 95.1 & 2.006 (0.029) & 2.006 (0.029) & 1.922 (0.095) & 2.033 (0.033) \\
		$\sigma^2_e=1$ & 30:5 & 6 & 95.9 & 95.6 & 94.7 & 95.7  & 2.175 (0.017) & 2.175 (0.017) & 2.152 (0.031) & 2.187 (0.017) \\
		$\sigma^2_u=1$ & 60:5 & 12 &96.5 & 96.1 & 94.6 & 96.1 & 2.350 (0.009) & 2.350 (0.009) & 2.280 (0.020) & 2.358 (0.009) \\
		& 90:5 & 18 & 95.0 & 94.6 & 90.6 & 94.6 & 2.455 (0.007) & 2.455 (0.007) & 2.317 (0.019) & 2.457 (0.007)\\\hline
	\end{tabular}
	\end{adjustbox}
	\caption{ECP (in \%), WS and VS for a subset of $D'$ areas under the NERM with normal errors and random effects. The nominal coverage probability is $95\%$.}
	\label{tab:simulations_subset}
\end{table}

\begin{table}[htb]
	\centering 								\setlength{\tabcolsep}{3.5pt} 
	\begin{adjustbox}{max width=\textwidth}
	\begin{tabular}{|cccccccccc|}
		\hline
		&    & \multicolumn{4}{c}{ECP}   & \multicolumn{4}{c|}{WS (VS)}                      \\
		& $D:n_d$  & BS   & MC   & BE   & BO   & BS            & MC            & BE            & BO            \\\hline
		& 15:5 & 92.8 & 88.0 & 79.8 & 92.4 & 2.322 (0.086) & 2.322 (0.086) & 2.190 (0.262) & 2.476 (0.103) \\
		$e_{dj}\sim \chi_5(1)$   & 30:10 & 91.4 & 90.2 & 85.7 & 91.2 & 1.878 (0.012) & 1.876 (0.012) & 1.794 (0.021) & 1.899 (0.012) \\
		$u_{d}\sim \chi_5(0.5)$ & 60:20 & 91.3 & 90.5 & 80.6 & 91.1 & 1.455 (0.002) & 1.455 (0.002) & 1.334 (0.008) & 1.460 (0.002) \\
		& 90:30 & 92.3 & 92.5 & 75.1 & 92.8 & 1.238 (0.001) & 1.238 (0.001) & 1.096 (0.005) & 1.241 (0.001) \\\hline\hline
		& 15:5 & 93.1 & 89.9 & 87.6 & 91.6 & 1.742 (0.046) & 1.742 (0.046) & 1.738 (0.076) & 1.783 (0.050) \\
		$e_{dj}\sim\chi_5(0.5)$ & 30:10 & 90.8 & 90.4 & 85.6 & 90.4 & 1.370 (0.007) & 1.370 (0.007) & 1.301 (0.011) & 1.378 (0.007) \\
		$u_d\sim N(1)$ & 60:20 & 91.5 & 90.9 & 81.5 & 91.3 & 1.043 (0.001) & 1.043 (0.001) & 0.955 (0.004) & 1.046 (0.001) \\
		& 90:30 & 92.9 & 92.3 & 75.8 & 92.6 & 0.883 (0.000) & 0.883 (0.000) & 0.781 (0.003) & 0.885 (0.000)\\\hline\hline
		& 15:5 & 90.5 & 83.5 & 54.9 & 95.1 & 2.111 (0.105) & 2.111 (0.105) & 1.629 (0.412) & 2.492 (0.164) \\
		$e_{dj}\sim t_6(1)$   & 30:10 & 92.5 & 89.7 & 86.4 & 91.9 & 1.794 (0.014) & 1.794 (0.014) & 1.723 (0.030) & 1.843 (0.014) \\
		$u_{d}\sim t_6(0.5)$   & 60:20 & 91.6 & 91.2 & 79.7 & 91.9 & 1.419 (0.002) & 1.419 (0.002) & 1.304 (0.008) & 1.428 (0.002) \\
		& 90:30 & 92.1 & 91.9 & 73.6 & 92.2 & 1.217 (0.001) & 1.217 (0.001) & 1.079 (0.005) & 1.222 (0.001) \\\hline\hline
		& 15:5 & 92.7 & 89.0 & 87.3 & 91.1 & 1.750 (0.043) & 1.750 (0.043) & 1.740 (0.064) & 1.791 (0.049) \\
		$e_{dj}\sim t_6(0.5)$ & 30:10 & 92.4 & 91.4 & 85.6 & 91.9 & 1.365 (0.007) & 1.365 (0.007) & 1.285 (0.012) & 1.373 (0.007) \\
		$u_{d}\sim N(1)$      & 60:20 & 93.5 & 93.7 & 80.7 & 93.9 & 1.041 (0.001) & 1.041 (0.001) & 0.945 (0.004) & 1.044 (0.001) \\
		& 90:30 & 94.2 & 93.8 & 72.3 & 94.0 & 0.882 (0.000) & 0.882 (0.000) & 0.773 (0.003) & 0.884 (0.000)\\\hline\hline
		$e_{dj}\sim st_{5, 1.25}(2)$ & 26:50 & 95.5  & 93.9 & 89.2 & 96.5 & 1.257 (0.002)  & 1.257 (0.002) & 1.194 (0.009) & 1.328 (0.001) \\
		$u_d\sim N(0.25)$  & 52:100 & 94.3 & 93.8 & 81.4 & 94.6 & 0.992 (0.000) & 0.992 (0.000) & 0.892 (0.003) & 1.002 (0.000) \\\hline 
	\end{tabular}	
	\end{adjustbox}
	\caption{ECP (in \%), WS and VS under NERM with chi-square, t-distributed and skewed t-distributed departures from normality, centred and rescaled to variances given in parentheses. The nominal coverage probability is $95\%$.}
	\label{tab:simulations_chi_dist}
\end{table}

\begin{table}[htb]
	\centering
	\begin{adjustbox}{max width=\textwidth}
	\begin{tabular}{|cc cccc cccc|}\hline
		&    & \multicolumn{4}{c}{ECP (in \%)}  & \multicolumn{4}{c|}{WS (VS)}                               \\
		& D  & BS    & MC   & BE   & BO   & BS             & MC            & BE            & BO            \\ \hline
		S.1 & 15 & 97.3 & 95.6 & 96.7 & 96.5 & 3.728 (0.016) & 3.516 (0.024) & 3.672 (0.040) & 3.691 (0.019) \\
		& 30 & 96.6 & 95.2 & 94.8 & 96.6 & 3.792 (0.017) & 3.664 (0.023) & 3.688 (0.045) & 3.818 (0.013) \\
		& 60 & 95.7 & 92.6 & 89.6 & 93.9 & 3.973 (0.014) & 3.804 (0.031) & 3.760 (0.103) & 3.873 (0.027) \\
		& 90 & 95.2 & 93.3 & 84   & 94.4 & 4.024 (0.016) & 3.920 (0.025) & 3.694 (0.090) & 3.970 (0.022) \\ \hline\hline
		
		S.2 & 15 & 98.0 & 95.7 & 97.6 & 95.9 & 4.073 (0.034) & 3.749 (0.061) & 4.005 (0.076) & 3.962 (0.096) \\
		& 30 & 97.1 & 95.6 & 94.8 & 96.1 & 3.795 (0.017) & 3.667 (0.023) & 3.690 (0.046) & 4.028 (0.040) \\
		& 60 & 97.4 & 93.4 & 91.6 & 94.9 & 4.198 (0.035) & 3.956 (0.067) & 3.981 (0.147) & 4.029 (0.064) \\
		& 90 & 96.6 & 93.9 & 86.1 & 94.6 & 4.218 (0.037) & 4.006 (0.054) & 3.880 (0.123) & 4.119 (0.053)\\\hline
	\end{tabular}
	\end{adjustbox}
	\caption{ECP (in \%), WS and VS under FHM with normal errors and random effects. The nominal coverage probability is $95\%$.}
	\label{tab:simulationsFH}
\end{table}

\begin{table}[htb]
	\centering
	\begin{adjustbox}{max width=\textwidth}	
	\begin{tabular}{|ccc ccc ccc ccc c|}\hline
		\multicolumn{6}{|c}{A Coruña} & & 
		\multicolumn{6}{c|}{Lugo}                  \\\hline
		Min & $Q_1$ & $Q_2$ & $Q_3$ & Max & Total & & Min & $Q_1$ & $Q_2$ & $Q_3$  & Max  & Total \\
		18  & 36    & 90    & 197   & 930 & 3231 & & 18  & 76.5  & 90    & 193.5  & 449  & 1619  \\\hline
		\multicolumn{6}{|c}{Ourense}     &    & \multicolumn{6}{c|}{Pontevedra}             \\\hline
		Min & $Q_1$ & $Q_2$ & $Q_3$ & Max & Total & & Min & $Q_1$ & $Q_2$ & $Q_3$  & Max  & Total \\      18  & 22.5  & 90    & 158   & 683 & 1637 & & 36  & 94.5  & 162   & 368    & 1008 & 2716 \\\hline
	\end{tabular}
\end{adjustbox}
	\caption{Descriptive statistics of the number of units across \textit{comarcas} in provinces of Galicia. Statistics: Min - minimum, Q1 - first quartile, Q2 - median, Q3 - third quartile, Max - maximum.}
	\label{tab:Desc_Stats}
\end{table}

\begin{table}[htb]
	\centering 											\setlength{\tabcolsep}{2pt} 
	\begin{adjustbox}{max width=\textwidth}
	\begin{tabular}{|lllllll|}
		\hline
		\multicolumn{2}{|l}{Dependent variable}       & \multicolumn{2}{l}{Dir Mean} & \multicolumn{3}{l|}{Dir Stdev}    \\ \hline
		Inc & Monthly household income  &  \multicolumn{2}{l}{1914.884}      & \multicolumn{3}{l|}{13.766}  \\ \hline
		\multicolumn{2}{|l}{Characteristics of the household} & Mean    & Stdev  & $\hat{\beta}$ & S.E.  & p-value     \\ \hline
		Type1  & = 1 if households consists of 1 person & 0.208   & 0.406 & -1616.177 & 33.191 & 0.000\\
		Type2  & = 1 if households consists of more than 1 person  & 0.023   & 0.149  & -933.004  & 65.622 & 0.000\\
		Type3 & = 1 if households consists of a couple with children & 0.304 & 0.460  & -601.977  & 31.416 & 0.000\\
		Type4 & = 1 if households consists of a couple  without children & 0.246 & 0.431  & -983.258  & 31.972 & 0.000\\
		Type5  & = 1 if households consists of a single parent & 0.093 & 0.290 & -1056.531 & 39.714 & 0.000\\
		Type67  & = 1 if households consists of one or several centres or other  & \multicolumn{5}{l|}{Benchmark variable: dropped in fitting}\\
		Dif1 & = 1 if a lot of difficulties coming to the end of a month & 0.123 & 0.328 & -982.786  & 31.132 & 0.000 \\
		Dif2 & = 1 if some difficulties coming to the end of a month & 0.445 & 0.497  & -514.372  & 20.266 & 0.000 \\
		Dif3  & = 1 if no difficulties coming to the end of a month & \multicolumn{5}{l|}{Benchmark variable: dropped in fitting}\\
		Ten1 & =1 if property without mortgage & 0.663 & 0.473 & 301.229 & 26.218 & 0.000\\
		Ten2 & =1 if property with mortgage & 0.168 & 0.374 & 417.126 & 32.072 & 0.000 \\
		Ten34 & =1 if ceded, rental or another type of property & \multicolumn{5}{l|}{Benchmark variable: dropped in fitting}\\\hline
		\multicolumn{2}{|l}{Characteristics of the household head} & & & & &\\ \hline
		Educ1 & = 1 if primary education & 0.232 & 0.422 & -902.255  & 30.763 & 0.000 \\
		Educ2 & = 1 if  secondary education & 0.515 & 0.500 & -731.925  & 23.455 & 0.000 \\
		Educ3 & = 1 if higher education & \multicolumn{5}{l|}{Benchmark variable: dropped in fitting}\\
		Age1 & = 1 if 45 $\leqslant$ age $\leqslant$ 64 & 0.377 & 0.485 & 206.039   & 19.988 & 0.000 \\
		Age2 & = 1 if age $<$ 45 or age $>$ 64 & \multicolumn{5}{l|}{Benchmark variable: dropped in fitting}\\ \hline
		Intercept &  & - & - & 3308.762 & 46.481  &0.000 \\ \hline
	\end{tabular}
	\end{adjustbox}
	\caption{Descriptive statistics and coefficient estimates with standard errors and p-values.}	\label{tab:desc_stat}
\end{table} 

\section{Supplementary material}\label{sec:SM}
This section contains supplementary material to the main article which introduced simultaneous prediction intervals and multiple testing procedures for a joint, comparative analysis of mixed parameters under linear mixed models. We introduce an extension of the multiple testing procedurein order to guarantee a strong control of a family-wise error. We further state the theorem of \cite{chatterjee2008parametric} together with its assumptions and some additional details. In addition, we present a proof of Proposition 2. Finally, we discuss and analyze the consequences of deviations from standard model assumptions. The extension of the multiple testing procedure and the analysis of the departures from the model assumptions are studied empirically in simulations which complement numerical results presented in the main document.

\section{The step-down testing procedure: description and numerical results}\label{sec:step_down}

The testing procedure described in Section 3 of the main document controls weakly for the family-wise error rate (FWER). If one aims at testing multiple hypotheses like 
\begin{equation}\label{eq:test_ind}
	H_{0d}:\mu_d=h_d \text{ for all } d\in [D] \text{ vs. } H_{1d}:\mu_d\neq h_d, \text{ for some } d \in [D],
\end{equation}
then we can use the step-down technique of \cite{romano2005exact} which assures a strong control of the FWER. Originally, \cite{romano2005exact} developed his multiple testing procedure for fixed effects in a regression context. We adapt it to make it suitable for mixed parameters within the framework of LMM. 
%, i.e., it would reject at least one true null hypothesis with a probability not greater than a specified level $\alpha$ over the data-generating precess, especially, over the set of the true null hypotheses. 
More specifically, let $\Omega$ be the space for all data-generating processes and $w$ be the true one. Each $H_{0d}$ is equivalent to $\omega\in\Omega_d$ for some $\Omega_d \subseteq \Omega$. For any $\vartheta\subseteq[D]$, denote $\Omega^{\vartheta}=(\cap_{d\in \vartheta}\Omega_d)\cap 
(\cap_{d \notin \vartheta}\Omega^c_d)$ with $\Omega^c_d=
\Omega \setminus \Omega_d$. The strong control of the FWER implies that
\begin{equation}\label{eq:cond_Propo_RW}
	\sup\limits_{\vartheta \subseteq [D] }
	\sup\limits_{\omega \in \Omega^{\vartheta}} 
	P_{\omega}(\text{reject at least one hypothesis } H_{0d}, d \in \vartheta) \leqslant \alpha + o(1).
\end{equation}
We summarise the step-down procedure of \cite{romano2005exact} in the following way. Consider $t_{H_d}$, $d\in[D]$ defined in Section 3 of the main document and denote by $c_{\vartheta}(1-\alpha)$ an estimate for the $(1-\alpha)$-quantile of $\max_{d\in \vartheta} |t_{H_d}|$. Let  $\vartheta(1)=[D]$ and reject all hypotheses $H_{0d}$ for which  $t_{H_d}>c_{\vartheta(1)}(1-\alpha)$. If no hypothesis is rejected, then stop. Otherwise, let $\vartheta(2)$ be the set of null hypotheses not rejected at the first step. At step $k$, let $\vartheta(k)\subseteq[D]$ be the set of hypotheses not rejected at step $k-1$. Reject all hypotheses $H_{0d}$, $d\in \vartheta(k)$ for which $t_{H_d}>c_{\vartheta(k)}(1-\alpha)$. If no hypothesis is rejected, stop. Proceed in this way until the algorithm stops. \cite{romano2005exact} proved that their procedure controls strongly for FWER, that is
\begin{equation*}\label{eq:cond_1_RW}
	\begin{split}
		& c_{\vartheta}(1-\alpha)\leqslant c_{\vartheta'}(1-\alpha), \text{ for } \vartheta \subseteq \vartheta',\\
		\sup\limits_{\vartheta \subseteq [D] }
		&	\sup\limits_{\omega \in \Omega^{\vartheta}} 
		\mathrm{P}_{\omega}(\max\limits_{d \in \vartheta }|t_{H_d}| > c_{\vartheta}(1-\alpha)) \leqslant \alpha + o(1).
	\end{split}
\end{equation*}
A direct consequence of Proposition 1 from the main document 
is the following corollary.  
\begingroup
\renewcommand\thecorollary{3}
\begin{corollary}\label{corol:step_down}
	Under assumptions of Proposition 1, the step-down procedure of \cite{romano2005exact} with bootstrap approximations of the critical value $c_\vartheta(1-\alpha)$ provides a strong control of the FWER, satisfying condition in \eqref{eq:cond_Propo_RW}. 
\end{corollary}
In other words, Corollary \ref{corol:step_down} says that we can use
bootstrap approximations of the critical value defined in Section 4 of the main document in the step-down procedure of \cite{romano2005exact} which controls strongly for FWER. 
%algorithm  critical values approximated by 
%This means, Proposition 1 (of the main article) provides us with a practical method to obtain appropriate critical values $c_{\vartheta}(1-\alpha)$ in the above described step down-procedure to obtain a strong control of the FWER. 
%This will be verified now by additional simulations for the nested error regression model (NERM).

\begin{table}[hbt]
	\centering	
	\setlength{\tabcolsep}{5pt}
	\begin{tabular}{|ccccccccc|}		\hline
		& \multicolumn{4}{c}{BS}            & \multicolumn{4}{c|}{BO}           \\		& $D=15$ & $D=30$ & $D=60$ & $D=90$ & $D=15$ & $D=30$ & $D=60$ & $D=90$ \\ \hline
		$\sigma^2_e=0.5$, $\sigma^2_u=1$ & 0.043 & 0.042 & 0.049 & 0.048 & 0.048 & 0.044 & 0.508 & 0.046 \\
		$\sigma^2_e=1$, $\sigma^2_u=1$   & 0.039 & 0.037 & 0.045 & 0.042 & 0.042 & 0.042 & 0.046 & 0.044 \\
		$\sigma^2_e=1$, $\sigma^2_u=0.5$ & 0.047 & 0.040 & 0.038 & 0.037 & 0.025 & 0.039 & 0.044 & 0.040\\\hline
	\end{tabular}  
	\caption{Empirical FWER under NERM. The nominal FWER is $0.05$.}	\label{tab:RW_simulations}
\end{table}
We compare the finite sample performance of the step-down method of \cite{romano2005exact} with Bonferroni's procedure which is treated as a benchmark in the main document. To this end, we consider multiple two-sided testing hypotheses $H_{0,d}: \mu_d=h_d$ for all $d\in[D]$ for three scenarios under NERM without departures from normality. We assume that $\mu_d=h_d+1$ for $d=1,\dots,D/5$. The performance criterion is an empirical FWER while the nominal level of FWER is set to $\alpha=0.05$. The analysis of Table \ref{tab:RW_simulations} confirms that the bootstrap step-down procedure controls strongly for FWER, whereas Bonferroni's method fails to do it under $\sigma^2_e=0.5$, $\sigma^2_u=1$ for $D=60$.  Except for the third scenario and $D=15$, bootstrap procedure provides lower or equally good FWER as the Bonferroni's technique. Recall that the latter works relatively well, because we designed the simulations such that the simultaneous tests are asymptotically independent.   

%%%%%%%%%%%%%%%%%%%%%%%%%%%%%%%%
\section{Theorem CLL, its assumptions and additional details}\label{sec:CLL_assumptions}

To demonstrate Proposition 1, we made use of Theorem CLL of \cite{chatterjee2008parametric}. For the sake of completeness, we state it below with the technical assumptions adapted to our notation and modelling problem. Similarly as in the main document, we use some of the notation of \cite{chatterjee2008parametric} whenever it is not in conflict with ours. 
\begin{theorem} \label{theorem:CLL}
	(Theorem 3.1 of \cite{chatterjee2008parametric}). Let $T=\bm{f}^t(\bm{X\beta}+\bm{Zu})$, $\bm{f}\in\mathbb{R}^{n}$ a vector and a conditional distribution of $T$ given $\bm{y}$ be $N(\mu_T,\sigma^2_T)$, where
	\begin{eqnarray*}
		\mu_T&=&\bm{f}^t\bm{X}\bm{\beta}+\bm{f}^t\bm{ZG}\bm{Z}^t\bm{V}^{-1}(\bm{y}-\bm{X\bm{\beta}}),  \\
		\sigma^2_T&=&\bm{f}^t\bm{Z}(\bm{G}-\bm{GZ}^t\bm{V}^{-1}\bm{ZG})\bm{Z}^t\bm{f} \ .
	\end{eqnarray*}
	Let $T^*=\bm{f}^t(\bm{X}\hat{\bm{\beta}}+\bm{Z}\bm{u}^*)$ be a bootstrap equivalent of $T$, 
	$\mathcal{L}_n$ the cdf of $\hat{\sigma}_T^{-1}(T-\hat{\mu}_T)$ with $\bm{\theta}$
	replaced by $\hat{\bm{\theta}}$, and $\mathcal{L}^*_n$ the cdf of $\hat{\sigma}^{-1*}(T^*-\hat{\mu}^*_T)$.
	Suppose Assumptions 1-4 hold and $p+1+h=o(n)$.
	Then, for any continuity point $v$
	\begin{equation*}
		\sup\limits_{q\in\mathbb{R}} \left\lvert\mathcal{L}_n(v)-\mathcal{L}^*_n(v)\right\rvert=O_P\{(p+h+1)^3n^{-3/2}\}.
	\end{equation*} 
	Moreover, $\mathcal{L}_n(v)$ admits a short asymptotic expansion, i.e., 
	$\mathcal{L}_n(v)=\Phi(v)+h^2n^{-1}\gamma(q,\bm{\beta},\bm{\theta})+O(h^3n^{-3/2})$ where $\Phi$ stands for the cdf of a standard Gaussian random variable.  
\end{theorem}

Before stating the technical assumptions, recall that $\bm{V}=\bm{R} + \bm{Z G}\bm{Z}^t$ where $\bm{R}$ and $\bm{G}$ denote the covariance matrices of errors $\bm{e}$ and random effects $\bm{u}$, respectively. In addition, the vector of unknown variance components is $\bm{\theta}\in\mathbb{R}^{h}$. For the readers convenience, we keep the numbering of the assumptions as in \cite{chatterjee2008parametric}.

{\bf Assumption 1}: Conditions on vector $\bm{f}$.
\begin{eqnarray*} 
	&& \| \bm{X^tf} \| = O(1), \\  
	&& \| \bm{X^t V^{-1} ZGZ^t f} \| = O(1), \\  
	&& \bm{f^t ZGZ^t f}  = O(1), \\  
	&& \bm{f^t ZGZ^t V^{-1} ZGZ^t f}  = O(1). 
\end{eqnarray*}
Furthermore, $\sigma_T^2=\bm{f}^t\bm{Z}(\bm{G}-\bm{GZ}^t\bm{V}^{-1}\bm{ZG})\bm{Z}^t\bm{f}>M>0$ for a constant $M>0$.

{\bf Assumption 2}: Regularity conditions on design matrix $\bm{X}$.
\begin{eqnarray*} 
	\sup_{1\le i\le n} \sum_{j=1}^{p+1} \left( \sum_{l=1}^n x_{jl} \bm{V}_{li}^{1/2} \right)^2 = O\left(\frac{p+1}{n}\right).
\end{eqnarray*}
In addition,  the smallest eigenvalue of $\bm{X^tX}/n$ is bounded away from zero.

{\bf Assumption 3}: Conditions on the variance matrices $\bm{G}$, $\bm{R}$ and their estimates $\bm{G}(\hat{\bm{\theta}})$, $\bm{R}(\hat{\bm{\theta}})$.

\noindent The eigenvalues of $\bm{G}$, $\bm{R}$ lie in $(L^{-1},L)$ for some $L>1$, and the eigenvalues of $\bm{G}(\hat{\bm{\theta}})$, $\bm{R}(\hat{\bm{\theta}})$ lie in $(L^{-1}/2,2L)$. Furthermore, the eigenvalues of $\bm{V}$ lie in a compact set on the positive half of the real line. Finally, let $\bm{\Lambda}_G(\hat{\bm{\theta}})$,  $\bm{\Lambda}_R(\hat{\bm{\theta}})$ be diagonal matrices implicitly defined via $\bm{G}(\hat{\bm{\theta}}) = \bm{G}_1^{1/2} \bm{\Lambda}_{G}(\hat{\bm{\theta}}) \bm{G}_2^{t/2}$ with $\bm{G}= \bm{G}_1^{1/2} \bm{G}_2^{t/2}$ and $\bm{R}(\hat{\bm{\theta}}) = \bm{R}_1^{1/2} \bm{\Lambda}_{R}(\hat{\bm{\theta}}) \bm{R}_2^{t/2}$ with $\bm{R}= \bm{R}_1^{1/2} \bm{R}_2^{t/2}$ depending only on $\bm{\theta}\in\mathbb{R}^{h}$. It holds that all entries of $\bm{\Lambda}_G(\hat{\bm{\theta}})$,  $\bm{\Lambda}_R(\hat{\bm{\theta}})$ have three bounded continuous derivatives (partial derivatives with respect to the elements of $\hat{\bm{\theta}}$). 

{\bf Assumption 4}: Conditions on the estimator of the variance parameters $\hat{\bm{\theta}}$.

Define $\bm{E}=\sqrt{h/n} (\hat{\bm{\theta}} -\bm{\theta})$, then all moments of $\|\bm{E} \|$ are of order $O(1)$ and for the elements $E_j$ of vector $E$, it holds that $\forall\ j,l=1,...,h$
\begin{eqnarray*} 
	%	&& \mbox{Let  } \bm{E}=\sqrt{h/n} (\hat{\bm{\theta}} -\bm{\theta}), \mbox{ then all moments of } \| \bm{E} \| 
	%	\mbox{ are of order } O(1)  \mbox{  and }  \forall j,l=1,...,h, \text{we have:}\\
	&&  \mathbb{E} (E_j) = O(\sqrt{h/n}), \qquad \mathbb{E} (E_jE_l) = O(\sqrt{h/n}), \\  
	&&   \mathbb{E} \{E_j(\bm{Zu}+\bm{e})_i\} = O(\sqrt{h/n}), \qquad \mathbb{E} \{E_jE_l(\bm{Zu}+\bm{e})_i\} = O(\sqrt{h/n}) \ , \ i=1,...,n .
\end{eqnarray*}

\section{Alternative argument for consistency of bootstrap SPI}
%\begin{remark}
The consistency of $\mathcal{I}_{1-\alpha}^{BS}$ and Proposition 1 can be also based on a heuristic argument of \cite{hall1990simultaneous} combined with the asymptotic expansion in Theorem \ref{theorem:CLL}.
Ideally, $c_{S_0}(1-\alpha)$ would be determined from 
\begin{equation*}
	\pi\left\{c_{S_0}(1-\alpha)\right\}=P\left\{ -c_{S_0}(1-\alpha)\leqslant S_{0d} \leqslant c_{S_0}(1-\alpha) \; \forall d \in [D]\right\}=1-\alpha.
\end{equation*} 
Since $\pi\left\{c_{S_0}(1-\alpha)\right\}$ is unknown, we approximate it by bootstrap such that
\begin{equation*}
	\pi^*\{c_{BS}(1-\alpha)\}=P\left\{ -c_{BS}(1-\alpha)\leqslant S^*_{Bd} \leqslant c_{BS}(1-\alpha) \; \forall d \in [D]|\mathcal{W}\right\}=1-\alpha
\end{equation*}
with $\mathcal{W}=\{(y_{dj},\bm{x}_{dj},\bm{z}_{dj}),d\in[D],j\in[n_d] \} $. If we prove that $\pi$ and $\pi^*$ are asymptotically close up to order $O_P((h^2n^{-3/2})$, then it implies the same order of accuracy for $\mathcal{I}^{BS}_{1-\alpha}$. Define
\begin{equation*}
	R=\{\bm{x}\times \bm{z}\in\mathcal{X}\times\mathcal{Z}:-c_{S_0}(1-\alpha)\leqslant S_{0d} \leqslant c_{S_0}(1-\alpha) \, \forall d \in [D]\},
\end{equation*}
which can be represented as a finite number of unions and intersection of convex sets. This number is bounded uniformly for $D\geqslant2$ and $c_{S_0}(1-\alpha)>0$. Observe that $\pi=\int_{\mathbb{R}} d\mathcal{L}_n$. Theorem \ref{theorem:CLL} shows that for all continuity points $v$ the {cdf}'s $\mathcal{L}_n$ and $\mathcal{L}^*_n$ converge to the same limit at the desired speed, and the same speed is maintained in the convergence of $\pi$ and $\pi^*$. Since $\pi$ is defined as an integral of $d\mathcal{L}_n$ over $\mathbb{R}$, a direct consequence of Lemma \ref{theorem:CLL} is $\sup_{k\in\mathbb{R}}| \pi(k)-\pi^*(k)|=O_P(h^2n^{-3/2})$. 		
%\end{remark}

%%%%%%%%%%%%%%%%%%%%%%%%%%%%%%%%
\section{Proof of Proposition 2}\label{sec:prop_2}

To derive approximation formulas in Proposition 2, 
we suppose for the moment that the manifold $\mathcal{M}=\left\{\mathcal{Q}(\bm{c}),\bm{c}\in \mathcal{C}\right\}$ 
has no boundary, that is, Euler-Poincar\'{e} characteristic $\mathcal{E}=0$. 
In addition, $\bm{l}_M$, $\hat{\bm{l}}_M$, $\bm{e}_M$, $\hat{\bm{e}}_{M}$, 
$\lambda=\hat{\bm{e}}_M-\bm{e}_M$ and the other assumptions remain as defined in the main document. 
For some $\bm{x}\in \mathcal{X}$, $\bm{z}\in\mathcal{Z}$ and $\bm{c}\in\mathcal{C}$ 
the difference between $\bm{x}^t\bm{\beta}+\bm{z}^t\bm{u}$ and its estimate can be bounded by
\begin{equation}\label{eq:bound_dif}
	\left|\bm{x}^t\bm{\beta}+\bm{z}^t\bm{u}-\hat{\bm{l}}^t\bm{y}\right|=
	\left|\hat{\bm{l}}_M^t \hat{\bm{e}}_M\right|=\left|\bm{l}_M^t \bm{e}_M +
	\left(\hat{\bm{l}}_M-\bm{l}_M\right)^t\bm{e}_M + \hat{\bm{l}}^t_M\lambda_M 
	\right|\leqslant \left| \bm{l}_M^t \bm{e}_M \right| + \eta(\bm{x}),
\end{equation}
where $\eta(\bm{x})=\left|\left(\hat{\bm{l}}_M-\bm{l}_M\right)^t\bm{e}_M +
\hat{\bm{l}}^t_M\lambda_M \right|$. If we normalize the first term on the 
right hand side, it is straightforward to see that
\begin{equation*}
	Z=\frac{\left\langle\bm{l}_M,\bm{e}_M\right\rangle}{\sigma_e||\bm{l}_M||}=
	\left\langle\frac{\bm{l}}{||\bm{l}_M||},\frac{\bm{e}_M}{\sigma_{e}}\right\rangle=
	\left\langle\mathcal{Q},\frac{\bm{e}_M}{\sigma_{e}}\right\rangle,
\end{equation*}
which coincides with the formula of a Gaussian random variable in Section 5.1 of the main document. Here, $\bm{e}_M/\sigma_{e}$ is an $n$-vector of normally distributed random variables. Following the derivation of \cite{sun1994simultaneous} and \cite{sun1999confidence}, one needs to choose $c_{VT}(1-\alpha)$ such that 
\begin{equation*}
	\begin{split}
		\alpha&= P\left(\sup_{\bm{c}\in\mathcal{C}} \frac{\left|\bm{x}^t\bm{\beta}+\bm{z}^t\bm{u}-\hat{\bm{l}}^t\bm{y}\right|}{\hat{\sigma}_e||\hat{\bm{l}}_M||} > c_{VT}(1-\alpha) \right)\leqslant  P\left[    \sup_{\bm{c}\in\mathcal{C}}   \left[ \left\{  \left\lvert\frac{\left\langle\bm{l}_M,\bm{e}_M\right\rangle}{\sigma_e||\bm{l}_M||}\right\rvert + 
		\frac{\eta(\bm{x})}{ \sigma_e||\bm{l}_M||}   \right\}   \frac{||\bm{l}_M||}{||\hat{\bm{l}}_M||}  \right]   >  c_{VT}(1-\alpha)\frac{\hat{\sigma}_e}{\sigma_e} \right]\\
		&\leqslant P\left\{\sup_{\bm{c}\in\mathcal{C}} \left|Z\right| >
		c_{VT}(1-\alpha)\frac{\hat{\sigma}_e}{\sigma_e} \xi -\eta \right\} =2P\left\{\sup_{\bm{c}\in\mathcal{C}} Z >  c_{VT}(1-\alpha)\frac{\hat{\sigma}_e}{\sigma_e} \xi -\eta \right\}   , 
	\end{split}
\end{equation*}
where $\xi=\inf\limits_{\bm{c}\in\mathcal{C}}\frac{||\hat{\bm{l}}_M||}{||\bm{l}_M||}$ is the minimum of the ratio between estimated $\hat{\bm{l}}_M$ and the true $\bm{l}_M$, and $\eta=\sup\limits_{\bm{c}\in\mathcal{C}}\frac{\eta(\bm{x})}{ \sigma_e||\bm{l}_M||}$ accounts for the difference in the estimation of variance parameters. When $\bm{\theta}$ is obtained using some consistent estimator (for example REML), then \cite{jiang1998asymptotic} proved that $\hat{\sigma}_e$ is asymptotically independent of $Z$ and
\begin{equation*}
	\xi=1+o_p(1), \quad
	\eta=o_p(1) \quad \text{as} \quad 
	n\rightarrow\infty.
\end{equation*}
Therefore, $\xi$ and $\eta$ can be bounded by positive constants $\xi\leqslant\xi_0$ and $\eta\leqslant\eta_{0}$ in probability as $n\rightarrow\infty$, and we obtain the approximation
\begin{equation}\label{eq:Proba}
	\alpha\leqslant 2P\left(\sup_{\bm{c}\in\mathcal{C}} Z > \frac{c_{VT}(1-\alpha)}{\nu^{1/2}} 
	\frac{\hat{\sigma}_e\nu^{1/2}}{\sigma_e} \xi_0 -\eta_0 \right)\ + o(\alpha).
\end{equation}

Under our setting, the variable $\nu^{1/2}\hat{\sigma}/\sigma$ is approximately $\chi$ distributed with $\nu$ degrees of freedom and a pdf
\begin{equation*}
	f(w,\nu)=\frac{w^{\nu-1}e^{-w^2/2}}{2^{\nu/2-1}\Gamma(\nu/2)}.
\end{equation*}
Since we deal with a Guassian random variable, we can adjust the  approximations formulas of \cite{sun1993tail} to retrieve a value of $c_{VT}(1-\alpha)$ for $p=1$, $p=2$ and $p\geqslant3$. First of all, let  us focus on the cases $p=1$ and $p=2$, where we need to find a solution of
\begin{equation*}
	\alpha\leqslant 2\int_{0}^{\infty} P\left\{\sup_{\bm{c}\in\mathcal{C}} Z >
	\frac{c_{VT}(1-\alpha)\xi_0 }{\nu^{1/2}} w -\eta_0 \right\}\
	f(w,\nu)dw+o(\alpha).
\end{equation*}
We develop two expressions using Taylor expansions. Let $c'_{VT}(1-\alpha)=c_{VT}(1-\alpha)\xi_0$ and $\frac{c'_{VT}(1-\alpha)w}{\nu^{1/2}}=j(w)$. 
Then for any $\eta \rightarrow 0$ we have
\begin{eqnarray*}
	& & \exp\left\{-\frac{1}{2}\left(j(w)-\eta\right)^2\right\} =\exp\left\{-\frac{1}{2}j^2(w)\right\}+
	\exp\left\{-\frac{1}{2}j^2(w)\right\}j(w)\eta+o(\eta)  \\ & &
	=A_1+A_2\eta+o(\eta) , \mbox{  and } \\ & & 
	\left\{j(w)-\eta\right\}\exp\left\{-\frac{1}{2}\left(j(w)-\eta\right)^2\right\}
	=j(w)\exp\left\{-\frac{1}{2}j^2(w)\right\} + \left[j'(w)\exp\left\{-\frac{1}{2}j^2(w)\right\} \right. \\
	& & \left.  -j^2(w) \exp\left\{-\frac{1}{2}j^2(w)\right\} \right](-\eta)+o(\eta) =j(w)\exp\left\{-\frac{1}{2}j^2(w)\right\}-\frac{c'_{VT}(1-\alpha)\eta}{\nu^{1/2}}\exp\left\{-\frac{1}{2}j^2(w)\right\}\\
	& & +j^2(w)\exp\left\{-\frac{1}{2}j^2(w)\right\}\eta+o(\eta) =A_2-A_1 \frac{c'_{VT}(1-\alpha) \eta}{\nu^{1/2}} + A_3 \eta+o(\eta) .
\end{eqnarray*}
Using a $\chi$ distribution, we have for ${\cal A}_j := \int_{0}^{\infty} A_j f(w,\nu) dw$, $j=1,2,3$:
\begin{eqnarray*}
	{\cal A}_1 &=& \int_{0}^{\infty}\exp\left\{ -\frac{1}{2} j^2(w) \right\}   f(w,\nu) dw
	=   \int_{0}^{\infty}\exp\left\{ - \frac{ c'^2_{VT}(1-\alpha)}{2\nu}  w^2 \right\} f(w,\nu)dw 
	\\ & = & \left\{ 1+\frac{c_{VT}(1-\alpha)^{2}\xi_0^2}{\nu} \right\}^{-\nu/2},
	\\
	{\cal A}_2 &=&  \int_{0}^{\infty}\exp\left\{ -\frac{1}{2} j^2(w) \right\}  j(w)  f(w,\nu) dw
	=  \frac{ c'_{VT}(1-\alpha)}{\nu^{1/2}}\int_{0}^{\infty}\exp\left\{- \frac{ c'^2_{VT}(1-\alpha)}{2\nu}  w^2 \right\}  w f(w,\nu)dw  \\ & = &  \frac{ c_{VT}(1-\alpha)\xi_0}{\nu^{1/2}} \frac{2^{1/2} \Gamma\left(\frac{\nu+1}{2}\right) }{\Gamma\left(\frac{\nu}{2}\right) 
		\left\{1+\frac{c_{VT}(1-\alpha)^2\xi_{0}^{2}}{\nu} \right\}^{(\nu+1)/2}  },
	\\
	{\cal A}_3 &=&  \int_{0}^{\infty}\exp\left\{ -\frac{1}{2} j^2(w) \right\}  j^2(w)  f(w,\nu) dw
	=  \frac{ c'^2_{VT}(1-\alpha)}{\nu}\int_{0}^{\infty}\exp\left\{ - \frac{ c'^2_{VT}(1-\alpha)      }{2\nu}  w^2 \right\} w^2 f(w,\nu)dw 
	\\ & = &  \frac{ c^2_{VT}(1-\alpha) \xi_0^{2} }{\nu} \frac{2 \Gamma\left(\frac{\nu+2}{2}\right) }{\Gamma\left(\frac{\nu}{2}\right) 
		\left\{1+\frac{c_{VT}(1-\alpha)^2\xi_{0}^{2}}{\nu} \right\}^{(\nu+2)/2}  }  .
\end{eqnarray*}
To find an approximation for the model with $p\geqslant 3$, we modify a following expression
\begin{eqnarray*}
	\alpha&\leqslant & 2 P\left\{\sup_{\bm{c}\in\mathcal{C}} Z >
	\frac{c_{VT}(1-\alpha)}{\nu^{1/2} } \frac{\hat{\sigma}_e\nu^{1/2}}{\sigma_e} \xi_0  -\eta_0 \right\}
	=2 P\left\{\sup_{\bm{c}\in\mathcal{C}} Z >
	\frac{c_{VT}(1-\alpha)}{\nu^{1/2} } \frac{\hat{\sigma}_e\nu^{1/2}}{\sigma_e} \xi_0  -
	\eta_0 \frac{\nu^{1/2} \hat{\sigma}_e \sigma_e }{\nu^{1/2} \sigma_e \hat{\sigma}_e } \right\} \\
	\\ &=&2 \mathrm{P}\left[\sup_{\bm{c}\in\mathcal{C}} Z > \frac{\hat{\sigma}_e\nu^{1/2}}{\sigma_e \nu^{1/2} } 
	\left\{  c_{VT}(1-\alpha)  \xi_0  -
	\eta_0 \frac{\sigma_e }{ \hat{\sigma}_e }  \right\}  \right] .
\end{eqnarray*}
Having calculated all of the necessary terms, we obtain final approximations;\\
%\begin{enumerate}[topsep=0pt]
%	\setlength\itemsep{-0.5em}
%	\item for $p=1$:
for $p=1$:
\begin{eqnarray*}
	\alpha&\leqslant & 2\int_{0}^{\infty} P\left\{\sup_{\bm{c}\in\mathcal{C}} Z >
	\frac{c_{VT}(1-\alpha)\xi_0 }{\nu^{1/2}} w -\eta_0 \right\} \ f(w,\nu)dw+o(\alpha) 
	\\  & \approx &  \int_{0}^{\infty}\frac{\kappa_0}{\pi}\exp\left[ -\frac{1}{2} \left\{ \frac{ c'_{VT}(1-\alpha)w}{\nu^{1/2}} -\eta_0\right\}\right]f(w,\nu) 
	= \frac{\kappa_0}{\pi}\left[\left\{ 1+\frac{c_{VT}(1-\alpha)^{2}\xi_{0}^2}{\nu} \right\}^{-\nu/2}	\right.	\\ & &  + \left.	\eta_0\frac{2^{1/2}c_{VT}(1-\alpha)\xi_{0}\Gamma\left(\frac{\nu+1}{2}\right)} {\nu^{1/2}\Gamma\left(\frac{\nu}{2}\right)} \left\{1+\frac{c_{VT}(1-\alpha)^2\xi_{0}^{2}}{\nu} \right\}^{-(\nu+1)/2}\right];%+\mathcal{E}P(|t_v|\geqslant c_{VT}(1-\alpha))
\end{eqnarray*} 
for $p=2$:
\begin{eqnarray*}
	\alpha&\leqslant & 2\int_{0}^{\infty} P\left\{\sup_{\bm{c}\in\mathcal{C}} Z >
	\frac{c_{VT}(1-\alpha)\xi_0 }{\nu^{1/2}} w -\eta_0 \right\}\
	f(w,\nu)dw+o(\alpha) \\
	&\approx & \int_{0}^{\infty}  \left[
	\frac{\kappa_0 \left\{  c'_{VT}(1-\alpha) w / \nu^{1/2} -\eta_0 \right\} }{2^{1/2}  \pi^{3/2} }   
	\exp\left[ -\frac{1}{2} \left\{\frac{ c'_{VT}(1-\alpha)w}{\nu^{1/2}} -\eta_0\right\}^2\right] \right. \\  
	& &  \left.  + \frac{\zeta_0}{2\pi}\exp\left[ -\frac{1}{2} \left\{ \frac{ c'_{VT}(1-\alpha)w}{\nu^{1/2}} - \eta_0\right\}^2\right]\right]  f(w,\nu)  dw 
	\\	& =& \frac{\kappa_0}{ 2^{1/2} \pi^{3/2} }\left[ \frac{2^{1/2} c_{VT}(1-\alpha)\xi_{0}\Gamma\left(\frac{\nu+1}{2}\right)}	{\nu^{1/2}\Gamma\left(\frac{\nu}{2}\right)} \left\{1+\frac{c_{VT}(1-\alpha)^2\xi_{0}^{2}}{\nu} \right\}^{-(\nu+1)/2}   \right.  - \eta_0\frac{ c_{VT}(1-\alpha)\xi_{0}}{\nu^{1/2}} 		\\ & & \left\{ 1+\frac{c_{VT}(1-\alpha)^{2}\xi_{0}^2}{\nu} \right\}^{-\nu/2} +  \left. \eta_0 \frac{ 2 c_{VT}(1-\alpha)^{2} \xi_{0}^{2}  \Gamma\left(\frac{\nu+2}{2}\right) }{\nu \Gamma\left(\frac{\nu}{2}\right)} \left\{1+\frac{c_{VT}(1-\alpha)^2\xi_{0}^{2}}{\nu} \right\}^{-(\nu+2)/2}   \right]  \\ & & 	+\frac{\zeta_0}{2\pi} \left[  \left\{ 1+\frac{c_{VT}(1-\alpha)^{2}\xi_{0}^2}{\nu} \right\}^{-\nu/2} +\eta_0 \frac{2^{1/2} c_{VT}(1-\alpha)\xi_{0}\Gamma\left(\frac{\nu+1}{2}\right)}{\nu^{1/2}\Gamma\left(\frac{\nu}{2}\right)}  \right]; 
\end{eqnarray*}
%\begin{eqnarray*}
%\alpha	& \leqslant & \frac{\kappa_0}{ 2^{1/2} \pi^{3/2} }\left[ \frac{2^{1/2} c_{VT}(1-\alpha)\xi_{0}\Gamma\left(\frac{\nu+1}{2}\right)}
%{\nu^{1/2}\Gamma\left(\frac{\nu}{2}\right)} \left\{1+\frac{c_{VT}(1-\alpha)^2\xi_{0}^{2}}{\nu} \right\}^{-\frac{\nu+1}{2}}  \right.  \\ & -	& 
%\eta_0\frac{ c_{VT}(1-\alpha)\xi_{0}}{\nu^{1/2}}  \left\{ 1+\frac{c_{VT}(1-\alpha)^{2}\xi_{0}^2}{\nu} \right\}^{-\frac{\nu}{2}} + 
%\left. \eta_0 \frac{ 2 c_{VT}(1-\alpha) \xi_{0}^{2}  \Gamma\left(\frac{\nu+2}{2}\right) }{\nu \Gamma\left(\frac{\nu}{2}\right)} 
%\left\{1+\frac{c_{VT}(1-\alpha)^2\xi_{0}^{2}}{\nu} \right\}^{-\frac{\nu+2}{2}}  \right]    \\ & +&
%\frac{\zeta_0}{2\pi} \left[  \left\{ 1+\frac{c_{VT}(1-\alpha)^{2}\xi_{0}^2}{\nu} \right\}^{-\frac{\nu}{2}} +
%\eta_0 \frac{2^{1/2} c_{VT}(1-\alpha)\xi_{0}\Gamma\left(\frac{\nu+1}{2}\right)}
%{\nu^{1/2}\Gamma\left(\frac{\nu}{2}\right)}  \right] + 2\mathcal{E} P(|t_v|> c_{VT}(1-\alpha) \xi_0);  
%\end{eqnarray*}  
%	\item 
for $p\geqslant3$:.
\begin{eqnarray*}
	\alpha &\leqslant & 2 P\left[\sup_{\bm{c}\in\mathcal{C}} Z > \frac{\hat{\sigma}_e\nu^{1/2}}{\sigma_e\nu^{1/2}} 
	\left\{  c_{VT}(1-\alpha)   \xi_0  -
	\eta_0 \frac{\sigma_e }{  \hat{\sigma}_e }  \right\} \right] 
	\\ & \approx & \frac{\kappa_0 \Gamma ( (p+1)/2 )}{\pi^{(p+1)/2}}\mathrm{P} \left\{ \mathrm{F}_{p+1,\nu} \geqslant  \frac{\left(c_{VT}(1-\alpha) \xi_0 -\eta_{0} \frac{\sigma_e}{\hat{\sigma}_e}\right)^2}{p+1} \right\}\\ & & +  \frac{\zeta_0}{2} \frac{\Gamma ( p/2 )}{\pi^{p/2}}P \left[ \mathrm{F}_{p,\nu} \geqslant 		\frac{\left\{c_{VT}(1-\alpha) \xi_0 -\eta_{0} \frac{\sigma_e}{\hat{\sigma}_e}\right\}^2 }{p} \right] \\ & & + \frac{\kappa_2+\zeta_1+m_0}{2\pi} \frac{\Gamma ( (p-1)/2 )}{\pi^{(p-1)/2}}P \left[ \mathrm{F}_{p-1,\nu} \geqslant 	\frac{\left\{c_{VT}(1-\alpha) \xi_0 -\eta_{0} \frac{\sigma_e}{\hat{\sigma}_e}\right\}^2}{p-1} \right],
\end{eqnarray*}
%\end{enumerate}
%	for $p\geqslant3$
%	\begin{eqnarray*}
%		\alpha & \leqslant & \frac{\kappa_0 \Gamma ( \frac{p+1}{2} )}{\pi^{(p+1)/2}} P \left\{ \mathrm{F}_{p+1,\nu} \geqslant 
%		\frac{\left( c_{VT}(1-\alpha)\xi_0 -\eta_{0} \frac{\sigma_e}{\hat{\sigma}_e} \right)^2}{p+1} \right\} 	+  \frac{\zeta_0}{2} \frac{\Gamma ( \frac{p}{2} )}{\pi^{p/2}} P \left\{ \mathrm{F}_{p,\nu} \geqslant 
%		\frac{\left( c_{VT}(1-\alpha)\xi_0 -\eta_{0} \frac{\sigma_e}{\hat{\sigma}_e} \right)^2 }{p} \right\}   \\  &
%		& + \frac{\kappa_2+\zeta_1+m_0}{2\pi} \frac{\Gamma ( \frac{p-1}{2} )}{\pi^{(p-1)/2}} P \left\{\mathrm{F}_{p-1,\nu} \geqslant 
%		\frac{\left( c_{VT}(1-\alpha)\xi_0 -\eta_{0} \frac{\sigma_e}{\hat{\sigma}_e} \right)^2 }{p-1} \right\}    ;    
%	\end{eqnarray*} 
where $t_{\nu}$ is a t-distributed random variable with $\nu$ degrees of freedom, $F_{d_1,d_2}$ an F-distributed random variable with parameters $d_1$ and $d_2$,
$\kappa_0=\int_{\bm{c}\in\mathcal{C}}^{}\norm{\mathcal{Q}'(\bm{x})}d{\bm{x}}$ the volume of the manifold $\mathcal{M}=\left\{\mathcal{Q}(\bm{c}),\bm{c}\in \mathcal{C}\right\}$, and $\zeta_0$ the boundary area of $\mathcal{M}$. Furthermore, $\kappa_2$ and $\zeta_1$ measure the curvatures of $\mathcal{M}$ and $\partial\mathcal{M}$ respectively, whereas $m_0$ measures the rotation angles of $\partial^2 \mathcal{M}$. Finally, $\mathcal{E}$ is the Euler-Poincar\'{e} characteristic of $\mathcal{M}$. For the manifold $\mathcal{M}$ with the boundary we need to add the boundary's correction terms  $\mathcal{E}P\{|t_v|> c_{VT}(1-\alpha)\}$ and
$2\mathcal{E}P\{|t_v|>c_{VT}(1-\alpha)\}$ for $p=1$ and $p=2$ respectively.

%%%%%%%%%%%%%
\section{Departures from model assumptions}\label{sec:model_assump}

In what follows we first extend simulation results in order to complement the findings regarding departures from normality of the distribution of errors and random effects. Afterwards we discuss potential deviations from the mixed model assumptions, their consequences and remedies. 

Various authors \citep[see, among others, a review paper of][]{mcculloch2011misspecifying} found that the departures from normality of random effects do not have a strong effect. In contrast, the lack of normality of error terms can have a serious impact, for example on the coverage probability of prediction intervals. Nevertheless, the results of the simulation experiment in Table 3 of the main document indicate
that our bootstrap SPI works reasonably well under the departures from normality which
we seem to face in our case study. However, the performance of all methods deteriorates if we deal with more sever departures from normality. Therefore we studied this problem further.
%In particular, under the scenario with fat tails for $e_{dj}$ but (almost) normal $u_d$, the bootstrap intervals converge to the nominal coverage for moderate sample size (recall that we analyse the income distribution in $D=52$ counties in Galicia).
Tables \ref{tab:simulations_t_dist} and \ref{tab:simulations_chi_square} present additional simulation results of the evaluating criteria introduced in Section 6 of the main document. In the simulation experiments shown in the main document in
Table 3 we discovered the same phenomena.  %Similarly as in the main document for Tables 2 to 4, 
In particular, the coverage probability depends not just on the sample size (both, $D$ and $n_d$) but also on ICC and the combination of distributions for $e_{dj}$ and $u_d$. Furthermore, in accordance with the remarks of \cite{mcculloch2011misspecifying}, it seems that our SPI is more affected and exhibits more severe
undercoverage when error terms deviate from normality than when this happens for the random
effects $\bm{u}_d$. As we mentioned in the main document,  \cite{reluga2020thesis} studied the performance of alternative bootstrap procedures which seem to achieve a stronger robustness against non-normalities, but they require several changes of the proposed procedure. Moreover, they would require a different asymptotic theory. Furthermore, parametric bootstrap SPI fails when the distribution of errors and random effects is contaminated by outliers belonging to a different distribution with a large variance.
In fact, this was confirmed by some simulations not shown in this document. In case of contaminated data, outlier robust estimator should be applied; see \cite{Chamb2006,jacqmin2007robustness,SR09}. However, the development of simultaneous inference techniques for robust estimators is beyond the scope of this article.
\begin{table}[htb]
	\centering
	\setlength{\tabcolsep}{3.5pt} 
	\begin{tabular}{|cccccccccc|}
		\hline	
		&    & \multicolumn{4}{c}{ECP}   & \multicolumn{4}{c|}{WS (VS)} \\
		& $D:n_d$  & BS  & MC   & BE   & BO   & BS  & MC   & BE  & BO   \\\hline
		& 15:5 & 93.7 & 87.9 & 77.1 & 92.4 & 2.325 (0.079) & 2.325 (0.079) & 2.145 (0.279) & 2.484 (0.111) \\
		$e_{dj}\sim t_6(1)$    & 30:10 & 92.3 & 90.9 & 87.4 & 92.0 & 1.883 (0.015) & 1.883 (0.015) & 1.800 (0.025) & 1.905 (0.015) \\
		$u_{d}\sim t_6(1)$     & 60:20 & 91.1 & 90.7 & 80.4 & 90.9 & 1.454 (0.002) & 1.454 (0.002) & 1.333 (0.008) & 1.460 (0.002) \\
		& 90:30 & 91.4 & 91.3 & 72.2 & 91.5 & 1.238 (0.001) & 1.238 (0.001) & 1.096 (0.005) & 1.241 (0.001) \\\hline\hline
		& 15:5 & 92.0 & 85.7 & 58.3 & 95.1 & 2.150 (0.082) & 2.150 (0.082) & 1.709 (0.330) & 2.476 (0.145) \\
		$e_{dj}\sim t_6(1)$   & 30:10 & 94.0 & 91.0 & 86.6 & 93.1 & 1.797 (0.011) & 1.797 (0.011) & 1.720 (0.023) & 1.842 (0.012) \\
		$u_{d}\sim N(0.5)$    & 60:20 & 93.5 & 93.7 & 81.6 & 93.8 & 1.420 (0.002) & 1.420 (0.002) & 1.293 (0.008) & 1.429 (0.002) \\
		& 90:30 & 92.8 & 92.9 & 73.3 & 93.0 & 1.217 (0.001) & 1.217 (0.001) & 1.069 (0.005) & 1.222 (0.001) \\\hline\hline
		& 15:5 & 95.0 & 92.0 & 89.6 & 93.4 & 1.752 (0.023) & 1.752 (0.023) & 1.732 (0.048) & 1.795 (0.023) \\
		$e_{dj}\sim N(0.5)$   & 30:10 & 95.2 & 94.9 & 89.3 & 95.1 & 1.367 (0.003) & 1.367 (0.003) & 1.287 (0.008) & 1.376 (0.003) \\
		$u_{d}\sim t_6(1)$    & 60:20 & 94.1 & 94.0 & 82.5 & 94.4 & 1.042 (0.000) & 1.042 (0.000) & 0.945 (0.004) & 1.044 (0.000) \\
		& 90:30 & 95.4 & 95.3 & 76.6 & 95.5 & 0.883 (0.000) & 0.883 (0.000) & 0.774 (0.002) & 0.885 (0.001) \\\hline\hline
		& 15:5 & 92.5 & 88.7 & 86.2 & 90.6 & 1.740 (0.043) & 1.740 (0.043) & 1.726 (0.071) & 1.786 (0.048) \\
		$e_{dj}\sim t_6(0.5)$ & 30:10 & 93.5 & 92.4 & 86.8 & 92.8 & 1.362 (0.007) & 1.362 (0.007) & 1.295 (0.011) & 1.371 (0.007) \\
		$u_{d}\sim t_6(1)$   & 60:20 & 92.4 & 92.8 & 82.4 & 93.0 & 1.040 (0.001) & 1.040 (0.001) & 0.952 (0.004) & 1.043 (0.001) \\
		& 90:30 & 92.1 & 91.9 & 70.9 & 92.2 & 0.882 (0.000) & 0.882 (0.000) & 0.780 (0.003) & 0.884 (0.000)\\\hline
	\end{tabular}
	\caption{ECP (in \%), WS and VS under NERM with t-distributed errors and random effects. The nominal coverage probability is $95\%$.}
	\label{tab:simulations_t_dist}
\end{table}

\begin{table}[htb]
	\centering
	\setlength{\tabcolsep}{3.5pt} 
	\begin{tabular}{|cccccccccc|}
		\hline	
		&    & \multicolumn{4}{c}{ECP}   & \multicolumn{4}{c|}{WS (VS)} \\
		& $D:n_d$  & BS  & MC   & BE   & BO   & BS  & MC   & BE  & BO   \\\hline
		& 15:5 & 94.0 & 90.6 & 87.7 & 91.9 & 2.452 (0.088) & 2.452 (0.088) & 2.438 (0.148) & 2.516 (0.099) \\
		$e_{dj}\sim \chi_5(1)$ & 30:10 & 90.9 & 90.1 & 85.2 & 90.3 & 1.934 (0.014) & 1.934 (0.014) & 1.839 (0.024) & 1.946 (0.014) \\
		$u_{d}\sim \chi_5(1)$  & 60:20 & 91.2 & 91.0 & 80.0 & 91.1 & 1.474 (0.002) & 1.474 (0.002) & 1.351 (0.009) & 1.478 (0.002) \\
		& 90:30 & 92.2 & 91.9 & 75.1 & 92.0 & 1.248 (0.001) & 1.248 (0.001) & 1.105 (0.005) & 1.251 (0.001)\\\hline\hline
		& 15:5 & 91.4 & 85.8 & 59.7 & 94.2 & 2.148 (0.092) & 2.147 (0.092) & 1.717 (0.360) & 2.456 (0.121) \\
		$e_{dj}\sim \chi_5(1)$ & 30:10 & 92.0   & 89.3 & 83.7 & 91.2 & 1.803 (0.011) & 1.803 (0.011) & 1.725 (0.023) & 1.846 (0.011) \\
		$u_{d}\sim N(0.5)$     & 60:20 & 92.3 & 91.7 & 81.3 & 92.3 & 1.419 (0.002) & 1.419 (0.002) & 1.293 (0.008) & 1.428 (0.002) \\
		& 90:30 & 91.9 & 92.3 & 73.6 & 92.5 & 1.218 (0.001) & 1.218 (0.001) & 1.069 (0.005) & 1.222 (0.001) \\\hline\hline
		& 15:5 & 95.2 & 93.1 & 92.0 & 93.5 & 1.800 (0.024) & 1.800 (0.024) & 1.785 (0.035) & 1.823 (0.025) \\
		$e_{dj}\sim N(0.5)$    & 30:10 & 95.0 & 94.1 & 89.1 & 94.5 & 1.386 (0.003) & 1.386 (0.003) & 1.303 (0.008) & 1.392 (0.003) \\
		$u_{d}\sim \chi_5(1)$  & 60:20 & 94.1 & 94.0 & 82.5 & 94.4 & 1.048 (0.000) & 1.048 (0.000) & 0.951 (0.004) & 1.051 (0.000) \\
		& 90:30 & 95.4 & 95.3 & 76.6 & 95.5 & 0.886 (0.000) & 0.886 (0.000) & 0.777 (0.002) & 0.888 (0.000)\\\hline\hline& 15:5 & 93.5 & 91.1 & 90.4 & 91.5 & 1.787 (0.049) & 1.787 (0.049) & 1.787 (0.066) & 1.809 (0.052) \\
		$e_{dj}\sim \chi_5(0.5)$  & 30:10 & 90.7 & 90.0 & 85.4 & 90.2 & 1.383 (0.007) & 1.383 (0.007) & 1.312 (0.012) & 1.388 (0.007) \\
		$u_{d}\sim \chi_5(1)$   & 60:20 & 90.7 & 90.7 & 79.4 & 90.9 & 1.049 (0.001) & 1.049 (0.001) & 0.960 (0.004) & 1.051 (0.001) \\
		& 90:30 & 91.9 & 92.4 & 74.3 & 92.6 & 0.886 (0.000) & 0.886 (0.000) & 0.785 (0.003) & 0.888 (0.000)\\\hline
	\end{tabular}
	\caption{ECP (in \%), WS and VS under NERM with chi-square distributed errors and random effects. The nominal coverage probability is $95\%$.}
	\label{tab:simulations_chi_square}
\end{table}

Many other violations of our model assumptions are possible. The proposed methods for constructing simultaneous prediction intervals and tests are model based. In fact, this characteristic is widely shared by inferential methods in many fields and is not particular to the bootstrap based technique which we developed. Nonetheless it is worth to discuss the consequences of model violations. Apart from the most studied departures from distributional assumptions, the second interesting deviation might be the  existence of nonlinearities and interactions.

First, \cite{reluga2021} consider an extension to a nonlinear link function and generalised linear mixed
model. Assuming this modelling framework, the authors did not use EBLUP, but examined the empirical best predictor \citep{jiang2001empirical}. More sophisticated nonlinearities, potentially together with interactions, are in the focus of interest when switching to non- and semiparametric mixed models,  see, for example,  \cite{lombardiasperlich2008} and \cite{opsomer2008splines} for estimation with kernels and splines, respectively. Moreover,  \cite{gonzalez2013bootstrap} discussed in detail the application of bootstrap in samiparametric mixed effects models, whereas \cite{sperlichlombardia2010} investigated nonparametric specification testing for small area statistics. One could certainly imagine an extension of our methods to those models, but both, the statistical properties as well as the computational challenges would be more involved. Yet, we consider these to be interesting subjects for future research.

We should also mention model specifications which deal with potential dependencies between the random components, or dependencies between covariates and area effects. In the context of the former, we can think about models with temporal and/or spatial correlation, see \cite{singh2005spatial} and \cite{moura2002spatial}, whereas the latter is examined by \cite{lombardia2012new}. They propose a semiparametric device to control for potential correlation between covariates and cluster effects. Parametric counterparts to their semiparametric device fall into the class of models we consider in our main document.

Finally, the question of model (or variable) selection in mixed models is a related aspect to
consider once extending the scope of our theory \citep[see][for an extensive review of model selection techniques]{muller2013model}. Nevertheless, due to the additional variability coming from the
selection process, one would need to first theoretically investigate a post-selection distribution
of mixed effects in order to develop statistically valid simultaneous intervals and tests  \citep[see, for example][for a valid post-selection inference for fixed parameters]{charkhi2018asymptotic}.

\bibliography{cite}

\end{document}